\documentclass[sigconf, nonacm]{acmart}

\usepackage{makecell}
\usepackage{multirow}
\usepackage{hhline}
\usepackage{caption}
\usepackage{subcaption,graphicx}
\usepackage{xparse}
\usepackage{array}
\usepackage{enumitem}
\usepackage{balance}

\RequirePackage[normalem]{ulem} 
\RequirePackage{color}\definecolor{RED}{rgb}{1,0,0}\definecolor{BLUE}{rgb}{0,0,1} 

\usepackage{tikz}
\newcommand\encircle[1]{%
  \tikz[baseline=(X.base)] 
    \node (X) [draw, shape=circle, inner sep=0, fill=white, text=black] {\strut #1};%
}

\usepackage{xcolor}

\newcommand{\scream}[1]{{\color{red} \bf *** #1 ***}{\typeout{#1}}}
\newcommand{\eat}[1]{}
\newcommand{\infl}{INFL}
\newcommand{\inflg}{INFL-Y}
\newcommand{\inflo}{INFL-D}
\newcommand{\duti}{DUTI}

\newcommand{\diffone}{\text{Diff}_1}
\newcommand{\difftwo}{\text{Diff}_2}

\newcommand{\deltagrad}{DeltaGrad}
\newcommand{\deltagradl}{DeltaGrad-L}
\newcommand{\deltagradg}{Increm-INFL}

\newcommand{\mimicfull}{MIMIC-CXR-JPG}
\newcommand{\mimic}{MIMIC}
\newcommand{\retinafull}{Diabetic Retinopathy Detection}
\newcommand{\retina}{Retina}
\newcommand{\chexpert}{Chexpert}
\newcommand{\fashion}{Fashion}
\newcommand{\fashionfull}{Fashion 10000}
\newcommand{\twitter}{Twitter}
\newcommand{\twitterfull}{Twitter sentiment analysis}
\newcommand{\fact}{Fact}
\newcommand{\factfull}{Fact Evaluation Judgement}
\newcommand{\full}{Full}
\newcommand{\retrain}{Retrain}

\newcommand{\tars}{TARS}
\newcommand{\ou}{O2U}
\newcommand{\ac}{Active (one)}
\newcommand{\actwo}{Active (two)}

\newcommand{\cleanset}{\textit{Fully clean datasets}}
\newcommand{\crowdset}{\textit{Crowdsourced datasets}}

\newcommand{\cleanone}{{\infl\ (one)}}
\newcommand{\cleantwo}{{\infl\ (two)}}
\newcommand{\cleanthree}{{\infl\ (three)}}

\newcommand{\iw}{\textbf{w}^{I}}
\newcommand{\uw}{{\textbf{w}^{U}}}
\newcommand{\B}{\textbf{B}}

\newcommand{\w}{\textbf{w}}
\newcommand{\x}{\textbf{x}}
\newcommand{\y}{y}
\newcommand{\nx}{\tilde{\textbf{x}}}
\newcommand{\ny}{\tilde{y}}
\newcommand{\cleanz}{\mathcal{Z}_d}
\newcommand{\dirtyz}{\mathcal{Z}_p}
\newcommand{\z}{\textbf{z}}
\newcommand{\nz}{\tilde{\textbf{z}}}

\newcommand{\testz}{\textbf{z}_{\text{test}}}

\newcommand{\method}{CHEF}

\newcommand{\sgd}{SGD}
\newcommand{\miniB}{\mathscr{B}}
\newcommand{\bH}{\textbf{H}}
\usepackage{algorithm}
\usepackage{algorithmic}
\usepackage{eqparbox}

\usepackage{graphicx}
\newsavebox\CBox
\def\textBF#1{\sbox\CBox{#1}\resizebox{\wd\CBox}{\ht\CBox}{\textbf{#1}}}

\usepackage{amsmath}
\usepackage{setspace}

\newtheorem{theorem}{Theorem}

\newcommand\vldbdoi{10.14778/3476249.3476290}
\newcommand\vldbpages{XXX-XXX}
\newcommand\vldbvolume{14}
\newcommand\vldbissue{11}
\newcommand\vldbyear{2021}
\newcommand\vldbauthors{\authors}
\newcommand\vldbtitle{\shorttitle} 
\newcommand\vldbavailabilityurl{https://github.com/thuwuyinjun/Chef}
\newcommand\vldbpagestyle{empty} 

\usepackage{lipsum}

\makeatletter
\renewcommand\paragraph{\@startsection{paragraph}{4}{\parindent}%
  {0pt}
  {-\parindent}
  {\ACM@NRadjust{\@parfont\@adddotafter}}}
\makeatother

\begin{document}
\title{\method: A Cheap and Fast Pipeline for Iteratively Cleaning Label Uncertainties}
\author{Yinjun Wu}
\affiliation{%
  \institution{University of Pennsylvania}
}
\email{wuyinjun@seas.upenn.edu}

\author{James Weimer}
\affiliation{University of Pennsylvania}
\email{weimerj@seas.upenn.edu}

\author{Susan B. Davidson}
\affiliation{University of Pennsylvania}
\email{susan@cis.upenn.edu}




\begin{abstract}
High-quality labels are expensive to obtain for many machine learning tasks, such as medical image classification tasks.  Therefore, probabilistic (weak) labels produced by weak supervision tools are used to seed a process in which influential samples with weak labels are identified and cleaned by several human annotators to improve the model performance. To lower the overall cost and computational overhead of this process, we propose a solution called \method\ (CHEap and Fast label cleaning), which consists of the following three components. 
First, to reduce the cost of human annotators, we use \infl, which prioritizes the {\em most influential} training samples for cleaning and provides cleaned labels to save the cost of one human annotator. Second, to accelerate the sample selector phase and the model constructor phase, we use
\deltagradg\ to {\em incrementally} produce influential samples, and \deltagradl\ to {\em incrementally} update the model. Third, we redesign the typical label cleaning pipeline so that human annotators iteratively clean smaller batch of samples rather than one big batch of samples.
This yields better overall model performance and enables possible early termination when the expected model performance has been achieved. 
Extensive experiments show that our approach gives good model prediction performance while achieving significant speed-ups.

\end{abstract}

\maketitle

\pagestyle{\vldbpagestyle}
\begingroup\small\noindent\raggedright\textbf{PVLDB Reference Format:}\\
\vldbauthors. \vldbtitle. PVLDB, \vldbvolume(\vldbissue): \vldbpages, \vldbyear.\\
\href{https://doi.org/\vldbdoi}{doi:\vldbdoi}
\endgroup
\begingroup
\renewcommand\thefootnote{}\footnote{\noindent
This work is licensed under the Creative Commons BY-NC-ND 4.0 International License. Visit \url{https://creativecommons.org/licenses/by-nc-nd/4.0/} to view a copy of this license. For any use beyond those covered by this license, obtain permission by emailing \href{mailto:info@vldb.org}{info@vldb.org}. Copyright is held by the owner/author(s). Publication rights licensed to the VLDB Endowment. \\
\raggedright Proceedings of the VLDB Endowment, Vol. \vldbvolume, No. \vldbissue\ %
ISSN 2150-8097. \\
\href{https://doi.org/\vldbdoi}{doi:\vldbdoi} \\
}\addtocounter{footnote}{-1}\endgroup

\ifdefempty{\vldbavailabilityurl}{}{
\vspace{.3cm}
\begingroup\small\noindent\raggedright\textbf{PVLDB Artifact Availability:}\\
The source code, data, and/or other artifacts have been made available at \url{\vldbavailabilityurl}.
\endgroup
}

\section{Introduction}\label{sec: intro}
There is a general consensus that the success of advanced machine learning models depends on the availability of extremely large training sets with high-quality labels.  Unfortunately, obtaining high-quality labels may be prohibitively expensive.
For example, labeling medical images typically requires the effort of experts with domain knowledge.
To produce labels at large scale with low cost,
weak supervision tools\textemdash such as 
Snorkel \cite{ratner2017snorkel}\textemdash can be used to automatically generate {\em probabilistic labels} (or {\em weak labels}) for unlabeled training samples by leveraging labeling functions \cite{ratner2017snorkel}.  

It has been shown in \cite{smyth2020training, bach2019snorkel, nashaat2020wesal}, however, that imperfect labeling functions can produce inferior probabilistic labels,
thus hurting the downstream model quality.
Therefore, it is necessary to perform additional {\em cleaning operations} to clean such label uncertainties \cite{nashaat2020wesal}. 

\begin{figure}[!t]
 \includegraphics[width=1.0\columnwidth, height=0.5\columnwidth]{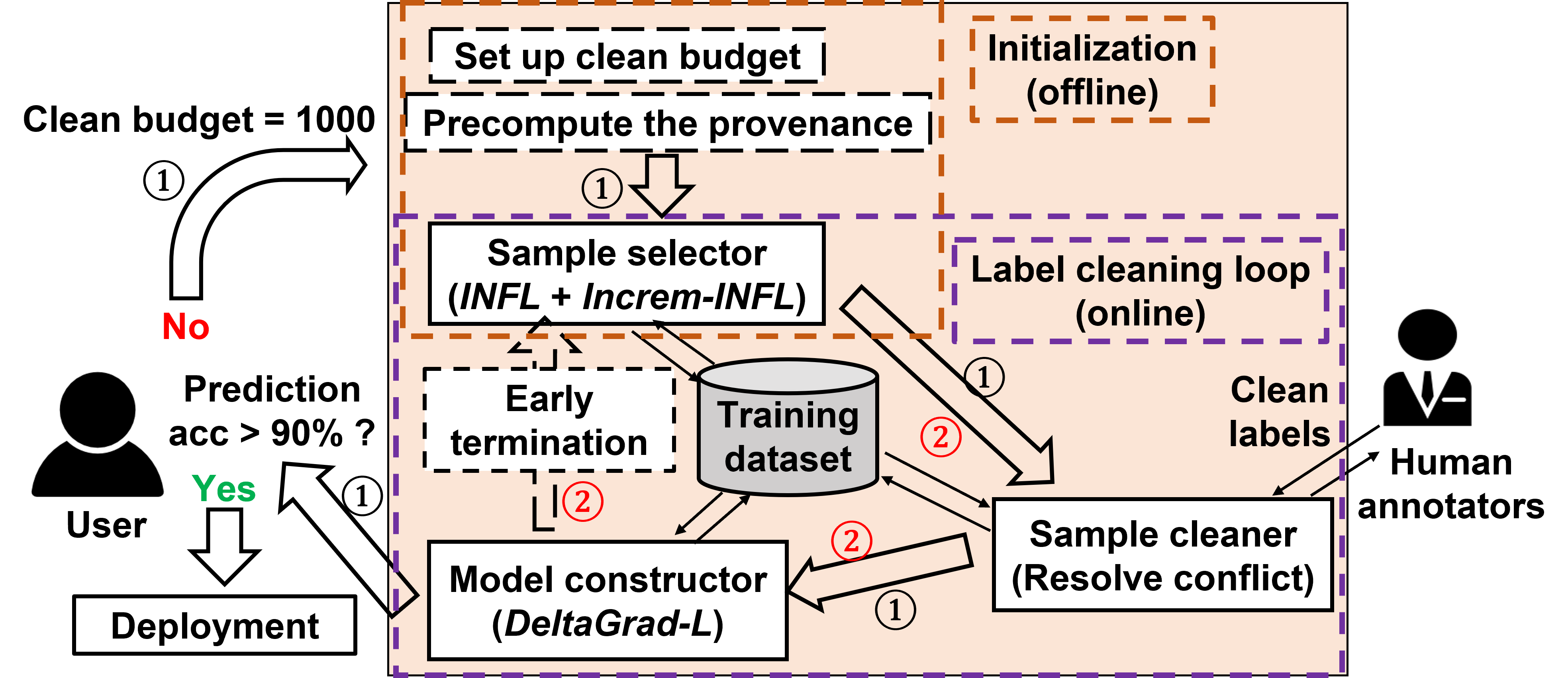}
 \caption{The iterative pipeline of cleaning uncertainties from the labels of training set. }\label{fig:pipeline}
\end{figure}

The label cleaning process is typically {\em iterative} \cite{mahdavi2019towards, krishnan2016activeclean}, and requires multiple rounds (see Figure \ref{fig:pipeline}, loop labeled \encircle{1}).
First, given a {\em cleaning budget} $B$, the top-$B$ influential training samples with probabilistic labels are selected (the {\em sample selector phase}).  Second, for those selected samples, cleaned labels are provided by human annotators (the {\em annotation phase}). Third, the ML model is calculated using the updated training set (the {\em model constructor phase}), and returned to the user.  If the resulting model performance is not good enough, the process is repeated with an additional budget $B'$. 
Otherwise, it is deployed.
Note that since each of these phases may be performed {\em repeatedly}, it is important that they be as efficient as possible. It is also noteworthy that for some applications---such as the medical image classification task---it is essential to have multiple human annotators for label cleaning to alleviate their labeling errors \cite{irvin2019chexpert} in the {\em annotation phase}, thus incurring substantial time overhead and financial cost.
{\bf In this paper, we propose a solution  called \method\ (CHEap and Fast label cleaning), to reduce the time overhead and cost of the label cleaning pipeline and 
simultaneously enhance the overall model performance.
}
Details of the overall design of \method\ are given next.

\paragraph{Sample selector phase.}
Finding the most influential training samples can be done with several different influence measures, e.g.,
the influence function \cite{koh2017understanding}, the Data Shapley values \cite{jia2019towards}, the noisy label detection algorithms \cite{huang2019o2u, dolatshah12cleaning}, the active learning technique \cite{settles2009active} or using a bi-level optimization solution \cite{zhang2018training}. 
Unfortunately, these 
do not work well for cleaning weak labels.
We therefore develop a variant of the influence function called \infl\ which can simultaneously detect the most influential samples and suggest cleaned labels.
\eat{However to our knowledge, apart from the influence function of \cite{zhang2018training} (which is not efficient enough for our purposes), most existing research on sample selection focuses on how to {\em detect} the most influential training samples rather than how to suggest possibly cleaned labels. In this paper, we proposed a variant of influence function called \infl\ to cover those two aspects simultaneously. }
One key technical challenge in the efficient implementation of \infl\ concerns the explicit evaluation of gradients on {\em every} training sample. 
{\bf We address this challenge by developing 
\deltagradg, which removes uninfluential training samples early and can thus {\em incrementally} recommend the most influential training samples to human annotators.}

\paragraph{Human annotation phase}
After influential samples are selected, 
the next step is for human annotators to clean the labels of those samples. 
Recall that 
{\em multiple} human annotators may be used to independently label each training sample, and inconsistencies between the labels are resolved, e.g., by majority vote \cite{irvin2019chexpert}. \textbf{To reduce the cost of the human annotation phase, we consider the suggested clean labels 
from the {\em sample selector phase} as one alternative labeler, 
which can be combined with results provided by the human annotators to reduce annotation cost.}

\paragraph{Model constructor phase.} 
In previous work \cite{wu2020deltagrad}, we developed a provenance-based algorithm called  \deltagrad\ for {\em incrementally updating model parameters} after the deletion or addition of a small subset of training samples, and showed that it was significantly faster than recalculating the model from scratch.
Since the result of the human annotation phase can be regarded as the deletion of top-$B$ samples with probabilistic labels, and insertion of those same samples with cleaned labels, we can adapt \deltagrad\ for this setting.  This algorithm is called \deltagradl.
\textbf{To accelerate the model constructor phase, rather than retraining from scratch after cleaning the labels of a small set of training samples, we {\em incrementally} update the model using \deltagradl.}  

\paragraph{Redesign of the cleaning pipeline} The final contribution of this paper, which is enabled by the reduced cost of the sample selection, human annotation, and model construction phases, is a {re-design of the pipeline in Figure \ref{fig:pipeline} (see the loop \encircle{2}).}
Rather than providing all top-$B$ influential training samples (and suggesting how to fix the label uncertainty) at once, the sample selector gives the human annotator the next top-$b$ influential training samples, where $b$ is 
smaller than $B$ and is specified by the user. The model is then {\em refreshed} using the cleaned labels, and the next top-$b$ samples to be given to the human annotator are calculated.  This continues until the initial budget $B$ has been exhausted or the expected prediction performance is reached (thus terminating early). 
\textbf{This can not only improve the overall model performance, but also lead to early termination,
thus further saving the cost of human annotation.} Note that to enable incremental computation by \deltagradg\ and \deltagradl, some ``provenance'' information is necessary, and can be pre-computed offline in an {\em Initialization step} prior to the start of loop \encircle{2}.
\eat{\textbf{Our experiments show that processing the label-cleaning budget in smaller batches rather than doing the label cleaning in one large batch can lead to significantly higher prediction accuracy.}}

We demonstrate the effectiveness of \method\ using several crowd-sourced datasets as well as real medical image datasets. Our experiments show that \method\ achieves up to 54.7x speed-up in the sample selector phase, and up to 7.5x speed-up in the model constructor phase. Furthermore, by using 
\infl\ and smaller batch sizes $b$, the overall model quality can be improved. 

Summarizing, the contributions of this paper include:
\begin{itemize}
\item A solution called \method\ which can significantly reduce the overall cost of label cleaning by 1) reducing the cost of 
the Sample selector phase, the Human annotation phase and the Model constructor phase
respectively and 2) redesigning the label cleaning pipeline to enable better model performance and early stopping in the human annotation phase. 
\item Extensive experiments 
which show the effectiveness of \method\ 
on real crowd-sourced datasets and medical image datasets.  
 \end{itemize}

The rest of this paper is organized as follows. In Section \ref{sec: related_work}, we summarize related work.  Preliminary notation, definitions and  assumptions are given in Section \ref{sec: prelim}, followed by our algorithms, \infl, \deltagradg\ and \deltagradl\ in Section \ref{sec: method}. Experimental results are discussed in Section \ref{sec: exp}, and we conclude in Section \ref{sec: conclusion}.

\section{Related work}\label{sec: related_work}

\textbf{Incremental updates on ML models} In the past few years, several approaches for incrementally maintaining different types of models have emerged \cite{ginart2019making,koh2017understanding,wu2020deltagrad,brophy2020dart,wu2020priu}, which address important practical problems such as GPDR \cite{ryz2016new} and training sample valuation \cite{ghorbani2019data}.  The \deltagradl\ algorithm in the model constructor phase is adapted from our \deltagrad\ algorithm \cite{wu2020deltagrad}, which addresses the problem of incrementally updating strongly convex models after a small subset of training samples are deleted or added. 
Note that this problem is related to the classical {\em materialized view maintenance problem} as mentioned in \cite{wu2020priu}, if we consider ML models as {\em views}. 

\textbf{Data cleaning for ML models} Diagnosing and cleaning errors or noises in training samples has attracted considerable attention \cite{huang2019o2u,dolatshah12cleaning}, and is typically addressed iteratively \cite{mahdavi2019towards,aguilar2021ease,krishnan2016activeclean}. For example, the authors of \cite{huang2019o2u} observed that the noisily labeled samples were memorized by the model in the overfitting phase, which can be detected through transferring the model status back to the underfitting phase. 
\cite{dolatshah12cleaning} identifies and fixes the noisy labels through jointly analyzing how probable one noisy label is flipped by the human annotators and how this label update influences the model performance. However, it explicitly assumes that the noisy labels are either 1 or 0, thus not applicable in the presence of probabilistic labels. 
The approach in \cite{krishnan2016activeclean} detects errors in both feature values and labels; 
\eat{in contrast, in this paper we assume that the feature values of training samples are clean but that the labels encode uncertainties. 
Unlike \cite{krishnan2016activeclean} }
But it explicitly assumes that the uncleaned samples are harmful
and thus excluded in the training process, we follow the principle of  \cite{ratner2017snorkel} by ``including'' the training samples with uncertain labels in the training phase. 

\textbf{Detecting the most influential training samples with uncertainties} As discussed in \cite{aguilar2021ease}, it is important to prioritize the most influential training samples for cleaning. This can depend on various influence measures, e.g.,
the uncertainty-based measures in active learning \cite{settles2009active}, the influence function \cite{koh2017understanding}, the data shapley value \cite{jia2019towards}, the training loss 
\cite{huang2019o2u, han2018co}, etc. However, to our knowledge, none of these techniques can be used to automatically suggest possibly cleaned labels, 
apart from \cite{zhang2018training}.  Furthermore, the applicability of \cite{zhang2018training} is limited due to its poor scalability and some of the above methods (including \cite{zhang2018training}) are not applicable in the presence of probabilistic labels
and the regularization on them.

\section{Preliminaries}\label{sec: prelim}
In this section, we introduce essential notation and  assumptions, and then describe the influence function and \deltagrad.

\subsection{Notation}\label{sec: notation}
A $C$-class classification task is a classification task in which the number of classes is $C$.  Suppose that the goal is to construct a machine learning model on a training set, $\mathcal{Z} = \cleanz \bigcup \dirtyz$, in which $\cleanz = \{\z_i\}_{i=1}^{N_d} = \{(\x_i, \y_i)\}_{i=1}^{N_d}$ and $\dirtyz = \{\nz_i\}_{i=1}^{N_p} = \{(\nx_i, \ny_i)\}_{i=1}^{N_p}$, denoting a set of $N_d$ training samples with deterministic labels and $N_p$ training samples with probabilistic labels, respectively.
A {\em probabilistic label}, $\ny_i$, is represented by a probabilistic vector of length $C$, in which the value in the $c_{th}$ entry ($c=1,2,\dots, C$) denotes the probability that $\nz_i$ belongs to the class $c$. The performance of the model constructed on $\mathcal{Z}$ is then validated on a validation dataset $\mathcal{Z}_{\text{val}}$ and tested on a test dataset $\mathcal{Z}_{\text{test}}$. Note that the size of $\mathcal{Z}_{\text{val}}$ and $\mathcal{Z}_{\text{test}}$ are typically small, consisting of samples with ground-truth labels or deterministic labels verified by the human annotators.
Due to the possibly negative effect brought by the uncleaned training samples with probabilistic labels, it is reasonable to regularize those samples
in the following objective function (e.g. see \cite{sukhbaatar2014learning}):
\begin{align}\label{eq: obj_function}
\small
\begin{split}
F\left(\w\right) = \frac{1}{N}[\sum\nolimits_{i=1}^{N_d} F\left(\w, \z_i\right) + \sum\nolimits_{i=1}^{N_p} \gamma F\left(\w, \nz_i\right)]
\end{split}
\end{align}

In the formula above, we use $\w$ to represent the model parameter, $F(\w,\z)$ to denote the loss incurred on a sample $\z$ with the model parameter $\w$ and $\gamma$ ($0 < \gamma < 1$, specified by users) to denote the weight on the uncleaned training samples. Furthermore, the first order gradient of this loss can be denoted by $\nabla_\w F\left(\w,\z\right)$, and the second order gradient (i.e. the Hessian matrix) by $\bH(\w, \z)$.
We further use $\nabla_\w F\left(\w\right)$ and $\bH(\w)$ to denote the first order gradient and the Hessian matrix averaged over all weighted training samples.

To optimize 
Equation \eqref{eq: obj_function}, 
stochastic Gradient Descent (\sgd) can be applied.
At each \sgd\ iteration $t$, one essential step is to evaluate the first-order gradients of a randomly sampled mini-batch of training samples, $\miniB_t$ (we denote the size of $\miniB_t$ as $|\miniB_t|$), i.e.:
\begin{align*}
\small
    \begin{split}
\nabla_\w F \left(\w,\miniB_{t}\right) = \frac{1}{|\miniB_t|}\sum\nolimits_{\z \in \miniB_{t}} \gamma_{\z} \nabla_\w F\left(\w,\z\right),
    \end{split}
\end{align*}
in which $\gamma_z$ is 1 if $z\in \cleanz$ and $\gamma$ otherwise. 

Plus, 
since loop \encircle{2} in Figure \ref{fig:pipeline} may be repeated for multiple rounds, 
we use $\mathcal{Z}^{(k)}$ to denote the updated training dataset after $k$ rounds and $\w^{(k)}$ to represent the model constructed on $\mathcal{Z}^{(k)}$. \eat{During the model training phase on $\mathcal{Z}^{(k)}$, we further use $\w^{(k)}_t$ to denote the model parameter at the $t_{th}$ SGD iteration. }



\subsection{Assumptions}\label{sec: assumption}
We make two assumptions: the {\em strong convexity assumption}, 
and the {\em small cleaning budget assumption}.

\textbf{Strong convexity assumption} 
Following \cite{wu2020deltagrad}, we focus on the models satisfying {\em $\mu-$strong convexity}, meaning that the minimal eigenvalue of each Hessian matrix $\bH(\w, \z)$ is always greater than a non-negative constant $\mu$ for arbitrary $\w$ and $\z$. 
\eat{One such model satisfying this property is the logistic regression model with L2 regularization.}
One such model is the logistic regression model with L2 regularization.

\textbf{Small cleaning budget assumption} Since manually cleaning labels   
is time-consuming and expensive, we assume that {\em the cleaning budget $B$ is far smaller than the size of training set, $\mathcal{Z}$.}


\subsection{Influence function}
The influence function method \cite{koh2017understanding} is originally proposed to
estimate how the prediction performance on one test sample $\testz$ is varied if we delete one training sample $\z$, or add an {\em infinitely small} perturbation on the feature of $\z$.
This is formulated as follows:
\begin{align*}
\small
\begin{split}
        &\mathcal{I}_{\text{del}}(\z) = -\nabla_\w F(\w,\testz)^\top \bH^{-1}(\w) \nabla_\w F(\w,\z)\\
    &\mathcal{I}_{\text{pert}}(\z) = -\nabla_\w F(\w,\testz)^\top \bH^{-1}(\w) \nabla_{\x}\nabla_\w F(\w,\z).
\end{split}
\end{align*}

We can then leverage $\mathcal{I}_{\text{del}}(\z)$ and $\mathcal{I}_{\text{pert}}(\z)\delta$ to approximate the additional errors incurred on the test sample $\testz$ after deleting the training sample $\z$, or perturbing the feature of $\z$ by $\delta$.

As \cite{koh2017understanding} indicates, by evaluating the training sample influence with the above influence function,
the ``harmful'' training samples on the model prediction (i.e. the one with negative influence) can be distinguished from the ``helpful'' ones (i.e. the one with positive influence). We can then prioritize the most ``harmful'' training samples with probabilistic labels for cleaning.
In practice, due to the invisibility of the test samples in most cases, the validation set is used instead, leading to the following modified influence functions:
\begin{small}
\begin{align}
    &\mathcal{I}_{\text{del}}(\z) = -\nabla_\w F(\w,\mathcal{Z}_{\text{val}})^\top \bH^{-1}(\w) \nabla_\w F(\w,\z) \label{eq: influence function}\\
    & \mathcal{I}_{\text{pert}}(\z) = -\nabla_\w F(\w,\mathcal{Z}_{\text{val}})^\top \bH^{-1}(\w) \nabla_{\x}\nabla_\w F(\w,\z)\label{eq: influence function pert}
\end{align}
\end{small}
The two formulas above also follow the modified influence function in \cite{zhang2018training} which uses a set of trusted validation samples instead of test samples to estimate the influence of each training sample.

\subsection{\deltagrad}
As introduced in \cite{wu2020deltagrad}, \deltagrad\ is used to incrementally update the parameters of a {\em strongly convex model} after the removal of a small subset of training samples, $\mathcal{R}$ ($|\mathcal{R}| \ll N$), and the addition of another small subset of training samples, $\mathcal{A}$ ($|\mathcal{A}| \ll N$), on the training dataset $\mathcal{Z}$; both $\mathcal{R}$ and $\mathcal{A}$ can be empty.
Before the above modifications on the training dataset $\mathcal{Z}$, suppose we derive the gradients on a randomly sampled mini-batch $\miniB_t$ and calculate the model parameter, $\w_t$, at the $t_{th}$ SGD iteration, 
Then after $\mathcal{R}$ is deleted and $\mathcal{A}$ is added, to obtain the updated model parameter $\iw_t$ at the $t_{th}$ SGD iteration, it is essential to evaluate the gradients on the following updated mini-batch $\miniB_t'$, i.e.,
$(\miniB_t - \mathcal{R}) \cup \mathcal{A}_{t}$.
Here,
$\miniB_t - \mathcal{R}$ represents the remaining training samples in $\miniB_t$ after $\mathcal{R}$ is deleted, while $\mathcal{A}_{t}$ denotes a randomly sampled mini-batch from 
$\mathcal{A}$. Note that $\miniB_t'$ can be further rewritten as $(\miniB_t - (\miniB_t\bigcap\mathcal{R})) \cup \mathcal{A}_{t}$.
As a result, the gradient on $\miniB_t'$ can be evaluated as follows:
\begin{align}\label{eq: updated_gradient}
\small
\begin{split}
    & \nabla_\w F (\iw_t,\miniB_t') = \frac{1}{|\miniB_t'|} \left[|\miniB_{t}| \nabla_\w F (\iw_t,\miniB_{t})\right.\\
    &\left. - |\miniB_{t} \cap \mathcal{R}| \nabla_\w F (\iw_t,\miniB_{t}\cap \mathcal{R}) + |\mathcal{A}_t|\nabla_\w F (\iw_t,\mathcal{A}_t)\right],
\end{split}
\end{align}

The latter two gradients in the above formula, $\nabla_\w F (\iw_t,\mathcal{A}_t)$ and $\nabla_\w F (\iw_t,\miniB_{t}\cap \mathcal{R})$, can be efficiently calculated due to the small size of $\mathcal{R}$ and $\mathcal{A}$. 
As a result, computing $\nabla_\w F (\iw_t,\miniB_{t})$ becomes the dominant overhead in evaluating Equation \eqref{eq: updated_gradient} when the mini-batch size is large.
Hence, \deltagrad\ aims to reduce the overhead of 
this term by incrementally computing it using the Cauchy-mean value theorem \cite{leach1978extended} with the approximate Hessian matrix, $\B_t$, as follows: 
\begin{align}\label{eq: sgd_lbfs}
\small
\begin{split}
    \nabla_\w F (\iw_t,\miniB_{t}) \approx \B_t(\iw_t - \w_t) + \nabla_\w F (\w_t,\miniB_{t}).
\end{split}
\end{align}
in which, the product 
$\B_t(\iw_t - \w_t)$ is calculated using the L-BFGS algorithm \cite{nocedal1980updating} while the gradient term $\nabla_\w F \left(\w_t,\miniB_{t}\right)$ is cached during the training phase on the original training dataset $\mathcal{Z}$. 

As described in \cite{wu2020deltagrad}, although this approximation is faster than computing $\nabla_\w F (\iw_t,\miniB_{t})$ explicitly, the approximation errors are not negligible. To balance between the approximation error and efficiency in \deltagrad, $\nabla_\w F (\iw_t,\miniB_{t})$ is explicitly evaluated in the first $j_0$ SGD iterations 
and every $T_0$ SGD iterations afterwards, where $T_0$ and $j_0$ are pre-specified hyper-parameters. 
Note that the evaluation of Equation \eqref{eq: sgd_lbfs} also requires to cache and reuse the last $m_0$ explicitly computed gradients, in which $m_0$ is also a hyper-parameter.
In \cite{wu2020deltagrad}, $j_0$, $T_0$ and $m_0$ are referred to as the number of ``burn-in'' iterations, the period and the history size, respectively. We refer readers to \cite{wu2020deltagrad} for more details.

\section{methodology}\label{sec: method}
In this section, we describe the system design in detail for the sample selector phase (Section \ref{sec: sample_selector}), the model constructor phase (Section \ref{sec: model_constructor}) and the human annotation phase (Section \ref{sec: sample_cleaner}).

\subsection{The sample selector phase}\label{sec: sample_selector}

Sample selection accomplishes two things: 1) it calculates the training sample influence using \infl\ in order to prioritize the most influential uncleaned training samples for cleaning, and simultaneously suggests possibly cleaned labels for them (see Section \ref{sec: infl}); and 2) it filters out uninfluential training samples early using \deltagradg\ at each round of loop \encircle{2} (see Section \ref{sec: deltagradg}).

\subsubsection{\infl}\label{sec: influence-l}\label{sec: infl}

The goal of \infl\ is to calculate the influence of an uncleaned training sample, $\nz$, by {estimating how much additional error will be incurred on the validation set $\mathcal{Z}_{\text{val}}$ if 1) the probabilistic label of $\nz$ is updated to some deterministic label; and 2) $\nz$ is up-weighted to 1 after it is cleaned}, which is similar to (but fully not covered by) the intuition of the influence function method \cite{koh2017understanding}.
To capture this intuition, we propose the following modified influence function (see  Appendix \ref{sec: infl_derive} for the derivation):
\begin{align}\label{eq: our_influence}
\small
\begin{split}
    &\mathcal{I}_{\text{pert}}(\nz,\delta_y,\gamma) \approx N\cdot(F(\uw, \mathcal{Z}_{\text{val}}) - F(\w, \mathcal{Z}_{\text{val}}))\\
    &\hspace{-3mm} = -\nabla_\w F(\w,\mathcal{Z}_{\text{val}})^\top \bH^{-1}(\w)[\nabla_{\y}\nabla_\w F(\w,\nz)\delta_{y} + (1-\gamma) \nabla_\w F(\w,\nz)],
\end{split}
\end{align}
in which $\delta_{y}$ denotes the difference between the original probabilistic label of $\nz$ and one deterministic label (ranging from 1 to $C$) and $\uw$ denotes the updated model parameters after the label is cleaned and $\nz$ is up-weighted. To calculate $\delta_{y}$, the deterministic label is first converted to its one-hot representation, i.e. a vector of length $C$ taking 1 in the $c_{th}$ entry $(c=1,2,\dots, C)$ for the label $c$ and taking 0 in all other entries (recall that $C$ represents the number of classes). 

\eat{By following the analysis of \cite{koh2017understanding}, we can derive the following formula:
\begin{align*}
\small
\begin{split}
    &\mathcal{I}_{\text{pert}}(\z,\delta_y) \approx F(\w(\delta_y), \mathcal{Z}_{\text{val}}) - F(\w, \mathcal{Z}_{\text{val}}),
\end{split}
\end{align*}
in which $\w(\delta_y)$ represents the model parameter constructed on the updated training dataset with the label of $\z$ perturbed by $\delta_y$. Therefore, Equation \eqref{eq: our_influence} can accurately quantify how the label perturbation $\delta_y$ leads to the change of the model prediction performance (represented by the loss on the validation dataset).}

To recommend the most influential uncleaned training samples to the human annotators and suggest possibly cleaned labels,
we 1) explicitly evaluate Equation \eqref{eq: our_influence} for each uncleaned training sample for {\em all possible deterministic labels}, 2) prioritize the most ``harmful'' training samples for cleaning, i.e. the ones with the smallest negative influence values after their labels are updated to {\em some} deterministic labels, and 3) suggest those deterministic labels as the potentially cleaned labels for the human annotators.

\textbf{Comparison to \cite{zhang2018training}} 
As 
discussed earlier, \duti\ \cite{zhang2018training} can also recommend the most influential training samples for cleaning and suggest possibly cleaned labels, which is accomplished through solving 
a bi-level optimization problem.
However, solving this problem is computationally challenging, and therefore this method cannot be used in real-time over multiple rounds (i.e. in loop \encircle{2}).
\eat{to give human annotators the most influential training samples and suggestions on how to clean their uncertain labels in real-time over multiple rounds.} 

The authors of \cite{zhang2018training} also modified the influence function to reflect the perturbations of the training labels as follows:
\begin{align}\label{eq: prev_influence}
\small
\begin{split}
    &\mathcal{I}_{\text{pert}}(\nz) = -\nabla_\w F(\w,\mathcal{Z}_{\text{val}})^\top \bH^{-1}(\w) \nabla_{\y}\nabla_\w F(\w,\nz),
\end{split}
\end{align}
and compared it against \duti. 
Equation \eqref{eq: prev_influence}
is equivalent to removing $\delta_{y}$ (which quantifies the effect of label changes) and $(1-\gamma) \nabla_{\w} F(\w,\nz)$ from Equation \eqref{eq: our_influence}. 
\eat{Therefore, Equation \eqref{eq: prev_influence} may fail to quantify the model performance changes that are triggered by the label updates $\delta_y$ and is not applicable when the uncleaned training samples are down-weighted. }
As we will see in Section \ref{sec: exp}, ignoring $\delta_{y}$ in Equation \eqref{eq: prev_influence} can lead to worse performance than \infl\ even when all the training samples are equally weighted. 

\textbf{Computing $\nabla_{\y}\nabla_\w F(\w,\nz)$} At first glance, it seems that the term $\nabla_{\y}\nabla_\w F(\w,\nz)$ cannot be calculated using auto-differentiation packages such as Pytorch, since it involves the partial derivative with respect to the label of $\nz$. However, we notice that this partial derivative can be explicitly calculated when the loss function $F(\w, \nz)$ is the cross-entropy function, which is the most widely used objective function in the classification task.
Specifically, the instantiation of the loss function $F(\w,\nz)$ into the cross-entropy function becomes:
\begin{align}\label{eq: cross_entropy_loss}
\small
\begin{split}
    F(\w, \nz) = -\sum\nolimits_{k=1}^C \ny^{(k)}\log(p^{(k)}(\w, \nx)),
\end{split}
\end{align}

In this formula above, $\ny=[\ny^{(1)}, \ny^{(2)},\dots, \ny^{(C)}]$ is the label of an input sample $\nz=(\nx,\ny)$ and $[p^{(1)}(\w, \nx), p^{(2)}(\w, \nx), \dots, p^{(C)}(\w, \nx)]$ represents the model output given this input sample, which is a probabilistic vector of length $C$ depending on the model parameter $\w$ and the input feature $\nx$.
\eat{Specifically, since given an input sample $\nz=(\nx,\ny)$, the model output is a probabilistic vector of length $C$ depending on the model parameter $\w$ and the input feature $\nx$, then the model output can be represented by $[p^{(1)}(\w, \nx), p^{(2)}(\w, \nx), \dots, p^{(C)}(\w, \nx)]$. By further representing $\ny$ as $\ny=[\ny^{(1)}, \ny^{(2)},\dots, \ny^{(C)}]$, the instantiation of the loss function $F(\w,\nz)$ into the cross-entropy function can be expressed as: 
\begin{align*}
\small
\begin{split}
    F(\w, \nz) = -\sum\nolimits_{k=1}^C \ny^{(k)}\log(p^{(k)}(\w, \nx)),
\end{split}
\end{align*}}
Then we can observe that Equation \eqref{eq: cross_entropy_loss} is a linear function of the label $\ny$. 
Hence, $\nabla_{\y}\nabla_\w F(\w,\nz)$ can be explicitly evaluated as:
\begin{align}\label{eq: class_wise_grad}
\small
    \begin{split}
\nabla_{\y}\nabla_\w F(\w,\nz) = [-\nabla_{\w} \log(p^{(1)}(\w, \nx)), \dots, -\nabla_{\w} \log(p^{(C)}(\w, \nx))]        
    \end{split}
\end{align}
As a result, each $-\nabla_{\w} \log(p^{(c)}(\w, \nx)),c=1,2,\dots, C$ can be calculated with the auto-differentiation package.

\textbf{Computing $\bH^{-1}(\w)$} Recall that $\bH(\w)$ denotes the Hessian matrix averaged on all training samples. Rather than explicitly calculating its inverse, 
by following \cite{koh2017understanding}, we leverage the conjugate gradient method \cite{martens2010deep} to approximately compute the Matrix-vector product $\nabla_\w F(\w,\mathcal{Z}_{\text{val}})^\top \bH^{-1}(\w)$ in Equation \eqref{eq: our_influence}. 

\subsubsection{\deltagradg}\label{sec: deltagradg} The goal of using \infl\ is to quantify the influence of all uncleaned training samples and select the Top-$b$ influential training samples for cleaning. But in loop \encircle{2}, this search space could be reduced by employing \deltagradg. 
Specifically, other than the initialization step, we can leverage {\em \deltagradg} to prune away most of the uninfluential training samples early in following rounds, thus only evaluating the influence of a small set of candidate influential training samples in those rounds.
\eat{Specifically, we can evaluate the influence of every training sample in the initialization step with Equation \eqref{eq: our_influence} and then leverage {\em \deltagradg} to filter out the uninfluential training samples early, thus only evaluating the influence of a small set of candidate influential training samples in following rounds. }
Suppose this set of samples is denoted as $\mathcal{Z}_{inf}^{(k)}$ for the round $k$; the derivation of this set is outlined in Algorithm \ref{alg: deltagradg}. 
As this algorithm indicates, the first step is 
to effectively estimate the maximal perturbations of Equation \eqref{eq: our_influence} at the $k_{th}$ cleaning round for each uncleaned training sample $\nz$ and each possible label change $\delta_y$ (see line \ref{line: cal_perturbation}),
which are assumed to take $\mathcal{I}_0(\nz, \delta_y, \gamma)$ (see Theorem \ref{theorem: bound_influence} for its definition) as the perturbation center.
Then the first part of $\mathcal{Z}_{inf}^{(k)}$ consists of all the training samples which produce the Top-$b$ smallest values of $\mathcal{I}_0(\nz, \delta_y, \gamma)$ with a given $\delta_y$ (see line 6).
For those $b$ smallest values, we also collect the maximal value of their upper bound, $L$. We then include in $\mathcal{Z}^{(k)}_{inf}$ all the remaining training samples whose lower bound, is smaller than $L$ with certain $\delta_y$ (see line \ref{line: second_z_k}).
This indicates the possibility of those samples becoming the Top-$b$ influential samples. 
The process to obtain $\mathcal{Z}^{(k)}_{inf}$ is also intuitively explained in Appendix \ref{sec: increm_infl_explanations}. 

As described above, it is critical to estimate the maximal perturbation of Equation \eqref{eq: our_influence} for each uncleaned training sample, $\nz$, and each label perturbation, $\delta_y$,
which requires the following theorem. 
\begin{theorem}\label{theorem: bound_influence}
For a training sample $\nz={(\nx, \ny)}$ which has not been cleaned before the $k^{th}$ round of loop \encircle{2}, the following bounds hold for Equation \eqref{eq: our_influence} evaluated on the training sample $\nz$ and a label perturbation $\delta_y$:
\begin{align*}
\small
\begin{split}
    & |-\mathcal{I}^{(k)}_{\text{pert}}(\nz,\delta_y,\gamma) - \mathcal{I}_0(\nz, \delta_y, \gamma) - \frac{1-\gamma}{2}e_1\mu - \sum\nolimits_{j=1}^C \delta_{y,j}e_1 \|\bH^{(j)}(\w^{(k)}, \nz)\||\\
    & \leq \sum\nolimits_{j=1}^C |\delta_{y,j}|e_2
    \|\bH^{(j)}(\w^{(k)}, \nz)\| + \frac{1-\gamma}{2}e_2\mu\\
\end{split}
\end{align*}

in which, $\mathcal{I}_0(\nz, \delta_y, \gamma) = \textbf{v}^\top[\nabla_{\y}\nabla_\w F(\w^{(0)},\nz)\delta_{y} + (1-\gamma) \nabla_\w F(\w^{(0)},\nz)]$,
$\textbf{v}^\top = -\nabla_\w F(\w^{(k)},\mathcal{Z}_{\text{val}})^\top \bH^{-1}(\w^{(k)})$,
$\delta_{y} = [\delta_{y,1}, \delta_{y,2},\dots, \delta_{y,C}]$,\\
$\bH^{(j)}(\w^{(k)}, \nz) = \int_{0}^1 -\nabla_{\w}^2\log(p^{(j)}(\w^{(0)} + s(\w^{(k)} - \w^{(0)}), \nx)) ds$, \\
$\mu = \|\int_{0}^1 \bH(\w^{(0)} + s(\w^{(k)} - \w^{(0)}), \nz) ds\|$, and\\
$e_1 = \textbf{v}^\top(\w^{(k)} - \w^{(0)})$, $e_2 = \|\textbf{v}\|\|\w^{(k)} - \w^{(0)}\|$.
\end{theorem}

To reduce the computational overhead, 
the integrated Hessian matrices, $\int_{0}^1 \bH(\w^{(0)} + s(\w^{(k)} - \w^{(0)}), \nz) ds$ and $\bH^{(j)}(\w^{(k)}, \nz)$, are approximated by their counterparts evaluated at $\w^{(0)}$, i.e., $\bH(\w^{(0)}, \nz)$ and $-\nabla_{\w}^2\log(p^{(j)}(\w^{(0)}, \nx))$.
As a consequence, the bounds 
can be calculated by applying several linear algebraic operations on $\textbf{v}$, $\w^{(k)}$, $\w^{(0)}$ 
and some pre-computed formulas, i.e., the norm of the Hessian matrices,  $\|-\nabla_{\w}^2\log(p^{(j)}(\w^{(0)}, \nx))\|$ and $\|\bH(\w^{(0)}, \nz)\|$, and the gradients, $\nabla_{\y}\nabla_\w F(\w^{(0)},\nz)$ and $\nabla_\w F(\w^{(0)},\nz)$, which can be computed as ``provenance'' information in the initialization step. Note that pre-computing $\nabla_{\y}\nabla_\w F(\w^{(0)},\nz)$ and $\nabla_\w F(\w^{(0)},\nz)$ is quite straightforward by leveraging Equation \eqref{eq: class_wise_grad}. Then the remaining question is how to compute $\|-\nabla_{\w}^2\log(p^{(j)}(\w^{(0)}, \nx))\|$ and $\|\bH(\w^{(0)}, \nz)\|$ efficiently without explicitly evaluating the Hessian matrices. Since those two terms calculate the norm of one Hessian matrix, we therefore only take one of them as a running example to describe how to compute them in a feasible way, as shown below.

\textbf{Pre-computing $\|\bH(\w^{(0)}, \nz)\|$} 
Since 1) a Hessian matrix is symmetric (due to its positive definiteness); and 2) the L2-norm of a symmetric matrix is equivalent to its eigenvalue with the largest magnitude \cite{meyer2000matrix}, the L2 norm of one Hessian matrix is thus equivalent to its largest eigenvalue. 
To evaluate this eigenvalue, 
we use the Power method \cite{golub2000eigenvalue}, which is discussed in Appendix \ref{sec: hessian norm}. 
\eat{In addition, we note that the power method involves computing the product between the Hessian matrix 
$\bH(\w^{(0)}, \nz)$ and a vector $\textbf{g}$,
which can be effectively calculated with the auto-differentiation package, such as Pytorch (see Appendix \ref{sec: hessian-vec product} for details).  }
\eat{the following equation, without explicitly computing the full Hessian matrix:
\begin{align}\label{eq: hessian_vector_prod}
\small
\begin{split}
    & \bH(\w^{(0)}, \nz)\textbf{g}= \lim_{s\rightarrow 0}\frac{\bH(\w^{(0)} + s \textbf{g}, \nz) - \bH(\w^{(0)}, \nz)}{s},
\end{split}
\end{align}
in which $s$ is a scalar. In the implementation, we approximately evaluate Equation \eqref{eq: hessian_vector_prod} by setting $s$ as a near-zero constant (e.g. $1e^{-6}$). }

\textbf{Time complexity of \deltagradg} By assuming that there are $n$ samples left after \deltagradg\ is used, the dimension of vectorized $\w$ is $m$, and the running time of computing the vector $\textbf{v}$ and the gradient ($\nabla_{\y}\nabla_\w F(\w,\nz)$ or $\nabla_\w F(\w,\nz)$) is denoted by $O(v)$ and $O(\text{Grad})$ respectively, the time complexity of \deltagradg\ is $O(v) + NC(O(Cm) + O(m) + O(C)) + nc O(\text{Grad})$ (see Appendix \ref{appendix-sec: deltagradg} for detailed analysis).
\begin{algorithm}
\small
 \caption{\deltagradg}\label{alg: deltagradg}
 \begin{spacing}{1}
 \begin{flushleft}
 \hspace*{\algorithmicindent}
 \textbf{Input:} 
 The number of samples to be cleaned at the $k_{th}$ round: $b$ \\
\hspace*{\algorithmicindent} \textbf{Output:} A set of prioritized training samples for cleaning: $\mathcal{Z}^{(k)}_{inf}$ 
\end{flushleft}
 \begin{algorithmic}[1]
\STATE Initialize $\mathcal{Z}^{(k)}_{inf} = \{\}$
  \STATE Calculate $\mathcal{I}_0(\nz, \delta_y, \gamma)$ and the perturbation bound on this term by using Theorem \ref{theorem: bound_influence} for each uncleaned sample, $\nz=(\nx,\ny)$, and each label perturbation, $\delta_y=\ny - onehot(c), (c=1,2,\dots, C)$ \label{line: cal_perturbation}
   \STATE 
   Add the training samples producing Top-$b$ smallest $\mathcal{I}_0(\nz, \delta_y, \gamma)$ to $\mathcal{Z}^{(k)}_{inf}$
   \STATE obtain the largest perturbation upper bound, $L$, for all Top-$b$ smallest $\mathcal{I}_0(\nz, \delta_y, \gamma)$\label{line: first_z_k}
    \STATE For any remaining training sample, $\nz$, if its lower perturbation bound of $\mathcal{I}_0(\nz, \delta_y, \gamma)$ is smaller than $L$ with a certain $\delta_y$, add it to $\mathcal{Z}^{(k)}_{inf}$ \label{line: second_z_k}
    \STATE \textbf{Return} $\mathcal{Z}^{(k)}_{inf}$

  \end{algorithmic}
  \end{spacing}
  \end{algorithm}

\subsection{The model constructor phase (\deltagradl)}\label{sec: model_constructor}

At the $k_{th}$ round of loop \encircle{2}, after the human annotators clean the labels for a set of $b$ influential training samples, $\mathcal{R}^{(k)}$,
provided by the {Sample selector}, the next step is to update the model parameters in the {Model constructor}. To reduce the overhead of this step, we can regard the process of updating labels as two sub-processes, i.e. the deletions of the training samples, $\mathcal{R}^{(k)}$ (with the associated weight, $\gamma$), and the additions of those training samples with the cleaned labels (with the updated weight, 1), thus facilitating the use of \deltagrad. Specifically, the following modifications to Equation \eqref{eq: updated_gradient} are required: 1) the input deleted training samples should be $\mathcal{R}^{(k)}$; 2) the input cached model parameters and the corresponding gradients become the ones obtained at the $k-1_{st}$ round of the loop \encircle{2}; 3) instead of randomly sampling a mini-batch $\mathcal{A}_t$ from the added training samples $\mathcal{A}$, $\mathcal{A}_t$ should be replaced with the removed training samples from $\miniB_t$, i.e., $\miniB_t \bigcap \mathcal{R}^{(k)}$, but with updated labels; 4) the cleaned training samples and the uncleaned training samples so far are weighted by 1 and $\gamma$ respectively (this only slightly modifies how the loss is calculated for each mini-batch).

\subsection{The human annotation phase}\label{sec: sample_cleaner}
As discussed in Section \ref{sec: intro}, the Sample selector not only suggests which samples should be cleaned, but also suggests the candidate cleaned labels, which can be regarded as one independent label annotator. When multiple annotators exist, we 
aggregate their labels by using majority vote to resolve possible label conflicts.

\section{Experiments}\label{sec: exp}
We conducted extensive experiments in Python 3.6  and PyTorch 1.7.0 \cite{paszke2017automatic}. All experiments were conducted on a Linux server with an Intel(R) Xeon(R) CPU E5-2630 v4 @ 2.20GHz, 3 GeForce 2080 Titan RTX GPUs and 754GB of main memory. 

\subsection{Experimental setup}\label{sec: exp_setup}
Two types of datasets are used, one of which is annotated with ground-truth labels but no human generated labels, while the other is fully annotated with crowdsourced labels but only partially annotated by ground-truth labels. The former type (referred to as \textit{\cleanset}) is used to evaluate the quality of  labels suggested by \infl\ by comparing them against the ground-truth labels. The latter type (referred to as \textit{\crowdset}) is used for evaluating the performance of our methods in more realistic settings. The two types of datasets are briefly described next; more detailed descriptions 
can be found in Appendix \ref{appendix_sec: dataset}. 

\cleanset: Three real medical image datasets are used: \mimicfull\ (\mimic\ for short) \cite{johnson2019mimic}, \chexpert\ \cite{irvin2019chexpert} and \retinafull\ (\retina\ for short)  \cite{gulshan2016development}. The datasets are used to identify whether one or more diseases or findings exist for each image sample. In the experiments, we are interested in predicting the existence of findings called ``Lung Opacity'', ``the referable Diabetic Retinopathy'' and ``Cardiomegaly'' for \mimic, \retina\ and \chexpert,  respectively.

\crowdset: Three realistic crowdsourced datasets are used: \fashionfull\ (\fashion\ for short)\footnote{available at \url{http://skuld.cs.umass.edu/traces/mmsys/2014/user05.tar}} \cite{loni2014fashion}, \factfull\ (\fact\ for short)\footnote{available at \url{https://sites.google.com/site/crowdscale2013/shared-task/task-fact-eval}}, and \twitterfull\ (\twitter\ for short)\footnote{available at \url{https://github.com/naimulhuq/Capstone/blob/master/Data/Airline-Full-Non-Ag-DFE-Sentiment\%20(raw\%20data).csv}}. 
Only a small portion of samples in the datasets have ground-truth labels while the rest are labeled by 
crowdsourcing workers (e.g., the labels of the \fashion\ dataset are collected through the Amazon Mechanical Turk (AMT) crowdsourcing platform). The \fashion\ dataset is an image dataset for distinguishing fashionable images from unfashionable ones; the \fact\ dataset uses RDF triples to represent facts about public figures and the classification task is to judge whether or not each fact is true; and the \twitter\ dataset consists of plain-text tweets on major US airlines for sentiment analysis, i.e., identifying positive or negative tweets. For the \fashion\ and \fact\ datasets, extra text is also associated with each sample, e.g. user comments on each image in \fashion\ and the evidence for each fact in \fact, which is critical for producing probabilistic labels.
(see discussion below).





Since some samples in the datasets  have missing or unknown 
ground-truth labels,  
we remove them in the experiments. Also, except for \mimic\ which has 579 validation samples and 1628 test samples, other datasets do not have well-defined validation and test set. For example, as of the time the experiments were performed, the test samples of \chexpert\ had not been released. To remedy this, we partition the \chexpert\ validation set into two parts to create validation and test sets, each of which have 234 samples. Since there was no validation set for \retina, we randomly select roughly 10\% training samples, i.e., 3512 samples, as the validation set.
Similarly, for the \twitter\ and \fact\ datasets, we randomly partition the set of samples with ground-truth labels as the validation set and test set, and regard all the other samples as the training set.
Since ground-truth labels are not available in the \fashion\ dataset, we randomly select roughly 0.5\%\footnote{This ratio is determined based on the observation that in the \twitter\ and \fact\ datasets, the percentage of samples with ground-truth labels is less than 1\% of the size of the entire dataset.} of the sample samples as the validation set and test set, each containing 146 samples. The ``ground-truth'' labels for those samples are determined by aggregating human annotated labels using majority vote. The remaining samples in this dataset are then regarded as training samples. In the end, the six datasets, i.e., \mimic, \retina, \chexpert, \fashion, \fact\ and \twitter\ include $\sim$78k,$\sim$31k,$\sim$38k,$\sim$29k,$\sim$38k and $\sim$12k training samples. More detailed statistics of the six datasets are given in Appendix \ref{appendix_sec: dataset}.

\textbf{Producing probabilistic labels} 
Due to the lack of probabilistic labels or labeling functions \cite{ratner2017snorkel} for the datasets, 
we leverage \cite{boecking2020interactive},\cite{varma2018snuba} or \cite{das2020goggles} to {\em automatically} derive suitable labeling functions and thus probabilistic labels in the experiments.
\eat{Ideally, we expect to employ  the labeling functions \cite{ratner2017snorkel} provided by domain experts for producing probabilistic labels for those six datasets, which, unfortunately are not available. But we note that the labeling functions could be automatically derived, by utilizing such as \cite{boecking2020interactive},\cite{varma2018snuba} and \cite{das2020goggles}, which are utilized producing probabilistic labels in the experiments.}
Note that \cite{boecking2020interactive} and \cite{varma2018snuba} deal with text data (including the text associated with image data) while \cite{das2020goggles} targets pure image data. However, the time and space complexity of \cite{das2020goggles} is quadratic in the dataset size, and does not scale to large image datasets such as our \cleanset. Furthermore, no text information is available for images in \cleanset, so it is not feasible to use \cite{boecking2020interactive} or \cite{varma2018snuba}. As a result, 
random probabilistic labels are produced for  
all training samples. For \crowdset, we apply \cite{boecking2020interactive} on the extra text information in \fashion\ (e.g. user comments for each image) and the plain-text tweets in the \twitter\ dataset to produce probabilistic labels. For the \fact\ dataset, the two texts for each sample (i.e. the RDF triples and the associated evidence) are compared using \cite{chatterjee2019data} to generate labeling functions. 

\textbf{Human annotator setup}
For \crowdset, we can use the crowdsourced labels as the cleaned labels for the uncleaned training samples.
However, no such labels are available in \cleanset. To remedy this, we note that the error rate of manually labeling medical images is typically between 3\% and 5\%, but sometimes can be up to 30\% \cite{brady2017error}.
We therefore produce synthetic human annotated labels by flipping the ground truth labels of a randomly selected 5\% of the samples 
\footnote{Recall that although the samples have probabilistic labels, their true labels are known by construction.}.
We assume three independent annotators, and aggregate their labels as the cleaned labels using majority vote (denoted \cleanone).
Since \infl\ and \duti\ \cite{zhang2018training} can suggest cleaned labels, those labels can be used as cleaned labels by themselves for the uncleaned samples 
(denoted \cleantwo) or be combined with two other simulated human annotators for label cleaning (denoted \cleanthree).

\textbf{Model constructor setup}
Throughout the paper we assume that strong convexity holds on the ML models. Therefore, in this section, to justify the performance advantage of our design as a whole (including \deltagradg, \deltagradl\ and \infl), we focus on a scenario where pre-trained models are leveraged for feature transformation and 
then a logistic regression model is used for classification, which has emerged as a convention
for medical image classification tasks~\cite{raghu2019transfusion}. 
Specifically, in the experiments, we use a pre-trained ResNet50 \cite{he2016deep} for the image datasets (\cleanset\ and \fashion), and use a pre-trained BERT-based transformer \cite{devlin2018bert} for the text datasets (\fact\ and \twitter). 
Stochastic gradient descent (\sgd) is then used in the subsequent training process with a mini-batch size of 2000, 
and weight $\gamma= 0.8$ on the uncleaned samples. Early stopping is also applied to avoid overfitting. Other hyper-parameters are varied across different datasets and are included in Appendix \ref{appendix-sec: hyper}.
As discussed in Section \ref{sec: intro}, other than the initialization step, we can construct the models by either retraining from scratch (denoted \retrain) or leveraging \deltagrad\ for incremental updates. 

\eat{When the logistic regression models are incrementally updated using \deltagradl\  (see Algorithm \ref{alg: deltagrad}), its hyper-parameters are configured as $m=2$, $j_0=10$ and $T_0= 20$ for all six datasets.}

However, the strong convexity assumption on model type is only required for \deltagradg\ and \deltagradl, but not for \infl. 
Hence, extra experiments are conducted using convolutional neural networks (CNNs), which are presented in the Appendix \ref{sec: exp_cnn_model}. The results demonstrate the performance advantage of \infl\ in more general settings.

\textbf{Sample selector setup} 
We assume that the clean budge $B=100$, meaning that 100 training samples are cleaned in total.
We further vary the number of samples to be cleaned at each round, i.e. the value of $b$.

\textbf{Baseline against \infl} We compare \infl\ against several baseline methods, including other versions of the influence function, i.e. Equation \eqref{eq: influence function} \cite{koh2017understanding} (denoted by \inflo) and Equation \eqref{eq: prev_influence} \cite{zhang2018training} (denoted by \inflg) and \duti. 
Since solving the bi-level optimization problem in \duti\ is extremely expensive, 
we only run \duti\ once to identify the Top-100 influential training samples.

Since active learning and noisy sample detection algorithms can prioritize the most influential samples for label cleaning, they are also compared against \infl. Specifically, we consider two active learning methods, i.e., least confidence based sampling method (denoted by \ac) and entropy based sampling method (denoted by \actwo) \cite{settles2009active}, and two noisy sample detection algorithms, i.e., \ou\ \cite{huang2019o2u} and \tars\ \cite{dolatshah12cleaning}.

Note that many of these baseline methods are not applicable in the presence of either probabilistic labels or regularization on uncleaned training samples.
Hence, we modify the methods to handle these scenarios or adjust the experimental set-up to create a fair comparison. For example, in Appendix \ref{appendix-sec: duti}, we present necessary modifications to \duti\ so that it can handle probabilistic labels. However, it is not straightforward to modify \duti\ for quantifying the effect of up-weighting the training samples after they are cleaned. We therefore only compare \duti\ against \infl\ when all the training samples are equally weighted (i.e. $\gamma=1$ in Equation \eqref{eq: obj_function}), which is presented in Appendix \ref{sec: exp_vary_weight}. Similarly, \tars\ is only applicable when the noisy labels are either 0 or 1 rather than probabilistic ones. Therefore, to compare \infl\ and \tars, we round the probabilistic labels 
to their nearest deterministic labels for a fair comparison (see Appendix \ref{sec: exp_tars} for details). For other methods such as \ac, \actwo, \ou\ and \inflo, no modifications are made other than using Equation \eqref{eq: obj_function} for model training.


\textbf{Baseline against \deltagradl\ and \deltagradg} Recall that \deltagradl\ incrementally updates the model after some training samples are cleaned.  We compare this with retraining the model from scratch (denoted as 
\retrain). 
We also compare the running time for selecting the influential training samples with and without \deltagradg. When \deltagradg\ is not used, it is denoted as \full.



\eat{\begin{figure*}[h]
\includegraphics[width=2.0\columnwidth]{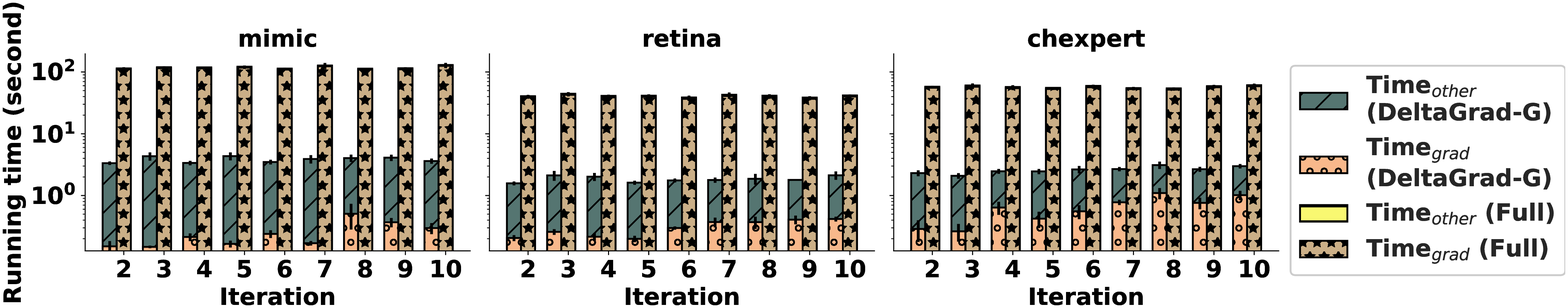}
\caption{Running time comparison between \deltagradg\ (green) and \full (yellow)}\label{fig:increm_time}
\end{figure*}}

\begin{figure}[ht]
\includegraphics[width=0.8\columnwidth, height=0.45\columnwidth]{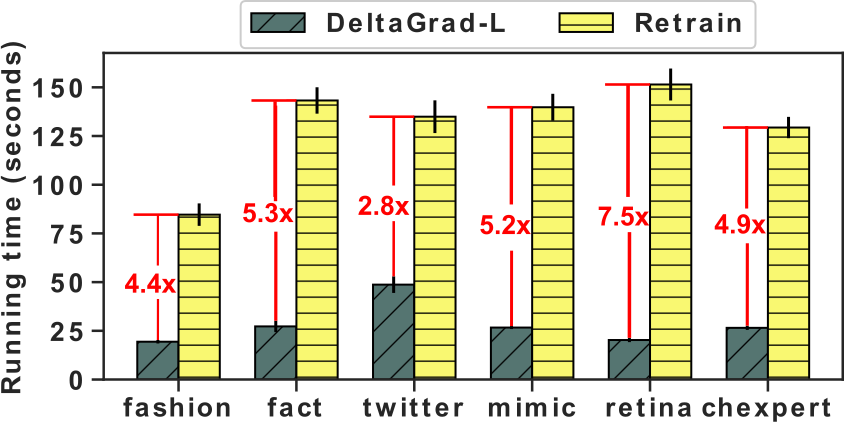}
\caption{Comparison of running time between \deltagradl\ and \retrain}\label{fig:training_time}
\end{figure}

\subsection{Experimental design}\label{sec: exp_design}

\begin{table*}
    \centering
    \small
    \caption{Comparison of the model prediction performance (F1 score) after 100 training samples are cleaned}\label{Table: prediction_strategy}
    \vspace{-0.2cm}
    \begin{tabular}[!h]{>{\arraybackslash}p{1cm}||>{\arraybackslash}p{1cm}||>{\centering\arraybackslash}p{0.7cm}>{\centering\arraybackslash}p{0.7cm}>{\centering\arraybackslash}p{0.7cm}|>{\centering\arraybackslash}p{0.7cm}>{\centering\arraybackslash}p{0.7cm}>{\centering\arraybackslash}p{0.7cm}>{\centering\arraybackslash}p{0.7cm}>{\centering\arraybackslash}p{0.7cm}>{\centering\arraybackslash}p{1.5cm}>{\centering\arraybackslash}p{1.5cm}>{\centering\arraybackslash}p{0.7cm}} \hline
    && \multicolumn{3}{c|}{b=100} & \multicolumn{8}{c}{b=10} \\ \hhline{~~-----------}
         & uncleaned&\cleanone\ & \cleantwo\ & \cleanthree\  & \inflo & \ac\ & \actwo\ & \ou & \cleanone\ & \cleantwo\ & \cleantwo\ + \deltagrad & \cleanthree\ \\\hline
        \mimic&0.6284 & 0.6292 & 0.6293 &0.6293&0.6283&0.6287&0.6287&0.1850&0.6292&\textBF{\underline{0.6293$\pm$0.0011}}&\textBF{0.6292$\pm$0.0005}&0.6292 \\
        \retina&0.5565&0.5580&\textBF{0.5582}&0.5581&0.5556&0.5568&0.5568&0.1314&0.5579&\textBF{0.5582$\pm$0.0003}&\textBF{\underline{0.5610$\pm$0.0010}}&0.5581\\ 
        \chexpert&0.5244& 0.5286&0.5297&0.5289&0.5246&0.5246&0.5246&0.5281&0.5287&\textBF{\underline{0.5300$\pm$0.0018}}&\textBF{0.5295$\pm$0.0030}&0.5291 \\ 
        \fashion&0.5140 & 0.5178&0.5177&0.5177&0.5143&0.5145&0.5145&0.5152      &0.5178&\textBF{0.5181$\pm$0.0131}&\textBF{\underline{0.5195$\pm$0.0144}}&0.5180 \\
        \fact& 0.6595 & 0.6601 & \textBF{0.6609} &0.6603&0.6596&0.6600&0.6600&0.6598&0.6601&\textBF{\underline{0.6609$\pm$0.0043}}&\textBF{\underline{0.6609$\pm$0.0065}}&0.6602 \\
        \twitter& 0.6485& 0.6594&0.6680&0.6594&0.6518&0.6515&0.6515&0.6490&0.6578&\textBF{\underline{0.6697$\pm$0.0058}}&\textBF{0.6597$\pm$0.0027}&0.6586 \\ \hline
    \end{tabular}
\end{table*}

In this section, we design the following three experiments:

\begin{table}[h]
\centering
\small
\caption{Running time of \deltagradg\ and \full}\label{Table: running_time_last}
\vspace{-0.2cm}
\begin{tabular}[!h]{>{\arraybackslash}p{0.9cm}||>{\centering\arraybackslash}p{0.8cm}>{\centering\arraybackslash}p{2.3cm}|>{\centering\arraybackslash}p{0.8cm}>{\centering\arraybackslash}p{2.2cm}} \hline
\multirow{2}{*}{} &\multicolumn{2}{c|}{$\text{Time}_{inf}$ (s)}&\multicolumn{2}{c}{$\text{Time}_{grad}$ (s)} \\ \hhline{~----}
& \full\ & \deltagradg\ & \full\ & \deltagradg\ \\ \hline
\mimic  & 151.4$\pm$0.5 & \textbf{2.77$\pm$0.03 \textcolor{red}{(54.7x)}} &145.4$\pm$0.7  & \textbf{0.17$\pm$0.03 \textcolor{red}{(855x)}} \\
\retina &74.0$\pm$0.6  & \textbf{1.36$\pm$0.04 \textcolor{red}{(54.4x)}} & 70.8$\pm$0.6  &  \textbf{0.21$\pm$0.03 \textcolor{red}{(337x)}} \\
\chexpert &72.5$\pm$0.2  & \textbf{17.9$\pm$1.9 \textcolor{red}{(4.1x)}} &69.3$\pm$0.2 & \textbf{14.7$\pm$1.5 \textcolor{red}{(4.7x)}} \\
\fashion &66.4$\pm$3.6 &\textbf{8.7$\pm$0.6 \textcolor{red}{(7.6x)}} &57.1$\pm$3.3 & \textbf{0.81$\pm$0.07 \textcolor{red}{(70.5x)}}\\
\fact &73.8$\pm$4.0 &\textbf{6.1$\pm$0.8 \textcolor{red}{(12.1x)}} &72.5$\pm$6.0 & \textbf{4.7$\pm$0.1 \textcolor{red}{(15.4x)}}  \\
\twitter &33.1$\pm$2.3 &\textbf{14.1$\pm$0.4 \textcolor{red}{(2.3x)}} &30.2$\pm$1.1 &\textbf{12.7$\pm$0.1 \textcolor{red}{(2.4x)}} \\ \hline
\end{tabular}
\end{table}

\textbf{Exp1} 
In this experiment, we compared the model prediction performance after \infl\ and other baseline methods (including \inflo, \ac, \actwo, \ou) are applied to select 100 training samples for cleaning. Recall that there are three different strategies of providing cleaned labels by \infl\ and their performance is compared. To show the benefit of using a smaller batch size $b$, we choose two different values for $b$, i.e. 100 and 10. 
Since the ground-truth labels are available for all samples in \cleanset, we count how many of them match the labels suggested by \infl.
We also vary $\gamma$ for a more extensive comparison 
(see Appendix \ref{sec: exp_supple}).

\textbf{Exp2} \eat{Recall that \deltagradg\ can be coupled with \infl\ to 
filter out most of the uninfluential samples early.  
Therefore, }
This experiment compares the time of selecting the Top-$b$ (with $b=10$) influential training samples (denoted $\text{Time}_{inf}$) with and without using \deltagradg\ at each round in the {\em Sample selector}  phase. 
\eat{When \deltagradg\ is not used, it means that the influence of {\em all} training samples are always re-evaluated, which is denoted as \full.}
Recall that the most time-consuming step to evaluate Equation \eqref{eq: our_influence} is to compute the class-wise gradients for each sample and the sample-wise gradients.
Therefore, its running time 
(denoted as $\text{Time}_{grad}$) is also recorded. For \deltagradg, the time to compute the bounds in Theorem \ref{theorem: bound_influence} is also included in $\text{Time}_{inf}$.

\textbf{Exp3} 
The main goal of this experiment is to explore the difference in running time between \retrain\ and \deltagradl\ for updating the model parameters in the {\em Model constructor} phase.
In addition,  the model parameters produced by \deltagradl\ and \retrain\ are not exactly the same 
\cite{wu2020deltagrad}, which could lead to different influence values for each training sample and
thus produce different models
in subsequent cleaning rounds.
Hence, we also explore whether such differences
produce divergent prediction performance for \deltagradl\ and \retrain.

\begin{figure}[h]
  \centering
  \begin{subfigure}{.45\columnwidth}
    \centering
    \includegraphics[width=\linewidth]{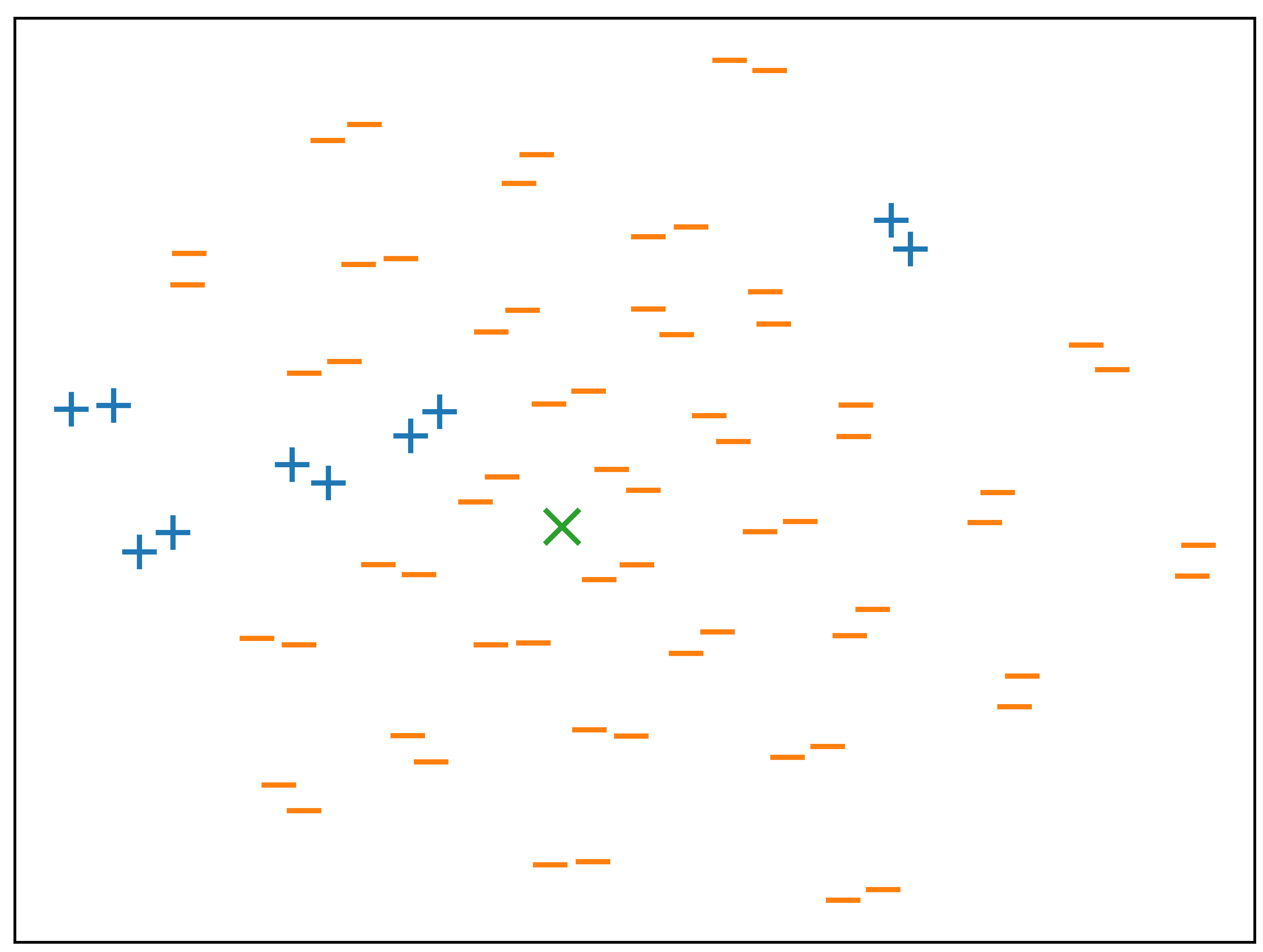}
    \caption{\twitter}
  \end{subfigure}%
  \hfill
  \begin{subfigure}{.45\columnwidth}
    \centering
    \includegraphics[width=\linewidth]{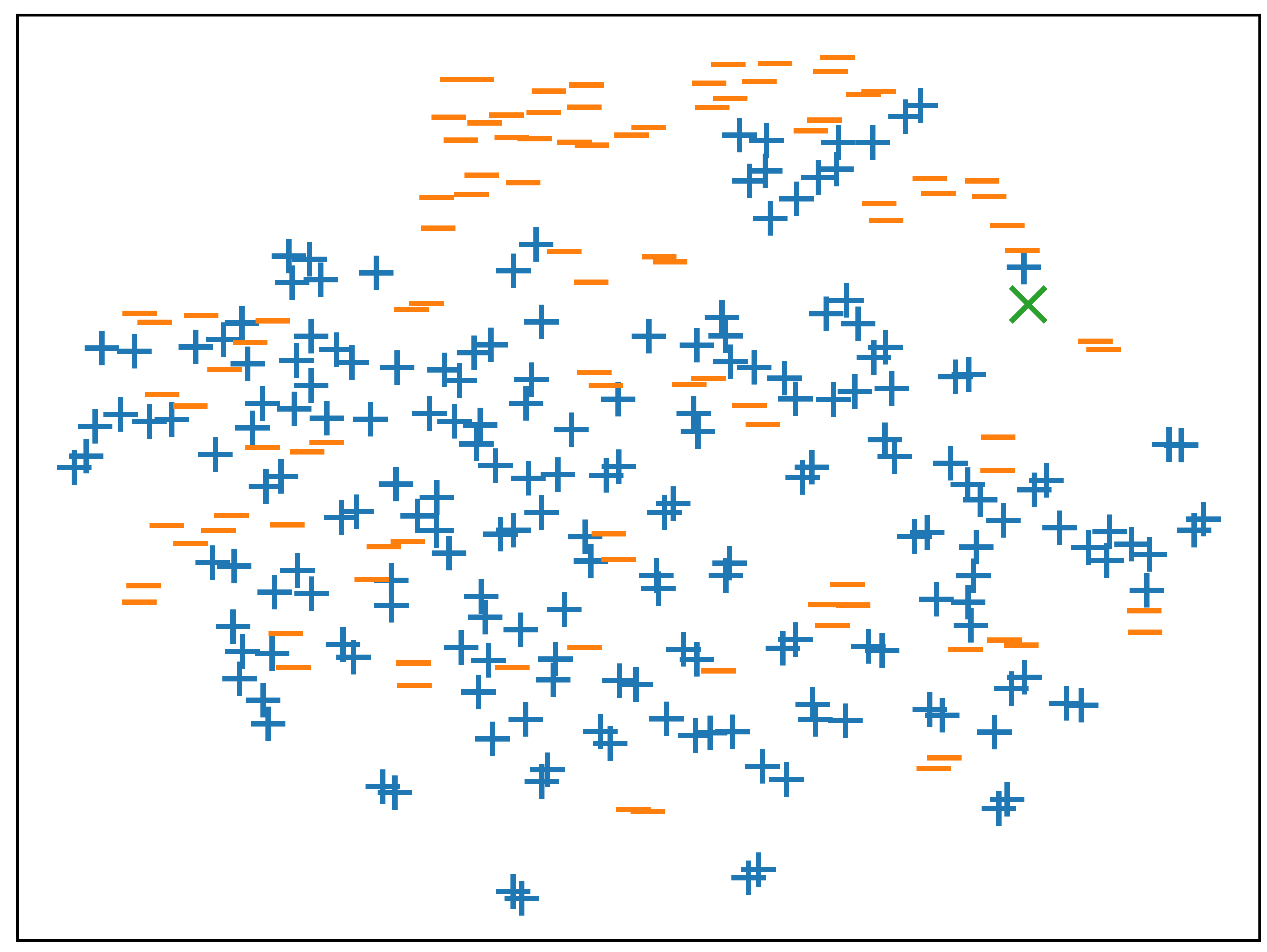}
    \caption{\fashion}
  \end{subfigure}%
  \caption{Visualization of the validation samples, test samples and the most influential training sample $S$ (`+', `-' and `X' denote the positive ground-truth samples, negative ground-truth samples and the sample $S$ respectively)}\label{fig:twitter_visual}
\end{figure}



\subsection{Experimental results}\label{sec: exp_res}

\textbf{Exp1} Experimental results 
are given in Table \ref{Table: prediction_strategy}\footnote{Except \cleantwo, only the averaged F1 scores are given. Due to space limit, the error bars of the F1 scores are included in Appendix \ref{sec: detailed_exp1}}.
We observe that with fixed $b$, e.g., 10, \cleantwo\ performs best across almost all datasets. Recall that \cleantwo\ 
uses the derived labels produced by \infl\ as the cleaned labels without additional human annotated labels. Due to its superior performance, 
especially on \crowdset,
this implies that the quality of the labels provided by \infl\ could actually be better than that of the human annotated labels.

To further understand 
this phenomenon, we compared the labels suggested by \infl\ against their ground-truth labels for \cleanset.  It turns out that over 70\% are equivalent (89 for \retina, 79 for \chexpert\ and 95 for \mimic). Note that even the ground-truth labels of these three datasets are not 100\% accurate. In the \chexpert\ dataset, for example, the ground-truth labels are generated through an automate labeling tool rather than being labeled by human annotators,
thereby leading to possible labeling errors. Those erroneous labels may not match the labels provided by \infl, thus leading to worse model performance (see the performance difference between \cleanone\ and \cleantwo\ for \chexpert\ dataset). 

However, the above comparison could not be done for \crowdset\ due to the lack of ground-truth labels. We therefore investigate the relationship between the samples with ground-truth labels and the influential samples identified by \infl. Specifically, 
we use t-SNE \cite{van2008visualizing} to visualize the 
samples with ground-truth labels
for the \twitter\ and \fashion\ datasets after feature transformation using the pre-trained models (see Figure \ref{fig:twitter_visual}). As described in Section \ref{sec: exp_setup}, those samples belong to validation or test set.
In addition, in this figure, we indicate the position of the most influential training sample $S$ identified by \infl.
As this figure indicates, the sample $S$ is proximal to the samples with negative ground-truth labels for the \twitter\ dataset (positive for the \fashion\ dataset). To guarantee the accurate predictions on those nearby ground-truth samples, it is therefore reasonable to label $S$ as negative (positive for \fashion\ dataset), which matches the labels provided by \infl\ but differs from ones given by the human annotators. This indicates the high quality of the labels given by \infl. 
Thus, when high-quality human labelers are unavailable, \infl\ can be an alternative labeler for reducing the labeling cost without harming the labeling quality.

Table \ref{Table: prediction_strategy} also exhibits the benefit of using smaller batch sizes $b$ since it results in better model performance when \infl, especially \cleantwo, is used for some datasets (e.g, see its model performance comparison between $b=100$ and $b=10$ for \twitter\ dataset in Table \ref{Table: prediction_strategy}). 
Intuitively, \infl\ only quantifies the influence of 
cleaning {\em single} training sample rather than multiple ones. Therefore, the larger $b$ is, the more likely that \infl\ selects a sub-optimal set of $b$ samples for cleaning. Ideally, $b$ should be one, meaning that one training sample is cleaned at each round. However, this can inevitably increase 
the number of rounds and thus 
the overall overhead. In Appendix \ref{sec: vary_b}, we empirically explored how to choose an appropriate $b$ to balance the model performance and the total running time.

\textbf{Exp2} In this experiment, we compare the running time of \deltagradg\ and \full\ in selecting the Top-10 influential training samples ($\text{Time}_{inf}$) at each cleaning round of the loop \encircle{2} (with $b=10$). Due to space, we only include results for the last round in Table \ref{Table: running_time_last}, which are similar to results in other rounds.
As Table \ref{Table: running_time_last} indicates,  \deltagradg\ is up to 54.7x faster than \full,
which is due to the significantly decreased overhead of computing the class-wise gradients for each sample (i.e. $\text{Time}_{grad}$) when \deltagradg\ is used. To further illustrate this point, we 
also record the number of candidate influential training samples whose influence values are explicitly evaluated with and without using \deltagradg.
The result indicates that due to the early removal of uninfluential training samples using \deltagradg, we only need to evaluate the influence of a small portion of training samples, thus reducing $\text{Time}_{grad}$ by up to two orders of magnitude and thereby significantly reducing the total running time, $\text{Time}_{inf}$. 
In addition, we observe that \deltagradg\ always returns the same set of influential training samples as \full, which thus guarantees the correctness of \deltagradg.

\textbf{Exp3} 
Experimental results of \textBF{Exp3} are shown in Figure \ref{fig:training_time}. The first observation is that \deltagradl\ can achieve up to 7.5x speed-up with respect to \retrain\ on updating the model parameters. As shown in Section \ref{sec: exp_design}, the models updated by \deltagradl\ are not exactly the same as those produced by \retrain, which might 
cause different model performance between the two methods. 
However, we observe that 
the models constructed by those two methods have almost equivalent prediction performance (see the second to last column in Table \ref{Table: prediction_strategy}).
\eat{To further understand the effect of the model differences produced by those two methods, we collected the sets of influential training samples from the sample selector with \deltagradl\ and \retrain\ as implemented in the model constructor, and observe substantial overlap for all three datasets. }
This indicates that it is worthwhile to leverage \deltagradl\ for speeding up the model constructor.
\vspace{-0.2cm}
\section{Conclusions}\label{sec: conclusion}
\eat{To mitigate the lack of  high-quality labels in medical image classification tasks, weak labels are widely used, which, however, often need to be further cleaned by human annotators to improve model performance. The pipeline of cleaning uncertain labels is typically iterative, consisting of a {\em sample selector phase}, {\em human annotation phase} and  {\em model constructor phase}. To reduce the overall cost and time overhead of each phase of this label cleaning process, }
In this paper, we propose \method, which can reduce the overall cost of the label cleaning pipeline and achieve better model performance than other approaches; it may also allow early termination in the human annotation phase. 
\eat{1) reduce the cost of the human annotation phase through a modified version of the influence function, \infl; 2) decrease the time overhead of the sample selector phase by incrementally identifying the most influential training samples with \deltagradg; and 3) speed up the model constructor phase by incrementally updating the model using \deltagradl. }
Extensive experimental studies show the effectiveness of our solution over a broad spectrum of real datasets when strongly convex models are used.
\eat{show that our approach can guarantee the model prediction performance while achieving significant speed-ups.}
\eat{We also evaluate the performance of \infl\ for CNN models. }
How to extend \method, in particular, \deltagradg\ and \deltagradl, beyond strongly convex models is left as future work.
\eat{ One limitation of this work is that other than \infl, it can only support strongly convex models. Exploring how to generalize the above techniques for more advanced ML models is left as the future work.}

\begin{acks}
This material is based upon work that is in part supported by the Defense Advanced Research Projects Agency (DARPA) under Contract No. HR001117C0047.
\end{acks}

\newpage
\balance
\bibliographystyle{ACM-Reference-Format}
\bibliography{reference}


\begin{thebibliography}{41}


\ifx \showCODEN    \undefined \def \showCODEN     #1{\unskip}     \fi
\ifx \showDOI      \undefined \def \showDOI       #1{#1}\fi
\ifx \showISBNx    \undefined \def \showISBNx     #1{\unskip}     \fi
\ifx \showISBNxiii \undefined \def \showISBNxiii  #1{\unskip}     \fi
\ifx \showISSN     \undefined \def \showISSN      #1{\unskip}     \fi
\ifx \showLCCN     \undefined \def \showLCCN      #1{\unskip}     \fi
\ifx \shownote     \undefined \def \shownote      #1{#1}          \fi
\ifx \showarticletitle \undefined \def \showarticletitle #1{#1}   \fi
\ifx \showURL      \undefined \def \showURL       {\relax}        \fi
\providecommand\bibfield[2]{#2}
\providecommand\bibinfo[2]{#2}
\providecommand\natexlab[1]{#1}
\providecommand\showeprint[2][]{arXiv:#2}

\bibitem[\protect\citeauthoryear{Aguilar~Melgar, Dao, Gan, G{\"u}rel,
  Hollenstein, Jiang, Karla{\v{s}}, Lemmin, Li, Li,
  et~al\mbox{.}}{Aguilar~Melgar et~al\mbox{.}}{2021}]%
        {aguilar2021ease}
\bibfield{author}{\bibinfo{person}{Leonel Aguilar~Melgar},
  \bibinfo{person}{David Dao}, \bibinfo{person}{Shaoduo Gan},
  \bibinfo{person}{Nezihe~M G{\"u}rel}, \bibinfo{person}{Nora Hollenstein},
  \bibinfo{person}{Jiawei Jiang}, \bibinfo{person}{Bojan Karla{\v{s}}},
  \bibinfo{person}{Thomas Lemmin}, \bibinfo{person}{Tian Li},
  \bibinfo{person}{Yang Li}, {et~al\mbox{.}}} \bibinfo{year}{2021}\natexlab{}.
\newblock \showarticletitle{Ease. ML: A Lifecycle Management System for Machine
  Learning}. In \bibinfo{booktitle}{\emph{11th Annual Conference on Innovative
  Data Systems Research (CIDR 2021)(virtual)}}. CIDR.
\newblock


\bibitem[\protect\citeauthoryear{Bach, Rodriguez, Liu, Luo, Shao, Xia, Sen,
  Ratner, Hancock, Alborzi, et~al\mbox{.}}{Bach et~al\mbox{.}}{2019}]%
        {bach2019snorkel}
\bibfield{author}{\bibinfo{person}{Stephen~H Bach}, \bibinfo{person}{Daniel
  Rodriguez}, \bibinfo{person}{Yintao Liu}, \bibinfo{person}{Chong Luo},
  \bibinfo{person}{Haidong Shao}, \bibinfo{person}{Cassandra Xia},
  \bibinfo{person}{Souvik Sen}, \bibinfo{person}{Alex Ratner},
  \bibinfo{person}{Braden Hancock}, \bibinfo{person}{Houman Alborzi},
  {et~al\mbox{.}}} \bibinfo{year}{2019}\natexlab{}.
\newblock \showarticletitle{Snorkel drybell: A case study in deploying weak
  supervision at industrial scale}. In \bibinfo{booktitle}{\emph{Proceedings of
  the 2019 International Conference on Management of Data}}.
  \bibinfo{pages}{362--375}.
\newblock


\bibitem[\protect\citeauthoryear{Boecking, Neiswanger, Xing, and
  Dubrawski}{Boecking et~al\mbox{.}}{2020}]%
        {boecking2020interactive}
\bibfield{author}{\bibinfo{person}{Benedikt Boecking}, \bibinfo{person}{Willie
  Neiswanger}, \bibinfo{person}{Eric Xing}, {and} \bibinfo{person}{Artur
  Dubrawski}.} \bibinfo{year}{2020}\natexlab{}.
\newblock \showarticletitle{Interactive Weak Supervision: Learning Useful
  Heuristics for Data Labeling}.
\newblock \bibinfo{journal}{\emph{arXiv preprint arXiv:2012.06046}}
  (\bibinfo{year}{2020}).
\newblock


\bibitem[\protect\citeauthoryear{Brady}{Brady}{2017}]%
        {brady2017error}
\bibfield{author}{\bibinfo{person}{Adrian~P Brady}.}
  \bibinfo{year}{2017}\natexlab{}.
\newblock \showarticletitle{Error and discrepancy in radiology: inevitable or
  avoidable?}
\newblock \bibinfo{journal}{\emph{Insights into imaging}} \bibinfo{volume}{8},
  \bibinfo{number}{1} (\bibinfo{year}{2017}), \bibinfo{pages}{171--182}.
\newblock


\bibitem[\protect\citeauthoryear{Brophy and Lowd}{Brophy and Lowd}{2020}]%
        {brophy2020dart}
\bibfield{author}{\bibinfo{person}{Jonathan Brophy} {and}
  \bibinfo{person}{Daniel Lowd}.} \bibinfo{year}{2020}\natexlab{}.
\newblock \showarticletitle{DART: Data Addition and Removal Trees}.
\newblock \bibinfo{journal}{\emph{arXiv preprint arXiv:2009.05567}}
  (\bibinfo{year}{2020}).
\newblock


\bibitem[\protect\citeauthoryear{Chatterjee, Ramakrishnan, and
  Sarawagi}{Chatterjee et~al\mbox{.}}{2019}]%
        {chatterjee2019data}
\bibfield{author}{\bibinfo{person}{Oishik Chatterjee}, \bibinfo{person}{Ganesh
  Ramakrishnan}, {and} \bibinfo{person}{Sunita Sarawagi}.}
  \bibinfo{year}{2019}\natexlab{}.
\newblock \showarticletitle{Data Programming using Continuous and
  Quality-Guided Labeling Functions}.
\newblock \bibinfo{journal}{\emph{arXiv preprint arXiv:1911.09860}}
  (\bibinfo{year}{2019}).
\newblock


\bibitem[\protect\citeauthoryear{Das, Chaba, Wu, Gandhi, Chau, and Chu}{Das
  et~al\mbox{.}}{2020}]%
        {das2020goggles}
\bibfield{author}{\bibinfo{person}{Nilaksh Das}, \bibinfo{person}{Sanya Chaba},
  \bibinfo{person}{Renzhi Wu}, \bibinfo{person}{Sakshi Gandhi},
  \bibinfo{person}{Duen~Horng Chau}, {and} \bibinfo{person}{Xu Chu}.}
  \bibinfo{year}{2020}\natexlab{}.
\newblock \showarticletitle{GOGGLES: Automatic Image Labeling with Affinity
  Coding}. In \bibinfo{booktitle}{\emph{Proceedings of the 2020 ACM SIGMOD
  International Conference on Management of Data}}.
  \bibinfo{pages}{1717--1732}.
\newblock


\bibitem[\protect\citeauthoryear{Devlin, Chang, Lee, and Toutanova}{Devlin
  et~al\mbox{.}}{2018}]%
        {devlin2018bert}
\bibfield{author}{\bibinfo{person}{Jacob Devlin}, \bibinfo{person}{Ming-Wei
  Chang}, \bibinfo{person}{Kenton Lee}, {and} \bibinfo{person}{Kristina
  Toutanova}.} \bibinfo{year}{2018}\natexlab{}.
\newblock \showarticletitle{Bert: Pre-training of deep bidirectional
  transformers for language understanding}.
\newblock \bibinfo{journal}{\emph{arXiv preprint arXiv:1810.04805}}
  (\bibinfo{year}{2018}).
\newblock


\bibitem[\protect\citeauthoryear{Dolatshah, Teoh, Wang, and Pei}{Dolatshah
  et~al\mbox{.}}{[n.d.]}]%
        {dolatshah12cleaning}
\bibfield{author}{\bibinfo{person}{Mohamad Dolatshah}, \bibinfo{person}{Mathew
  Teoh}, \bibinfo{person}{Jiannan Wang}, {and} \bibinfo{person}{Jian Pei}.}
  \bibinfo{year}{[n.d.]}\natexlab{}.
\newblock \showarticletitle{Cleaning Crowdsourced Labels Using Oracles for
  Statistical Classification}.
\newblock \bibinfo{journal}{\emph{Proceedings of the VLDB Endowment}}
  \bibinfo{volume}{12}, \bibinfo{number}{4} (\bibinfo{year}{[n.\,d.]}).
\newblock


\bibitem[\protect\citeauthoryear{Ghorbani and Zou}{Ghorbani and Zou}{2019}]%
        {ghorbani2019data}
\bibfield{author}{\bibinfo{person}{Amirata Ghorbani} {and}
  \bibinfo{person}{James Zou}.} \bibinfo{year}{2019}\natexlab{}.
\newblock \showarticletitle{Data shapley: Equitable valuation of data for
  machine learning}. In \bibinfo{booktitle}{\emph{International Conference on
  Machine Learning}}. PMLR, \bibinfo{pages}{2242--2251}.
\newblock


\bibitem[\protect\citeauthoryear{Ginart, Guan, Valiant, and Zou}{Ginart
  et~al\mbox{.}}{2019}]%
        {ginart2019making}
\bibfield{author}{\bibinfo{person}{Antonio Ginart}, \bibinfo{person}{Melody~Y
  Guan}, \bibinfo{person}{Gregory Valiant}, {and} \bibinfo{person}{James Zou}.}
  \bibinfo{year}{2019}\natexlab{}.
\newblock \showarticletitle{Making ai forget you: Data deletion in machine
  learning}.
\newblock \bibinfo{journal}{\emph{arXiv preprint arXiv:1907.05012}}
  (\bibinfo{year}{2019}).
\newblock


\bibitem[\protect\citeauthoryear{Golub and Van~der Vorst}{Golub and Van~der
  Vorst}{2000}]%
        {golub2000eigenvalue}
\bibfield{author}{\bibinfo{person}{Gene~H Golub} {and} \bibinfo{person}{Henk~A
  Van~der Vorst}.} \bibinfo{year}{2000}\natexlab{}.
\newblock \showarticletitle{Eigenvalue computation in the 20th century}.
\newblock \bibinfo{journal}{\emph{J. Comput. Appl. Math.}}
  \bibinfo{volume}{123}, \bibinfo{number}{1-2} (\bibinfo{year}{2000}),
  \bibinfo{pages}{35--65}.
\newblock


\bibitem[\protect\citeauthoryear{Gulshan, Peng, Coram, Stumpe, Wu,
  Narayanaswamy, Venugopalan, Widner, Madams, Cuadros, et~al\mbox{.}}{Gulshan
  et~al\mbox{.}}{2016}]%
        {gulshan2016development}
\bibfield{author}{\bibinfo{person}{Varun Gulshan}, \bibinfo{person}{Lily Peng},
  \bibinfo{person}{Marc Coram}, \bibinfo{person}{Martin~C Stumpe},
  \bibinfo{person}{Derek Wu}, \bibinfo{person}{Arunachalam Narayanaswamy},
  \bibinfo{person}{Subhashini Venugopalan}, \bibinfo{person}{Kasumi Widner},
  \bibinfo{person}{Tom Madams}, \bibinfo{person}{Jorge Cuadros},
  {et~al\mbox{.}}} \bibinfo{year}{2016}\natexlab{}.
\newblock \showarticletitle{Development and validation of a deep learning
  algorithm for detection of diabetic retinopathy in retinal fundus
  photographs}.
\newblock \bibinfo{journal}{\emph{Jama}} \bibinfo{volume}{316},
  \bibinfo{number}{22} (\bibinfo{year}{2016}), \bibinfo{pages}{2402--2410}.
\newblock


\bibitem[\protect\citeauthoryear{Han, Yao, Yu, Niu, Xu, Hu, Tsang, and
  Sugiyama}{Han et~al\mbox{.}}{2018}]%
        {han2018co}
\bibfield{author}{\bibinfo{person}{Bo Han}, \bibinfo{person}{Quanming Yao},
  \bibinfo{person}{Xingrui Yu}, \bibinfo{person}{Gang Niu},
  \bibinfo{person}{Miao Xu}, \bibinfo{person}{Weihua Hu}, \bibinfo{person}{Ivor
  Tsang}, {and} \bibinfo{person}{Masashi Sugiyama}.}
  \bibinfo{year}{2018}\natexlab{}.
\newblock \showarticletitle{Co-teaching: Robust training of deep neural
  networks with extremely noisy labels}. In \bibinfo{booktitle}{\emph{Advances
  in neural information processing systems}}. \bibinfo{pages}{8527--8537}.
\newblock


\bibitem[\protect\citeauthoryear{He, Zhang, Ren, and Sun}{He
  et~al\mbox{.}}{2016}]%
        {he2016deep}
\bibfield{author}{\bibinfo{person}{Kaiming He}, \bibinfo{person}{Xiangyu
  Zhang}, \bibinfo{person}{Shaoqing Ren}, {and} \bibinfo{person}{Jian Sun}.}
  \bibinfo{year}{2016}\natexlab{}.
\newblock \showarticletitle{Deep residual learning for image recognition}. In
  \bibinfo{booktitle}{\emph{Proceedings of the IEEE conference on computer
  vision and pattern recognition}}. \bibinfo{pages}{770--778}.
\newblock


\bibitem[\protect\citeauthoryear{Huang, Qu, Jia, and Zhao}{Huang
  et~al\mbox{.}}{2019}]%
        {huang2019o2u}
\bibfield{author}{\bibinfo{person}{Jinchi Huang}, \bibinfo{person}{Lie Qu},
  \bibinfo{person}{Rongfei Jia}, {and} \bibinfo{person}{Binqiang Zhao}.}
  \bibinfo{year}{2019}\natexlab{}.
\newblock \showarticletitle{O2u-net: A simple noisy label detection approach
  for deep neural networks}. In \bibinfo{booktitle}{\emph{Proceedings of the
  IEEE International Conference on Computer Vision}}.
  \bibinfo{pages}{3326--3334}.
\newblock


\bibitem[\protect\citeauthoryear{Irvin, Rajpurkar, Ko, Yu, Ciurea-Ilcus, Chute,
  Marklund, Haghgoo, Ball, Shpanskaya, et~al\mbox{.}}{Irvin
  et~al\mbox{.}}{2019}]%
        {irvin2019chexpert}
\bibfield{author}{\bibinfo{person}{Jeremy Irvin}, \bibinfo{person}{Pranav
  Rajpurkar}, \bibinfo{person}{Michael Ko}, \bibinfo{person}{Yifan Yu},
  \bibinfo{person}{Silviana Ciurea-Ilcus}, \bibinfo{person}{Chris Chute},
  \bibinfo{person}{Henrik Marklund}, \bibinfo{person}{Behzad Haghgoo},
  \bibinfo{person}{Robyn Ball}, \bibinfo{person}{Katie Shpanskaya},
  {et~al\mbox{.}}} \bibinfo{year}{2019}\natexlab{}.
\newblock \showarticletitle{CheXpert: A large chest radiograph dataset with
  uncertainty labels and expert comparison}. In
  \bibinfo{booktitle}{\emph{Thirty-Third AAAI Conference on Artificial
  Intelligence}}.
\newblock


\bibitem[\protect\citeauthoryear{Jia, Dao, Wang, Hubis, Hynes, G{\"u}rel, Li,
  Zhang, Song, and Spanos}{Jia et~al\mbox{.}}{2019}]%
        {jia2019towards}
\bibfield{author}{\bibinfo{person}{Ruoxi Jia}, \bibinfo{person}{David Dao},
  \bibinfo{person}{Boxin Wang}, \bibinfo{person}{Frances~Ann Hubis},
  \bibinfo{person}{Nick Hynes}, \bibinfo{person}{Nezihe~Merve G{\"u}rel},
  \bibinfo{person}{Bo Li}, \bibinfo{person}{Ce Zhang}, \bibinfo{person}{Dawn
  Song}, {and} \bibinfo{person}{Costas~J Spanos}.}
  \bibinfo{year}{2019}\natexlab{}.
\newblock \showarticletitle{Towards efficient data valuation based on the
  shapley value}. In \bibinfo{booktitle}{\emph{The 22nd International
  Conference on Artificial Intelligence and Statistics}}. PMLR,
  \bibinfo{pages}{1167--1176}.
\newblock


\bibitem[\protect\citeauthoryear{Johnson, Pollard, Greenbaum, Lungren, Deng,
  Peng, Lu, Mark, Berkowitz, and Horng}{Johnson et~al\mbox{.}}{2019}]%
        {johnson2019mimic}
\bibfield{author}{\bibinfo{person}{Alistair~EW Johnson}, \bibinfo{person}{Tom~J
  Pollard}, \bibinfo{person}{Nathaniel~R Greenbaum}, \bibinfo{person}{Matthew~P
  Lungren}, \bibinfo{person}{Chih-ying Deng}, \bibinfo{person}{Yifan Peng},
  \bibinfo{person}{Zhiyong Lu}, \bibinfo{person}{Roger~G Mark},
  \bibinfo{person}{Seth~J Berkowitz}, {and} \bibinfo{person}{Steven Horng}.}
  \bibinfo{year}{2019}\natexlab{}.
\newblock \showarticletitle{MIMIC-CXR-JPG, a large publicly available database
  of labeled chest radiographs}.
\newblock \bibinfo{journal}{\emph{arXiv preprint arXiv:1901.07042}}
  (\bibinfo{year}{2019}).
\newblock


\bibitem[\protect\citeauthoryear{Koh and Liang}{Koh and Liang}{2017}]%
        {koh2017understanding}
\bibfield{author}{\bibinfo{person}{Pang~Wei Koh} {and} \bibinfo{person}{Percy
  Liang}.} \bibinfo{year}{2017}\natexlab{}.
\newblock \showarticletitle{Understanding black-box predictions via influence
  functions}. In \bibinfo{booktitle}{\emph{Proceedings of the 34th
  International Conference on Machine Learning-Volume 70}}.
  \bibinfo{pages}{1885--1894}.
\newblock


\bibitem[\protect\citeauthoryear{Krishnan, Wang, Wu, Franklin, and
  Goldberg}{Krishnan et~al\mbox{.}}{2016}]%
        {krishnan2016activeclean}
\bibfield{author}{\bibinfo{person}{Sanjay Krishnan}, \bibinfo{person}{Jiannan
  Wang}, \bibinfo{person}{Eugene Wu}, \bibinfo{person}{Michael~J Franklin},
  {and} \bibinfo{person}{Ken Goldberg}.} \bibinfo{year}{2016}\natexlab{}.
\newblock \showarticletitle{Activeclean: Interactive data cleaning for
  statistical modeling}.
\newblock \bibinfo{journal}{\emph{Proceedings of the VLDB Endowment}}
  \bibinfo{volume}{9}, \bibinfo{number}{12} (\bibinfo{year}{2016}),
  \bibinfo{pages}{948--959}.
\newblock


\bibitem[\protect\citeauthoryear{Le~Cun, Jackel, Boser, Denker, Graf, Guyon,
  Henderson, Howard, and Hubbard}{Le~Cun et~al\mbox{.}}{1989}]%
        {le1989handwritten}
\bibfield{author}{\bibinfo{person}{Yann Le~Cun}, \bibinfo{person}{Lionel~D
  Jackel}, \bibinfo{person}{Brian Boser}, \bibinfo{person}{John~S Denker},
  \bibinfo{person}{Henry~P Graf}, \bibinfo{person}{Isabelle Guyon},
  \bibinfo{person}{Don Henderson}, \bibinfo{person}{Richard~E Howard}, {and}
  \bibinfo{person}{William Hubbard}.} \bibinfo{year}{1989}\natexlab{}.
\newblock \showarticletitle{Handwritten digit recognition: Applications of
  neural network chips and automatic learning}.
\newblock \bibinfo{journal}{\emph{IEEE Communications Magazine}}
  \bibinfo{volume}{27}, \bibinfo{number}{11} (\bibinfo{year}{1989}),
  \bibinfo{pages}{41--46}.
\newblock


\bibitem[\protect\citeauthoryear{Leach and Sholander}{Leach and
  Sholander}{1978}]%
        {leach1978extended}
\bibfield{author}{\bibinfo{person}{EB Leach} {and} \bibinfo{person}{MC
  Sholander}.} \bibinfo{year}{1978}\natexlab{}.
\newblock \showarticletitle{Extended mean values}.
\newblock \bibinfo{journal}{\emph{The American Mathematical Monthly}}
  \bibinfo{volume}{85}, \bibinfo{number}{2} (\bibinfo{year}{1978}),
  \bibinfo{pages}{84--90}.
\newblock


\bibitem[\protect\citeauthoryear{Loni, Cheung, Riegler, Bozzon, Gottlieb, and
  Larson}{Loni et~al\mbox{.}}{2014}]%
        {loni2014fashion}
\bibfield{author}{\bibinfo{person}{Babak Loni}, \bibinfo{person}{Lei~Yen
  Cheung}, \bibinfo{person}{Michael Riegler}, \bibinfo{person}{Alessandro
  Bozzon}, \bibinfo{person}{Luke Gottlieb}, {and} \bibinfo{person}{Martha
  Larson}.} \bibinfo{year}{2014}\natexlab{}.
\newblock \showarticletitle{Fashion 10000: an enriched social image dataset for
  fashion and clothing}. In \bibinfo{booktitle}{\emph{Proceedings of the 5th
  acm multimedia systems conference}}. \bibinfo{pages}{41--46}.
\newblock


\bibitem[\protect\citeauthoryear{Mahdavi, Neutatz, Visengeriyeva, and
  Abedjan}{Mahdavi et~al\mbox{.}}{2019}]%
        {mahdavi2019towards}
\bibfield{author}{\bibinfo{person}{Mohammad Mahdavi}, \bibinfo{person}{Felix
  Neutatz}, \bibinfo{person}{Larysa Visengeriyeva}, {and}
  \bibinfo{person}{Ziawasch Abedjan}.} \bibinfo{year}{2019}\natexlab{}.
\newblock \showarticletitle{Towards automated data cleaning workflows}.
\newblock \bibinfo{journal}{\emph{Machine Learning}}  \bibinfo{volume}{15}
  (\bibinfo{year}{2019}), \bibinfo{pages}{16}.
\newblock


\bibitem[\protect\citeauthoryear{Martens}{Martens}{2010}]%
        {martens2010deep}
\bibfield{author}{\bibinfo{person}{James Martens}.}
  \bibinfo{year}{2010}\natexlab{}.
\newblock \showarticletitle{Deep learning via hessian-free optimization.}. In
  \bibinfo{booktitle}{\emph{ICML}}, Vol.~\bibinfo{volume}{27}.
  \bibinfo{pages}{735--742}.
\newblock


\bibitem[\protect\citeauthoryear{Meyer}{Meyer}{2000}]%
        {meyer2000matrix}
\bibfield{author}{\bibinfo{person}{Carl~D Meyer}.}
  \bibinfo{year}{2000}\natexlab{}.
\newblock \bibinfo{booktitle}{\emph{Matrix analysis and applied linear
  algebra}}. Vol.~\bibinfo{volume}{71}.
\newblock \bibinfo{publisher}{Siam}.
\newblock


\bibitem[\protect\citeauthoryear{Nashaat, Ghosh, Miller, and Quader}{Nashaat
  et~al\mbox{.}}{2020}]%
        {nashaat2020wesal}
\bibfield{author}{\bibinfo{person}{Mona Nashaat}, \bibinfo{person}{Aindrila
  Ghosh}, \bibinfo{person}{James Miller}, {and} \bibinfo{person}{Shaikh
  Quader}.} \bibinfo{year}{2020}\natexlab{}.
\newblock \showarticletitle{WeSAL: Applying active supervision to find
  high-quality labels at industrial scale}. In
  \bibinfo{booktitle}{\emph{Proceedings of the 53rd Hawaii International
  Conference on System Sciences}}.
\newblock


\bibitem[\protect\citeauthoryear{Nocedal}{Nocedal}{1980}]%
        {nocedal1980updating}
\bibfield{author}{\bibinfo{person}{Jorge Nocedal}.}
  \bibinfo{year}{1980}\natexlab{}.
\newblock \showarticletitle{Updating quasi-Newton matrices with limited
  storage}.
\newblock \bibinfo{journal}{\emph{Mathematics of computation}}
  \bibinfo{volume}{35}, \bibinfo{number}{151} (\bibinfo{year}{1980}),
  \bibinfo{pages}{773--782}.
\newblock


\bibitem[\protect\citeauthoryear{Paszke, Gross, Chintala, Chanan, Yang, DeVito,
  Lin, Desmaison, Antiga, and Lerer}{Paszke et~al\mbox{.}}{2017}]%
        {paszke2017automatic}
\bibfield{author}{\bibinfo{person}{Adam Paszke}, \bibinfo{person}{Sam Gross},
  \bibinfo{person}{Soumith Chintala}, \bibinfo{person}{Gregory Chanan},
  \bibinfo{person}{Edward Yang}, \bibinfo{person}{Zachary DeVito},
  \bibinfo{person}{Zeming Lin}, \bibinfo{person}{Alban Desmaison},
  \bibinfo{person}{Luca Antiga}, {and} \bibinfo{person}{Adam Lerer}.}
  \bibinfo{year}{2017}\natexlab{}.
\newblock \showarticletitle{Automatic differentiation in pytorch}.
\newblock  (\bibinfo{year}{2017}).
\newblock


\bibitem[\protect\citeauthoryear{Raghu, Zhang, Kleinberg, and Bengio}{Raghu
  et~al\mbox{.}}{2019}]%
        {raghu2019transfusion}
\bibfield{author}{\bibinfo{person}{Maithra Raghu}, \bibinfo{person}{Chiyuan
  Zhang}, \bibinfo{person}{Jon Kleinberg}, {and} \bibinfo{person}{Samy
  Bengio}.} \bibinfo{year}{2019}\natexlab{}.
\newblock \showarticletitle{Transfusion: Understanding transfer learning for
  medical imaging}.
\newblock \bibinfo{journal}{\emph{arXiv preprint arXiv:1902.07208}}
  (\bibinfo{year}{2019}).
\newblock


\bibitem[\protect\citeauthoryear{Ratner, Bach, Ehrenberg, Fries, Wu, and
  R{\'e}}{Ratner et~al\mbox{.}}{2017}]%
        {ratner2017snorkel}
\bibfield{author}{\bibinfo{person}{Alexander Ratner},
  \bibinfo{person}{Stephen~H Bach}, \bibinfo{person}{Henry Ehrenberg},
  \bibinfo{person}{Jason Fries}, \bibinfo{person}{Sen Wu}, {and}
  \bibinfo{person}{Christopher R{\'e}}.} \bibinfo{year}{2017}\natexlab{}.
\newblock \showarticletitle{Snorkel: Rapid training data creation with weak
  supervision}. In \bibinfo{booktitle}{\emph{Proceedings of the VLDB Endowment.
  International Conference on Very Large Data Bases}},
  Vol.~\bibinfo{volume}{11}. NIH Public Access, \bibinfo{pages}{269}.
\newblock


\bibitem[\protect\citeauthoryear{Ryz and Grest}{Ryz and Grest}{2016}]%
        {ryz2016new}
\bibfield{author}{\bibinfo{person}{Lawrence Ryz} {and} \bibinfo{person}{Lauren
  Grest}.} \bibinfo{year}{2016}\natexlab{}.
\newblock \showarticletitle{A new era in data protection}.
\newblock \bibinfo{journal}{\emph{Computer Fraud \& Security}}
  \bibinfo{volume}{2016}, \bibinfo{number}{3} (\bibinfo{year}{2016}),
  \bibinfo{pages}{18--20}.
\newblock


\bibitem[\protect\citeauthoryear{Settles}{Settles}{2009}]%
        {settles2009active}
\bibfield{author}{\bibinfo{person}{Burr Settles}.}
  \bibinfo{year}{2009}\natexlab{}.
\newblock \showarticletitle{Active learning literature survey}.
\newblock  (\bibinfo{year}{2009}).
\newblock


\bibitem[\protect\citeauthoryear{Smyth}{Smyth}{2020}]%
        {smyth2020training}
\bibfield{author}{\bibinfo{person}{L Smyth}.} \bibinfo{year}{2020}\natexlab{}.
\newblock \showarticletitle{Training-ValueNet: A new approach for label
  cleaning on weakly-supervised datasets}.
\newblock  (\bibinfo{year}{2020}).
\newblock


\bibitem[\protect\citeauthoryear{Sukhbaatar and Fergus}{Sukhbaatar and
  Fergus}{2014}]%
        {sukhbaatar2014learning}
\bibfield{author}{\bibinfo{person}{Sainbayar Sukhbaatar} {and}
  \bibinfo{person}{Rob Fergus}.} \bibinfo{year}{2014}\natexlab{}.
\newblock \showarticletitle{Learning from noisy labels with deep neural
  networks}.
\newblock \bibinfo{journal}{\emph{arXiv preprint arXiv:1406.2080}}
  \bibinfo{volume}{2}, \bibinfo{number}{3} (\bibinfo{year}{2014}),
  \bibinfo{pages}{4}.
\newblock


\bibitem[\protect\citeauthoryear{Van~der Maaten and Hinton}{Van~der Maaten and
  Hinton}{2008}]%
        {van2008visualizing}
\bibfield{author}{\bibinfo{person}{Laurens Van~der Maaten} {and}
  \bibinfo{person}{Geoffrey Hinton}.} \bibinfo{year}{2008}\natexlab{}.
\newblock \showarticletitle{Visualizing data using t-SNE.}
\newblock \bibinfo{journal}{\emph{Journal of machine learning research}}
  \bibinfo{volume}{9}, \bibinfo{number}{11} (\bibinfo{year}{2008}).
\newblock


\bibitem[\protect\citeauthoryear{Varma and R{\'e}}{Varma and R{\'e}}{2018}]%
        {varma2018snuba}
\bibfield{author}{\bibinfo{person}{Paroma Varma} {and}
  \bibinfo{person}{Christopher R{\'e}}.} \bibinfo{year}{2018}\natexlab{}.
\newblock \showarticletitle{Snuba: Automating weak supervision to label
  training data}. In \bibinfo{booktitle}{\emph{Proceedings of the VLDB
  Endowment. International Conference on Very Large Data Bases}},
  Vol.~\bibinfo{volume}{12}. NIH Public Access, \bibinfo{pages}{223}.
\newblock


\bibitem[\protect\citeauthoryear{Wu, Dobriban, and Davidson}{Wu
  et~al\mbox{.}}{2020a}]%
        {wu2020deltagrad}
\bibfield{author}{\bibinfo{person}{Yinjun Wu}, \bibinfo{person}{Edgar
  Dobriban}, {and} \bibinfo{person}{Susan Davidson}.}
  \bibinfo{year}{2020}\natexlab{a}.
\newblock \showarticletitle{DeltaGrad: Rapid retraining of machine learning
  models}. In \bibinfo{booktitle}{\emph{International Conference on Machine
  Learning}}. PMLR, \bibinfo{pages}{10355--10366}.
\newblock


\bibitem[\protect\citeauthoryear{Wu, Tannen, and Davidson}{Wu
  et~al\mbox{.}}{2020b}]%
        {wu2020priu}
\bibfield{author}{\bibinfo{person}{Yinjun Wu}, \bibinfo{person}{Val Tannen},
  {and} \bibinfo{person}{Susan~B Davidson}.} \bibinfo{year}{2020}\natexlab{b}.
\newblock \showarticletitle{PrIU: A Provenance-Based Approach for Incrementally
  Updating Regression Models}. In \bibinfo{booktitle}{\emph{Proceedings of the
  2020 ACM SIGMOD International Conference on Management of Data}}.
  \bibinfo{pages}{447--462}.
\newblock


\bibitem[\protect\citeauthoryear{Zhang, Zhu, and Wright}{Zhang
  et~al\mbox{.}}{2018}]%
        {zhang2018training}
\bibfield{author}{\bibinfo{person}{Xuezhou Zhang}, \bibinfo{person}{Xiaojin
  Zhu}, {and} \bibinfo{person}{Stephen Wright}.}
  \bibinfo{year}{2018}\natexlab{}.
\newblock \showarticletitle{Training set debugging using trusted items}. In
  \bibinfo{booktitle}{\emph{Proceedings of the AAAI Conference on Artificial
  Intelligence}}, Vol.~\bibinfo{volume}{32}.
\newblock


\end{thebibliography}
\clearpage
\renewcommand{\theequation}{S\arabic{equation}}
\renewcommand{\thefigure}{S\arabic{figure}}
\renewcommand{\bibnumfmt}[1]{[S#1]}
\setcounter{section}{0}
\renewcommand\thesection{\Alph{section}}
\onecolumn
\section{Supplementary proofs}
\subsection{Derivation of Equation \eqref{eq: our_influence}}\label{sec: infl_derive}
\begin{proof}
According to \cite{koh2017understanding}, to analyze the influence of the label changes on one training sample $\nz$ as well as re-weighting this sample, we need to consider the following objective function:
\begin{align}\label{eq: obj_function_epsilon}
\begin{split}
F_{\epsilon_1, \epsilon_2, \nz}\left(\w\right) = \frac{1}{N}[\sum\nolimits_{i=1}^{N_d} F\left(\w, \z_i\right) + \sum\nolimits_{i=1}^{N_p} \gamma F\left(\w, \nz_i\right)] + \epsilon_1 F\left(\w, \nz(\delta_y)\right) - \epsilon_2 F\left(\w, \nz\right)
\end{split}
\end{align}

in which $\nz = (\nx,\ny) \in \dirtyz = \{\nz_i\}_{i=1}^{N_p}$, $\nz(\delta_y) = (\nx, \ny + \delta_y)$, representing the $\nz$ with the cleaned label $\ny + \delta_y$, and $\epsilon_1$ and $\epsilon_2$ are two small weights. We can adjust the values of $\epsilon_1$ and $\epsilon_2$ to obtain a new objective function such that the effect of the $\nz$ is cancelled out and its cleaned version is up-weighted. To achieve this, we can set $\epsilon_1 = \frac{1}{N}$ and $\epsilon_2 = \frac{\gamma}{N}$.

Then when Equation \eqref{eq: obj_function_epsilon} is minimized, its gradient should be zero. Then by denoting its minimizer as $\hat{\w}_{\epsilon_1, \epsilon_2, \nz}$, the following equation holds:
\begin{align*}
\begin{split}
\nabla_{\w} F_{\epsilon_1, \epsilon_2, \nz}\left(\hat{\w}_{\epsilon_1, \epsilon_2, \nz}\right) = \frac{1}{N}[\sum\nolimits_{i=1}^{N_d} \nabla_{\w} F\left(\hat{\w}_{\epsilon_1, \epsilon_2, \nz}, \z_i\right) + \sum\nolimits_{i=1}^{N_p} \gamma \nabla_{\w} F\left(\hat{\w}_{\epsilon_1, \epsilon_2, \nz}, \nz_i\right)] + \epsilon_1 \nabla_{\w} F\left(\hat{\w}_{\epsilon_1, \epsilon_2, \nz}, \nz(\delta_y)\right) - \epsilon_2 \nabla_{\w} F\left(\hat{\w}_{\epsilon_1, \epsilon_2, \nz}, \nz\right) = 0
\end{split}
\end{align*}

We also denote the minimizer of $\text{argmin}_{\w} F_{0,0,\nz}\left(\w\right)$ as $\hat{\w}$, which is also the minimizer of Equation \eqref{eq: obj_function} and is derived before any training sample is cleaned. Due to the closeness of $\hat{\w}_{\epsilon_1, \epsilon_2, \nz}$\footnote{this is one implicit assumption of the influence function method} and $\hat{\w}$ as both $\epsilon_1$ and $\epsilon_2$ are near-zero values, we can then apply Taylor expansion on $\nabla_{\w} F\left(\hat{\w}_{\epsilon_1, \epsilon_2, \nz}, \epsilon_1, \epsilon_2\right)$, i.e.:
\begin{align}\label{eq: infl_taylor}
\begin{split}
    & 0 = \nabla_{\w} F\left(\hat{\w}_{\epsilon_1, \epsilon_2, \nz}, \epsilon_1, \epsilon_2\right) \approx \nabla_{\w} F\left(\hat{\w}, \epsilon_1, \epsilon_2\right) + \bH_{\epsilon_1, \epsilon_2, \nz}\left(\hat{\w}\right)(\hat{\w}_{\epsilon_1, \epsilon_2, \nz} - \hat{\w})\\
    & = \frac{1}{N}[\sum\nolimits_{i=1}^{N_d} \nabla_{\w} F\left(\hat{\w}, \z_i\right) + \sum\nolimits_{i=1}^{N_p} \gamma \nabla_{\w} F\left(\hat{\w}, \nz_i\right)] + \epsilon_1 \nabla_{\w} F\left(\hat{\w}, \nz(\delta_y)\right) - \epsilon_2 \nabla_{\w} F\left(\hat{\w}, \nz\right) + \bH_{\epsilon_1, \epsilon_2, \nz}\left(\hat{\w}\right)(\hat{\w}_{\epsilon_1, \epsilon_2, \nz} - \hat{\w}),
\end{split}
\end{align}

in which $\bH_{\epsilon_1, \epsilon_2, \nz}\left(*\right)$ denotes the Hessian matrix of $F_{\epsilon_1, \epsilon_2, \nz}\left(\w\right)$. Then by using the fact that $\frac{1}{N}[\sum\nolimits_{i=1}^{N_d} \nabla_{\w} F\left(\hat{\w}, \z_i\right) + \sum\nolimits_{i=1}^{N_p} \gamma \nabla_{\w} F\left(\hat{\w}, \nz_i\right)] = 0$ (since $\hat{\w}$ is the minimizer of $F_{0, 0, \nz}\left(\w\right)$) and $\bH_{\epsilon_1, \epsilon_2, \nz}\left(\hat{\w}\right) \approx \bH_{0, 0, \nz}\left(\hat{\w}\right) = \bH\left(\hat{\w}\right)$ (since $\epsilon_1$ and $\epsilon_2$ are near zero, recall that $\bH\left(*\right)$ is the Hessian matrix of Equation \eqref{eq: obj_function}), the formula above is derived as:
\begin{align*}
    \hat{\w}_{\epsilon_1, \epsilon_2, \nz} - \hat{\w} = -\bH_{\epsilon_1, \epsilon_2, \nz}\left(\hat{\w}\right)^{-1}[\epsilon_1 \nabla_{\w} F\left(\hat{\w}, \nz(\delta_y)\right) - \epsilon_2 \nabla_{\w} F\left(\hat{\w}, \nz\right)]
\end{align*}

Recall that $\epsilon_1 = \frac{1}{N}$ and $\epsilon_2 = \frac{\gamma}{N}$ for the purpose of cleaning the labels of $\nz$ and re-weighting it afterwards. Then the formula above is further reformulated as:
\begin{align*}
    \hat{\w}_{\frac{1}{N}, \frac{\gamma}{N}, \nz} - \hat{\w} = -\bH_{\frac{1}{N}, \frac{\gamma}{N}, \nz}\left(\hat{\w}\right)^{-1}[\frac{1}{N} \nabla_{\w} F\left(\hat{\w}, \nz(\delta_y)\right) - \frac{\gamma}{N} \nabla_{\w} F\left(\hat{\w}, \nz\right)]
\end{align*}

By further reorganizing the formula above and utilize the Cauchy mean value theorem, we can get:
\begin{align}\label{eq: infl_w_diff}
\begin{split}
    & \hat{\w}_{\frac{1}{N}, \frac{\gamma}{N}, \nz} - \hat{\w} = -\bH_{\frac{1}{N}, \frac{\gamma}{N}, \nz}\left(\hat{\w}\right)^{-1}[\frac{1}{N} \nabla_{\w} F\left(\hat{\w}, \nz(\delta_y)\right) - \frac{\gamma}{N} \nabla_{\w} F\left(\hat{\w}, \nz\right)] \\
    & = -\bH_{\frac{1}{N}, \frac{\gamma}{N}, \nz}\left(\hat{\w}\right)^{-1}[\frac{1}{N} \nabla_{\w} F\left(\hat{\w}, \nz(\delta_y)\right) - \frac{1}{N} \nabla_{\w} F\left(\hat{\w}, \nz\right) + \frac{1}{N} \nabla_{\w} F\left(\hat{\w}, \nz\right) - \frac{\gamma}{N} \nabla_{\w} F\left(\hat{\w}, \nz\right)]\\
    & = -\bH_{\frac{1}{N}, \frac{\gamma}{N}, \nz}\left(\hat{\w}\right)^{-1}[\frac{1}{N} \nabla_{\w}\nabla_y F\left(\hat{\w}, \nz\right)\delta_y + \frac{1}{N} \nabla_{\w} F\left(\hat{\w}, \nz\right) - \frac{\gamma}{N} \nabla_{\w} F\left(\hat{\w}, \nz\right)]\\
    & = -\bH_{\frac{1}{N}, \frac{\gamma}{N}, \nz}\left(\hat{\w}\right)^{-1}[\frac{1}{N} \nabla_{\w}\nabla_y F\left(\hat{\w}, \nz\right)\delta_y + \frac{1 - \gamma}{N} \nabla_{\w} F\left(\hat{\w}, \nz\right)]
\end{split}
\end{align}

Recall that the influence function is to quantify how much the loss on the validation dataset varies after $\nz$ is cleaned and re-weighted. Therefore, we can obtain this version of the influence function as:
\begin{align*}
    & \mathcal{I}_{\text{pert}}(\z,\delta_y, \gamma) = N\cdot (F(\hat{\w}_{\frac{1}{N}, \frac{\gamma}{N}, \nz},\mathcal{Z}_{\text{val}}) - F(\hat{\w},\mathcal{Z}_{\text{val}}))\\
    & \approx N\cdot (\nabla_\w F(\hat{\w},\mathcal{Z}_{\text{val}})(\hat{\w}_{\frac{1}{N}, \frac{\gamma}{N}, \nz} - \hat{\w}))\\
    & = -\nabla_\w F(\hat{\w},\mathcal{Z}_{\text{val}})^\top \bH^{-1}(\hat{\w})[\nabla_{\y}\nabla_\w F(\hat{\w},\z)\delta_{y} + (1-\gamma) \nabla_\w F(\hat{\w},\z)],
\end{align*}

\end{proof}

\subsection{Proof of Theorem \ref{theorem: bound_influence}}
\begin{proof}
Recall that $\mathcal{I}_0(\nz, \delta_y, \gamma) = \textbf{v}^\top\nabla_{\y}\nabla_\w F(\w^{(0)},\z)\delta_{y}$, then the following equation holds:
\begin{align}\label{eq: diff_term}
\begin{split}
    & (-\mathcal{I}^{(k)}_{\text{pert}}(\nz,\delta_y,\gamma) - \mathcal{I}_0(\nz, \delta_y, \gamma))\\
    & = \textbf{v}^\top
    [\nabla_{\y}\nabla_\w F(\w^{(k)},\z)\delta_{y} + (1-\lambda) \nabla_\w F(\w^{(k)},\z)]
    - \textbf{v}^\top[\nabla_{\y}\nabla_\w F(\w^{(0)},\z)\delta_{y} + (1-\lambda) \nabla_\w F(\w^{(0)},\z)]\\
    & = \underbrace{\textbf{v}^\top
    [\nabla_{\y}\nabla_\w F(\w^{(k)},\z)\delta_{y} - \nabla_{\y}\nabla_\w F(\w^{(0)},\z)\delta_{y}]}_{\diffone} + 
    (1-\lambda)\underbrace{\textbf{v}^\top[\nabla_\w F(\w^{(k)},\z) - \nabla_\w F(\w^{(0)},\z)]}_{\difftwo}
\end{split}
\end{align}

Then by plugging the definition of $\nabla_{\y}\nabla_\w F(\w^{(k)},\z)$ into the formula $\diffone$ above, we can get:
\begin{align*}
    & \diffone  = \textbf{v}^\top[-[\nabla_{\w} \log(p^{(1)}(\w^{(k)}, \x)) - \nabla_{\w} \log(p^{(1)}(\w^{(0)}, \x))]\\
    & , \dots, -[\nabla_{\w} \log(p^{(C)}(\w^{(k)}, \x)) - \nabla_{\w} \log(p^{(C)}(\w^{(0)}, \x))]] \delta_y
\end{align*}

Then by utilizing the Cauchy mean value theorem, the formula above can be rewritten as:

\begin{align*}
    &\diffone = \textbf{v}^\top[-[\nabla_{\w} \log(p^{(1)}(\w^{(k)}, \x)) - \nabla_{\w} \log(p^{(1)}(\w^{(0)}, \x))]\\
    & , \dots, -[\nabla_{\w} \log(p^{(C)}(\w^{(k)}, \x)) - \nabla_{\w} \log(p^{(C)}(\w^{(0)}, \x))]] \delta_y\\
    & = \textbf{v}^\top[\int_{0}^1 -\nabla_{\w}^2\log(p^{(1)}(\w^{(0)} + s(\w^{(k)} - \w^{(0)}), \x)) ds (\w^{(k)} - \w^{(0)})\\
    & , \dots, \int_{0}^1 -\nabla_{\w}^2\log(p^{(C)}(\w^{(0)} + s(\w^{(k)} - \w^{(0)}), \x)) ds (\w^{(k)} - \w^{(0)})] \delta_y\\
    & = \textbf{v}^\top[\bH^{(1)}(\w^{(k)}, \z) (\w^{(k)} - \w^{(0)}), \dots, \bH^{(C)}(\w^{(k)}, \z) (\w^{(k)} - \w^{(0)})] \delta_y
\end{align*}

Then by using the definition of $\delta_y$, i.e. $\delta_y = [\delta_{y,1}, \delta_{y,2},\dots, \delta_{y,C}]$, the formula above can be further derived as:
\begin{align}\label{eq: theorem_2_derive_0}
\begin{split}
    & \diffone = \sum_{j=1}^C \delta_{y,j}\textbf{v}^\top \bH^{(j)}(\w^{(k)}, \z) (\w^{(k)}-\w^{(0)}).
\end{split}
\end{align}

Note that since each $\bH^{(j)}(\w^{(k)}, \z), j=1,2,\dots, C$ is a semi-positive definite matrix for strongly convex models, it can thus be decomposed with its eigenvalues and eigenvectors, i.e.:
\begin{align*}
    \bH^{(j)}(\w^{(k)}, \z) = \sum_{s=1}^m \sigma_s \textbf{u}_s\textbf{u}_s^\top
\end{align*}

Therefore, for each summed term in Equation \eqref{eq: theorem_2_derive_0}, it can be rewritten as below by using the formula above:
\begin{align}\label{eq: single_derive_term}
\begin{split}
    & \textbf{v}^\top\bH^{(j)}(\w^{(k)}, \z)(\w^{(k)} - \w^{(0)}) = \textbf{v}^\top(\sum_{s=1}^m \sigma_s \textbf{u}_s\textbf{u}_s^\top)(\w^{(k)} - \w^{(0)}) = \sum_{s=1}^m \sigma_s \textbf{v}^\top\textbf{u}_s\textbf{u}_s^\top(\w^{(k)} - \w^{(0)})
\end{split}
\end{align}

Since $\textbf{v}^\top\textbf{u}_s$ and $\textbf{u}_s^\top(\w^{(k)} - \w^{(0)})$ are two scalars, they can be rewritten as $\textbf{u}_s^\top\textbf{v}$ and $(\w^{(k)} - \w^{(0)})^\top\textbf{u}_s$ respectively. As a result, the formula above can be rewritten as:
\begin{align}\label{eq: theorem_2_derive_1}
\begin{split}
    & \textbf{v}^\top\bH^{(j)}(\w^{(k)}, \z)(\w^{(k)} - \w^{(0)}) = \sum_{s=1}^m \sigma_s \textbf{u}_s^\top\textbf{v}(\w^{(k)} - \w^{(0)})^\top\textbf{u}_s,
\end{split}
\end{align}

Then for each summed term above, it is still a scalar. Therefore, we can also rewrite it as follows by introducing its transpose:
\begin{align}\label{eq: theorem_2_derive_2}
\begin{split}
    &\textbf{u}_s^\top\textbf{v}(\w^{(k)} - \w^{(0)})^\top\textbf{u}_s  = \frac{1}{2}\left[(\textbf{u}_s^\top\textbf{v}(\w^{(k)} - \w^{(0)})^\top\textbf{u}_s + (\textbf{u}_s^\top\textbf{v}(\w^{(k)} - \w^{(0)})^\top\textbf{u}_s)^\top\right]\\
    & = \frac{1}{2}\left[\textbf{u}_s^\top\textbf{v}(\w^{(k)} - \w^{(0)})^\top\textbf{u}_s + \textbf{u}_s^\top(\w^{(k)} - \w^{(0)})\textbf{v}^\top\textbf{u}_s\right] = \frac{1}{2}\textbf{u}_s^\top[\textbf{v}(\w^{(k)} - \w^{(0)})^\top + (\w^{(k)} - \w^{(0)})\textbf{v}^\top]\textbf{u}_s
\end{split}
\end{align}

Note that $[\textbf{v}(\w^{(k)} - \w^{(0)})^\top + (\w^{(k)} - \w^{(0)})\textbf{v}^\top]$ is a symmetric matrix, which has orthogonal eigenvectors and thus can be decomposed with its eigenvectors as follows:
\begin{align*}
    [\textbf{v}(\w^{(k)} - \w^{(0)})^\top + (\w^{(k)} - \w^{(0)})\textbf{v}^\top] = \tilde{\textbf{U}}\textbf{A}\tilde{\textbf{U}}^\top = \sum_{t=1}^m a_t \tilde{\textbf{u}}_t \tilde{\textbf{u}}_t^\top
\end{align*}

in which $a_1 \geq a_2 \geq \dots \geq a_m$ are the eigenvalues and each $\tilde{\textbf{u}}_t$ is a mutually orthogonal eigenvector. The formula above is then plugged into Equation \eqref{eq: theorem_2_derive_2}, which results in:
\begin{align*}
\begin{split}
    &\textbf{u}_s^\top\textbf{v}(\w^{(k)} - \w^{(0)})^\top\textbf{u}_s  =\frac{1}{2} \left[\textbf{u}_s^\top\textbf{v}(\w^{(k)} - \w^{(0)})^\top\textbf{u}_s + (\textbf{u}_s^\top\textbf{v}(\w^{(k)} - \w^{(0)})^\top\textbf{u}_s)^\top\right]\\
    & = \frac{1}{2}\left[\textbf{u}_s^\top\textbf{v}(\w^{(k)} - \w^{(0)})^\top\textbf{u}_s + \textbf{u}_s^\top(\w^{(k)} - \w^{(0)})\textbf{v}^\top\textbf{u}_s\right] = \frac{1}{2}\textbf{u}_s^\top[\textbf{v}(\w^{(k)} - \w^{(0)})^\top + (\w^{(k)} - \w^{(0)})\textbf{v}^\top]\textbf{u}_s\\
    & =\frac{1}{2}\textbf{u}_s^\top [\sum_{t=1}^m a_t \tilde{\textbf{u}}_t \tilde{\textbf{u}}_t^\top]\textbf{u}_s
\end{split}
\end{align*}

This formula is then plugged into Equation \eqref{eq: theorem_2_derive_1}, leading to:
\begin{align}\label{eq: theorem_2_derive_3}
\begin{split}
    & \textbf{v}^\top\bH^{(j)}(\w^{(k)}, \z)(\w^{(k)} - \w^{(0)}) = \sum_{s=1}^m \sigma_s \textbf{u}_s^\top\textbf{v}(\w^{(k)} - \w^{(0)})^\top\textbf{u}_s\\
    & = \sum_{s=1}^m \sigma_s \left[\frac{1}{2}\textbf{u}_s^\top [\sum_{t=1}^m \textbf{a}_t \tilde{\textbf{u}}_t \tilde{\textbf{u}}_t^\top]\textbf{u}_s\right] = \frac{1}{2}\sum_{s=1}^m \sum_{t=1}^m \sigma_s a_t \textbf{u}_s^\top \tilde{\textbf{u}}_t \tilde{\textbf{u}}_t^\top \textbf{u}_s\\
    & = \frac{1}{2}\sum_{s=1}^m \sum_{t=1}^m \sigma_s a_t \tilde{\textbf{u}}_t^\top \textbf{u}_s \textbf{u}_s^\top \tilde{\textbf{u}}_t = \frac{1}{2}\sum_{t=1}^m a_t \tilde{\textbf{u}}_t^\top \left[\sum_{s=1}^m \sigma_s \textbf{u}_s \textbf{u}_s^\top\right]\tilde{\textbf{u}}_t
\end{split}
\end{align}

Recall that $$\bH^{(j)}(\w^{(k)}, \z) = \sum_{s=1}^m \sigma_s \textbf{u}_s\textbf{u}_s^\top$$, which is a semi-definite positive matrix. As a result, the following inequality holds for arbitrary vector $\textbf{u}$:
\begin{align*}
\begin{split}
    &0 \leq \textbf{u}^\top\bH^{(j)}(\w^{(k)}, \z)\textbf{u}\leq \|\bH^{(j)}(\w^{(k)}, \z)\| \textbf{u}^\top \textbf{u} = \|\bH^{(j)}(\w^{(k)}, \z)\| \|\textbf{u}\|^2
\end{split}
\end{align*}

Therefore, Equation \eqref{eq: theorem_2_derive_3} can be bounded as:
\begin{align}\label{eq: theorem_2_derive_4}
    \begin{split}
    & \textbf{v}^\top\bH^{(j)}(\w^{(k)}, \z)(\w^{(k)} - \w^{(0)}) \leq \frac{1}{2}\sum_{a_t \geq 0} a_t \|\bH^{(j)}(\w^{(k)}, \z)\|\|\tilde{\textbf{u}}_t\|^2  + \frac{1}{2}\sum_{a_t < 0}0  \\
    & = \frac{1}{2}\sum_{a_t \geq 0} a_t \|\bH^{(j)}(\w^{(k)}, \z)\| = \|\bH^{(j)}(\w^{(k)}, \z)\| \frac{1}{2}\sum_{a_t \geq 0} a_t
    \end{split}
\end{align}

and:
\begin{align}\label{eq: theorem_2_derive_4_2}
    \begin{split}
    & \textbf{v}^\top\bH^{(j)}(\w^{(k)}, \z)(\w^{(k)} - \w^{(0)}) \geq \frac{1}{2}\sum_{a_t <0} a_t \|\bH^{(j)}(\w^{(k)}, \z)\|\|\tilde{\textbf{u}}_t\|^2  + \frac{1}{2}\sum_{a_t \geq 0}0  \\
    & = \frac{1}{2}\sum_{a_t < 0} a_t \|\bH^{(j)}(\w^{(k)}, \z)\| = \|\bH^{(j)}(\w^{(k)}, \z)\| \frac{1}{2}\sum_{a_t < 0} a_t
    \end{split}
\end{align}

Note that the two non-zero eigenvalues of $\textbf{v}(\w^{(k)} - \w^{(0)})^\top + (\w^{(k)} - \w^{(0)})\textbf{v}^\top$ are $\textbf{v}^\top(\w^{(k)} - \w^{(0)}) \pm \|\textbf{v}\|\|(\w^{(k)} - \w^{(0)})\|$, which correspond to the eigenvectors $\|\textbf{v}\|(\w^{(k)} - \w^{(0)}) \pm \|(\w^{(k)} - \w^{(0)})\|\textbf{v}$. For those two non-zero eigenvalues, $\textbf{v}^\top(\w^{(k)} - \w^{(0)}) + \|\textbf{v}\|\|(\w^{(k)} - \w^{(0)})\|$ is greater than 0 while $\textbf{v}^\top(\w^{(k)} - \w^{(0)}) - \|\textbf{v}\|\|(\w^{(k)} - \w^{(0)})\|$ is smaller than 0. Therefore, we can explicitly derive $\frac{1}{2}\sum_{a_t \geq 0} a_t$ and $\frac{1}{2}\sum_{a_t < 0} a_t$ as follows:
\begin{align*}
    \frac{1}{2}\sum_{a_t \geq 0} a_t = \frac{1}{2}[\textbf{v}^\top(\w^{(k)} - \w^{(0)}) + \|\textbf{v}\|\|(\w^{(k)} - \w^{(0)})\|]\\
    \frac{1}{2}\sum_{a_t < 0} a_t =\frac{1}{2} [\textbf{v}^\top(\w^{(k)} - \w^{(0)}) - \|\textbf{v}\|\|(\w^{(k)} - \w^{(0)})\|]
\end{align*}

As a result, Equation \eqref{eq: theorem_2_derive_4} and Equation \eqref{eq: theorem_2_derive_4_2} can be further bounded as:
\begin{align*}
\begin{split}
    & \textbf{v}^\top\bH^{(j)}(\w^{(k)}, \z)(\w^{(k)} - \w^{(0)}) \leq \frac{1}{2}[\textbf{v}^\top(\w^{(k)} - \w^{(0)}) + \|\textbf{v}\|\|(\w^{(k)} - \w^{(0)})\|]\|\bH^{(j)}(\w^{(k)}, \z)\|
\end{split}
\end{align*}

and:
\begin{align*}
\begin{split}
    & \textbf{v}^\top\bH^{(j)}(\w^{(k)}, \z)(\w^{(k)} - \w^{(0)}) \geq \frac{1}{2}[\textbf{v}^\top(\w^{(k)} - \w^{(0)}) - \|\textbf{v}\|\|(\w^{(k)} - \w^{(0)})\|]\|\bH^{(j)}(\w^{(k)}, \z)\|
\end{split}
\end{align*}



Based on the results above, we can then derive the upper bound of Equation \eqref{eq: theorem_2_derive_0} as follows:
\begin{align}\label{eq: diff_term_one_bound1}
    \begin{split}
    &\diffone = \sum_{j=1}^C \delta_{y,j}\textbf{v}^\top\bH^{(j)}(\w^{(k)}, \z)(\w^{(k)}-\w^{(0)})\\
    & \leq \sum_{\delta_{y,j} \geq 0} \delta_{y,j}\|\bH^{(j)}(\w^{(k)}, \z)\|[\frac{1}{2}[\textbf{v}^\top(\w^{(k)} - \w^{(0)}) + \|\textbf{v}\|\|(\w^{(k)} - \w^{(0)})\|]]\\
    & + \sum_{\delta_{y,j} < 0} \delta_{y,j}\|\bH^{(j)}(\w^{(k)}, \z)\| [\frac{1}{2}[\textbf{v}^\top(\w^{(k)} - \w^{(0)}) - \|\textbf{v}\|\|(\w^{(k)} - \w^{(0)})\|]]
    \end{split}
\end{align}

Similarly, the lower bound of Equation \eqref{eq: theorem_2_derive_0} is derived as:
\begin{align}\label{eq: diff_term_one_bound2}
    \begin{split}
        &\diffone \geq \sum_{\delta_{y,j} < 0} \delta_{y,j}\|\bH^{(j)}(\w^{(k)}, \z)\|[\frac{1}{2}[\textbf{v}^\top(\w^{(k)} - \w^{(0)}) + \|\textbf{v}\|\|(\w^{(k)} - \w^{(0)})\|]]\\
    & + \sum_{\delta_{y,j} \geq 0} \delta_{y,j}\|\bH^{(j)}(\w^{(k)}, \z)\| [\frac{1}{2}[\textbf{v}^\top(\w^{(k)} - \w^{(0)}) - \|\textbf{v}\|\|(\w^{(k)} - \w^{(0)})\|]]
    \end{split}
\end{align}

Then we move on to derive the bounds on $\difftwo$ in Equation \eqref{eq: diff_term}. As the first step, we utilize the Cauchy mean value theorem on this term as follows:
\begin{align*}
    \difftwo=\textbf{v}^\top[\nabla_\w F(\w^{(k)},\z) - \nabla_\w F(\w^{(0)},\z)] = \textbf{v}^{\top} [\int_{0}^1 \bH(\w^{(0)} + s(\w^{(k)} - \w^{(0)}), \z) ds] (\w^{(k)} - \w^{(0)})
\end{align*}

which thus follows the same form as Equation \eqref{eq: single_derive_term}. Therefore, by following the same derivation of the bounds on Equation \eqref{eq: single_derive_term}, the formula above is bounded as:
\begin{align}\label{eq: diff_term_two_bound}
\begin{split}
    &\difftwo = \textbf{v}^\top[\nabla_\w F(\w^{(k)},\z) - \nabla_\w F(\w^{(0)},\z)]\\
    & \in \left[\frac{1}{2}[\textbf{v}^\top(\w^{(k)} - \w^{(0)}) - \|\textbf{v}\|\|\w^{(k)} - \w^{(0)}\|]\|\int_{0}^1 \bH(\w^{(0)} + s(\w^{(k)} - \w^{(0)}), \z) ds\|,\right.\\
    &\left.\frac{1}{2}[\textbf{v}^\top(\w^{(k)} - \w^{(0)}) + \|\textbf{v}\|\|\w^{(k)} - \w^{(0)}\|]\|\int_{0}^1 \bH(\w^{(0)} + s(\w^{(k)} - \w^{(0)}), \z) ds\|\right]
\end{split}
\end{align}

As a consequence, by utilizing the results in Equation \eqref{eq: diff_term_one_bound1}, \eqref{eq: diff_term_one_bound2} and \eqref{eq: diff_term_two_bound}, Equation \eqref{eq: diff_term} is bounded as:
\begin{align*}
    & \mathcal{I}^{(k)}_{\text{pert}}(\nz,\delta_y,\gamma) - \mathcal{I}_0(\nz, \delta_y, \gamma)\\
    & \leq \sum_{\delta_{y,j} \geq 0} \delta_{y,j}\|\bH^{(j)}(\w^{(k)}, \z)\|[\frac{1}{2}[\textbf{v}^\top(\w^{(k)} - \w^{(0)}) + \|\textbf{v}\|\|(\w^{(k)} - \w^{(0)})\|]]\\
    & + \sum_{\delta_{y,j} < 0} \delta_{y,j}\|\bH^{(j)}(\w^{(k)}, \z)\| [\frac{1}{2}[\textbf{v}^\top(\w^{(k)} - \w^{(0)}) - \|\textbf{v}\|\|(\w^{(k)} - \w^{(0)})\|]]\\
    & + \frac{1-\gamma}{2}[\textbf{v}^\top(\w^{(k)} - \w^{(0)}) + \|\textbf{v}\|\|\w^{(k)} - \w^{(0)}\|]\|\int_{0}^1 \bH(\w^{(0)} + s(\w^{(k)} - \w^{(0)}), \z) ds\|
\end{align*}

Then by denoting $e_1 = \textbf{v}^\top(\w^{(k)} - \w^{(0)})$ and $e_2 = \|\textbf{v}\|\|(\w^{(k)} - \w^{(0)})\|$ 
, the upper bound of $\mathcal{I}^{(k)}_{\text{pert}}(\nz,\delta_y,\gamma) - \mathcal{I}_0(\nz, \delta_y, \gamma)$ can be denoted as:
\begin{align*}
    & \mathcal{I}^{(k)}_{\text{pert}}(\nz,\delta_y,\gamma) - \mathcal{I}_0(\nz, \delta_y, \gamma)\\
    &\leq \sum_{\delta_{y,j} \geq 0} \delta_{y,j}\|\bH^{(j)}(\w^{(k)}, \z)\|(e_1 + e_2) + \sum_{\delta_{y,j} < 0} \delta_{y,j}\|\bH^{(j)}(\w^{(k)}, \z)\| (e_1 - e_2) + \frac{1-\gamma}{2}(e_1 + e_2)\|\int_{0}^1 \bH(\w^{(0)} + s(\w^{(k)} - \w^{(0)}), \z) ds\| \\
    & = \sum_{j=1}^C [\delta_{y,j} e_1 +|\delta_{y,j}| e_2] \|\bH^{(j)}(\w^{(k)}, \z)\| + \frac{1-\gamma}{2}(e_1 + e_2)\|\int_{0}^1 \bH(\w^{(0)} + s(\w^{(k)} - \w^{(0)}), \z) ds\|
\end{align*}

Similarly, we can derive the lower bound of Equation \eqref{eq: diff_term}, i.e.:
\begin{align*}
    & \mathcal{I}^{(k)}_{\text{pert}}(\nz,\delta_y,\gamma) - \mathcal{I}_0(\nz, \delta_y, \gamma) \\
    & \geq \sum_{\delta_{y,j} < 0} \delta_{y,j}\|\bH^{(j)}(\w^{(k)}, \z)\|[\frac{1}{2}[\textbf{v}^\top(\w^{(k)} - \w^{(0)}) + \|\textbf{v}\|\|(\w^{(k)} - \w^{(0)})\|]]\\
    & + \sum_{\delta_{y,j} \geq 0} \delta_{y,j}\|\bH^{(j)}(\w^{(k)}, \z)\| [\frac{1}{2}[\textbf{v}^\top(\w^{(k)} - \w^{(0)}) - \|\textbf{v}\|\|(\w^{(k)} - \w^{(0)})\|]]\\
    & + \frac{1-\gamma}{2}[\textbf{v}^\top(\w^{(k)} - \w^{(0)}) - \|\textbf{v}\|\|\w^{(k)} - \w^{(0)}\|]\|\int_{0}^1 \bH(\w^{(0)} + s(\w^{(k)} - \w^{(0)}), \z) ds\|\\
    & = \sum_{\delta_{y,j} < 0} \delta_{y,j}\|\bH^{(j)}(\w^{(k)}, \z)\|(e_1 + e_2) + \sum_{\delta_{y,j} > 0} \delta_{y,j}\|\bH^{(j)}(\w^{(k)}, \z)\|(e_1 - e_2) + \frac{1-\gamma}{2}(e_1 - e_2)\|\int_{0}^1 \bH(\w^{(0)} + s(\w^{(k)} - \w^{(0)}), \z) ds\|\\
    & = \sum_{j=1}^C [\delta_{y,j} e_1 - |\delta_{y,j}|e_2]
    \|\bH^{(j)}(\w^{(k)}, \z)\| + \frac{1-\gamma}{2}(e_1 - e_2)\|\int_{0}^1 \bH(\w^{(0)} + s(\w^{(k)} - \w^{(0)}), \z) ds\|
\end{align*}

\end{proof}

\section{Intuitively explaining \deltagradg}\label{sec: increm_infl_explanations}

\begin{figure}[!t]
\includegraphics[width=0.5\columnwidth]{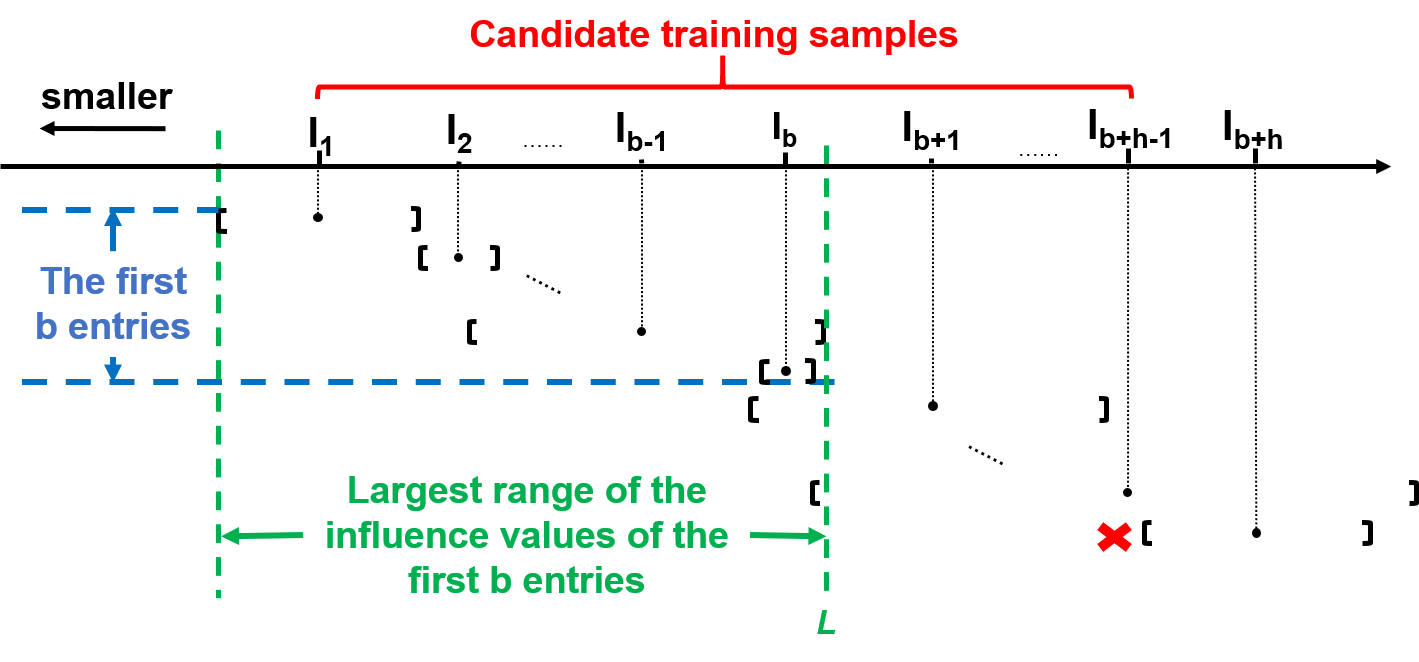}
\caption{Intuitive illustration of \deltagradg }\label{fig:deltagradg}
\end{figure}

We provided an intuitive explanation for \deltagradg\ in Figure \ref{fig:deltagradg}. In this figure, we use $\textbf{I}_1 \leq  \textbf{I}_2 \leq \textbf{I}_3 \leq \dots$ to denote the sorted list of $\mathcal{I}_0(\nz, \delta_y, \gamma)$. As described in Section \ref{sec: deltagradg}, the set of candidate influential training samples consists of two parts, one comprised of training samples producing Top-b smallest values of $\mathcal{I}_0(\nz, \delta_y, \gamma)$, i.e., the training samples generating the value $\textbf{I}_1, \textbf{I}_2, \textbf{I}_3, \dots, \textbf{I}_b$ for $\mathcal{I}_0(\nz, \delta_y, \gamma)$. The other part includes all the other training samples whose lower bound on $\mathcal{I}_0(\nz, \delta_y, \gamma)$ is smaller than the largest upper bound of the items, $\textbf{I}_1, \textbf{I}_2, \textbf{I}_3, \dots, \textbf{I}_b$. For example, in Figure \ref{fig:deltagradg}, the training samples corresponding to the value, $\textbf{I}_{b+1}, \textbf{I}_{b+2}, \textbf{I}_{b+3},\dots, \textbf{I}_{b+h-1}$ will become the candidate training samples while the sample producing value, $\textbf{I}_{b+h}$ for the term $\mathcal{I}_0(\nz, \delta_y, \gamma)$ will not be counted as the candidate influential sample. 

\section{Algorithmic details of \deltagrad}\label{sec: deltagrad_details}
The algorithmic details of \deltagrad\ are provided in Algorithm \ref{alg: deltagrad}. Note that to attain the approximate Hessian-vector product, $\B_t(\iw_t - \w_t)$, with the L-BFGS algorithm, we need to cache and reuse the last $m_0$ explicitly computed gradients (see line \ref{line: cache_grad} and line \ref{line: lbfgs} resp.), in which $m_0$ is also a hyper-parameter. 
See \cite{wu2020deltagrad} for more details.

\begin{algorithm}
\small
\begin{flushleft}
\textbf{Input: }{A training set $\mathcal{Z}$, a set of added training samples, $\mathcal{A}$, a set of deleted training samples, $\mathcal{R}$, total number of the SGD iterations, $T$, the model parameters and 
gradients cached before $\mathcal{Z}$ is updated, $\{\w_{t}\}_{t=1}^T$ and $\{\nabla F_{\w}\left(\w_{t},\miniB_t\right)\}_{t=1}^T$, and the hyper-parameters used in \deltagrad: $m_0$, $j_0$ and $T_0$}

\textbf{Output: }{Updated model parameter $\iw_{T}$}
\end{flushleft}
 \begin{algorithmic}[1]
\STATE Initialize $\iw_{0} \leftarrow \w_{0}$, $\Delta G = \left[\right]$, $\Delta W = \left[\right]$


\FOR{$t=0;t<T; t++$}{

\IF{$[((t-j_0) \mod T_0) == 0]$ or $t \leq j_0$}
{
    \STATE randomly sample a mini-batch, $\mathcal{A}_t$, from $\mathcal{A}$ \label{alg_line: r_2_t}

    \STATE explicitly compute $\nabla_{\w} F\left(\iw_{t};\miniB_t\right)$ 
    \STATE compute $\nabla_{\w} F\left(\iw_{t};(\miniB_t - \mathcal{R}) \cup  \mathcal{A}_t \right)$ by using Equation \eqref{eq: updated_gradient}
    
    
    \STATE set $\Delta G\left[r\right] = \nabla F\left(\iw_{t};\miniB_t\right) - \nabla F\left(\w_{t};\miniB_t\right)$, $\Delta W\left[r\right] = \iw_{t} - \w_{t}$\label{line: cache_grad}
    , $r\leftarrow r+1$
}
\ELSE
{
    \STATE pass 
    the last $m_0$ elements in $\Delta W$ and $\Delta G$, and 
    $\textbf{v} = \iw_{t} - \w_{t}$
    to the L-BFGFS Algorithm 
    to calculate the product,
    $\B_{t}\textbf{v}$ \label{line: lbfgs}
    
    \STATE approximate $\nabla_\w F \left(\iw_t,\miniB_{t}\right)$ by utilizing Equation \eqref{eq: sgd_lbfs}
    
    \STATE compute $\nabla_{\w} F\left(\iw_{t};(\miniB_t - \mathcal{R}) \cup  \mathcal{A}_t \right)$ by using Equation \eqref{eq: updated_gradient}
}
\ENDIF
}
\STATE update $\iw_t$ to $\iw_{t+1}$ with $\nabla_{\w} F\left(\iw_{t};(\miniB_t - \mathcal{R})\cup  \mathcal{A}_t \right)$

\ENDFOR
\STATE \textbf{Return} $\iw_{T}$
\end{algorithmic}
\caption{\deltagrad}
\label{alg: deltagrad}
\end{algorithm}

\section{Computing $\|\bH(\w^{(0)}, \nz)\|$ with the Power method}\label{sec: hessian norm}

Algorithm \ref{alg: pre_deltagrad} presents how to pre-compute $\|\bH(\w^{(0)}, \nz)\|$ in the initialization step.

\begin{algorithm}
\small
 \caption{Pre-compute $\|\bH(\w^{(0)}, \nz)\|$ in the initialization step}\label{alg: pre_deltagrad}
 \begin{spacing}{1}
 \begin{flushleft}
 \hspace*{\algorithmicindent}
 \textbf{Input:} A training sample $\z \in \mathcal{Z}$, the class $j$ and the model parameter obtained in the initialization step: $\w^{(0)}$\\
\hspace*{\algorithmicindent} \textbf{Output:} $\|\bH(\w^{(0)}, \nz)\|$. 
\end{flushleft}
 \begin{algorithmic}[1]

   \STATE Initialize $\textbf{g}$ as a random vector;
   \STATE \COMMENT{\textit{Power method below}}
   \WHILE{\textbf{g} is not converged} 
    \STATE Calculate $\bH(\w^{(0)}, \nz)\textbf{g}$ by using the auto-differentiation package
    \STATE Update $\textbf{g}$: $\textbf{g}=\frac{\bH(\w^{(0)}, \nz)\textbf{g}}{\|\bH(\w^{(0)}, \nz)\textbf{g}\|}$
   \ENDWHILE 
   \STATE Calculate the largest eigenvalue of $\bH(\w^{(0)}, \nz)$ in magnitude by using $ \frac{\textbf{g}^\top\bH(\w^{(0)}, \nz)\textbf{g}}{\|\textbf{g}\|}$, which is equivalent to $\|\bH(\w^{(0)}, \nz)\|$.
    \STATE \textbf{Return} $\|\bH(\w^{(0)}, \nz)\|$.

  \end{algorithmic}
  \end{spacing}
  \end{algorithm}

Note that the algorithm above relies on the auto-differentiation package for calculating the Hessian-vector product effectively. Specifically, for a Hessian-vector product $\bH(\w^{(0)}, \nz)\textbf{g}$, it can be further rewritten as follows:
\begin{align}
    \begin{split}
        & \bH(\w^{(0)}, \nz)\textbf{g} = \nabla^2_{\w} F(\w^{(0)}, \nz)\textbf{g} = \nabla_{\w}(\nabla_{\w} F(\w^{(0)}, \nz))\textbf{g} \\
        & = \nabla_{\w}(\nabla_{\w} F(\w^{(0)}, \nz)\textbf{g})
    \end{split}
\end{align}

in which the first equality utilizes the definition of the Hessian matrix while the last equality regards the vector $\textBF{g}$ as a constant with respect to $\w$ and utilizes the chain rule in reverse. Therefore, to obtain the result of $\bH(\w^{(0)}, \nz)\textbf{g}$, we can invoke the auto-differentiation package twice. The first one is on the loss $F(\w^{(0)}, \nz)$, resulting in the first order derivative $\nabla_{\w} F(\w^{(0)}, \nz)$, while the second one is on the product $\nabla_{\w} F(\w^{(0)}, \nz)\textbf{g}$, leading to the final result of $\bH(\w^{(0)}, \nz)\textbf{g}$.

\section{Time complexity of prioritizing the most influential training samples with \deltagradg}\label{appendix-sec: deltagradg}
According to Theorem \ref{theorem: bound_influence}, evaluating the bound on $\mathcal{I}^{(k)}_{\text{pert}}(\nz,\delta_y,\gamma)$ requires four major steps, including 1) computing the Hessian-vector product, $\textbf{v}$, by employing the solution shown in Section \ref{sec: hessian norm}, which can be computed once for all training samples (suppose the time complexity of this step is $O(v)$; 2) computing $\textbf{v}[\nabla_{\y}\nabla_\w F(\w^{(0)},\z)\delta_{y} + (1-\gamma)\nabla_\w F(\w^{(0)},\z)]$ in $\mathcal{I}_0(\nz, \delta_y, \gamma)$ with two matrix-vector multiplications (recall that $\nabla_{\y}\nabla_\w F(\w^{(0)},\z)$ and $\nabla_\w F(\w^{(0)},\z)$ are pre-computed), which requires $O(C m)$ operations ($m$ is used to denote the dimension of $\w$); 3) computing $\textbf{v}(\w^{(k)} - \w^{(0)})$ and $\|\textbf{v}\|\|\w^{(k)} - \w^{(0)}\|$, which requires $O(m)$ operations; 4) computing $\sum_{r=1}^C|\delta_{y,r}|\|\bH^{(r)}(\w^{(k)}, \z)\|$ and $\sum_{r=1}^C\delta_{y,r}\|\bH^{(r)}(\w^{(k)}, \z)\|$, which requires $O(C)$ operations (recall that $\|\bH^{(r)}(\w^{(k)}, \z)\|$ is also pre-computed). Hence, the overall overhead of evaluating the bound on $\mathcal{I}^{(k)}_{\text{pert}}(\nz,\delta_y,\gamma)$ for all $N$ training samples and all possible $C$ classes is $O(v) + NC(O(Cm) + O(m) + O(C))$. Suppose after Algorithm \ref{alg: deltagradg} is invoked, $n (\ll N)$ samples become the candidate influential training samples. Then the next step is to evaluate Equation \eqref{eq: our_influence} on each of those candidate samples for each possible deterministic class. Note that the main overhead of each invocation of Equation \eqref{eq: our_influence} comes from deriving the class-wise gradient $\nabla_{\y}\nabla_\w F(\w^{(k)},\z)$ and the sample-wise gradient $\nabla_\w F(\w^{(k)},\z)$, which is supposed to have time complexity $O(\text{Grad})$. Therefore, the total time complexity of utilizing the Algorithm \ref{alg: deltagradg} first and evaluating Equation \eqref{eq: our_influence} on $n$ candidate training samples afterwards is $O(v) + NC(O(Cm) + O(m) + O(C)) + nc O(\text{Grad})$. In contrast, without utilizing Algorithm \ref{alg: deltagradg}, it is essential to evaluate Equation \eqref{eq: our_influence} on every training sample which thus requires $O(v) + NC\cdot O(\text{Grad})$ operations. Considering the fact that the time overhead of single gradient computation is much larger than $O(Cm)$, $O(m)$ or $O(C)$, then we can expect that with small $n$, \deltagradg\ can lead to significant speed-ups. 


\section{Supplementary experimental setups}
\subsection{Details of the datasets}\label{appendix_sec: dataset}
\textbf{\mimic\ dataset} is a large chest radiograph dataset containing 377,110 images, which have been partitioned into training set, validation set and test set. There are 13 binary labels for each image, corresponding to the existence of 13 different findings. Those labels are automatically extracted from the text \cite{johnson2019mimic}, thus leading to possibly undetermined labels for some finds.
In the experiments, we focused on predicting whether the finding ``Lung Opacity'' exists for each image
and only retained those training samples with determined binary labels for this finding,
eventually producing 85046 samples, 579 samples and 1628 samples in the training set, validation set and test set respectively.

\textbf{\chexpert\ dataset} is another large chest radiograph dataset consisting of 223,415 X-ray images as the training set and another 234 images as the validation set. Since the test set is not publicly available yet, we regard the original validation set as the test set and randomly selected 10\% of the training samples as the validation set. This dataset is used to predict whether each of the 14 observations exists in each X-ray image. In the experiments, we focus on predicting the existence of the observation ``Cardiomegaly'' in each image. Similar to the pre-processing operations on \mimic, we removed the training samples and the validation samples with undetermined labels (labeled as -1) for this observation, leading to 38629 samples and 4251 samples in the training set and validation set respectively. All the test samples, i.e. the original validation samples, are fully labeled, which are all retained in the experiments.

\textbf{\retina\ dataset} is an image dataset consisting of fully labeled retinal fundus photographs \cite{gulshan2016development}. The target use of this dataset is to diagnose one eye disease called Diabetic Retinopathy (DR) for each image, which is classified into 5 categories based on severity. We followed \cite{raghu2019transfusion} to predict whether an image belongs to a referable DR, which regard the label 1 and 2 as the referable one and the label 3-5 as the non-referable one. As a consequence, the original five-class classification problem is transformed into a binary classification problem. In the original version of \retina\ dataset, there are 35127 samples and 53576 samples in the training set and test. We randomly select 10\% of the training samples as the validation samples and use the rest of them as the training set in the experiments.


\textbf{\fashion\ dataset} includes 30525 images and the label of each image represents whether it is fashionable or not, annotated by three different human annotators. In addition to those labels, some text information such as the users' comments is also associated with each image. However, ground-truth labels are not available in this dataset and simulated with the labels by aggregating the human annotated labels through majority vote. For the experiments in Section \ref{sec: exp}, similar to \cleanset, we apply ResNet50 for feature transformation and run logistic regression model afterwards.

\textbf{\fact\ dataset} Each sample in \fact\ dataset is an RDF triple for representing one fact and there are over 40000 of such facts. Each such fact is labeled as true, false or ambiguous by five different human annotators. But the total number of human annotators is 57. Among all the samples, only 577 samples have ground-truth labels. In the experiments, we removed the samples with the ground-truth label ``ambiguous'' and randomly partition the remaining samples with ground-truth labels into two parts, Although there are three different labels, we ignore the label `ambiguous', meaning that we only conduct a binary classification task on this dataset. However, it is likely that the aggregated label for some uncleaned training sample becomes `ambiguous' even after we resolve the labeling conflicts between different human annotators. To deal with this, the probabilistic labels of this sample is not updated for representing the labeling uncertainties from the human annotators. 

To facilitate the feature transformation as mentioned in Section \ref{sec: exp}, we concatenate each RDF triple as one sentence and then employ the pre-trained bert-based transformer \cite{devlin2018bert} for transforming each raw text sample into a sequence of embedding vectors. To guarantee batch training on this dataset, only the last 20 embedding vectors are used. If the total number of embedding sequence for a sample is smaller than 20, we pad this sequence with zero vectors. As introduced in Section \ref{sec: exp}, to identify whether a fact is true or not, it is essential to compare this RDF triple against the associated evidence (represented by a sentence). Therefore, by following the above principle, we transform each piece of evidence into a embedding sequence and trim the length of this sequence to 20 for accelerating the training process.


\textbf{\twitter\ dataset} is comprised of $\sim$12k tweets for sentimental analysis. In other words, the classification problem on this dataset is to judge whether the expression in each tweet is positive, negative or neutral. The labels of those samples are provided by a group of 507 human annotators and each individual tweet is labeled by three different human annotators.  Among all the samples, 577 of them have ground-truth labels. Similar to \fact\ dataset, only the positive label and the negative label are employed in the experiments. Therefore, the samples taking the neutral label as the ground truth are removed. Also, if the aggregated human annotated labels on one uncleaned sample is neural, then the probabilistic label on this sample is not updated
Plus, we generate a 768-D embedding sequence by running the pre-trained bert-based transformer on each tweet and trim the length of the resulting embedding sequence to 20. When logistic regression model is used, 

The detailed statistics of the above six datasets are included in Table \ref{Table: dataset}.

\begin{table}[h]
    \centering
    \small
    \caption{Sizes of \cleanset\ and \crowdset}\label{Table: dataset}
    \vspace{-0.2cm}
    \begin{tabular}[!h]{|>{\arraybackslash}p{3cm}|>{\centering\arraybackslash}p{0.7cm}|>{\centering\arraybackslash}p{0.7cm}|>{\centering\arraybackslash}p{0.9cm}|>{\centering\arraybackslash}p{0.7cm}|>{\centering\arraybackslash}p{0.7cm}|>{\centering\arraybackslash}p{0.7cm}|} \hline
        \makecell{Dataset} &\mimic & \retina & \chexpert & \fashion & \fact & \twitter \\\hline
        Training set& 78487 & 31615 & 37882 &29031 &38176 &11606\\\hline
        Validation set& 579 & 3512 & 234 &146 & 255&37\\ \hline
        Test set& 1628& 53576 & 234 & 146& 259&37\\ \hline
        \# of samples with ground truth & 80649 & 88703 & 43114 & 29323 & 514&74\\ \hline
    \end{tabular}
\end{table}

\subsection{Hyper-parameters for model training}\label{appendix-sec: hyper}
We included all the other hyper-parameters in Table \ref{Table: hyper_params0}, which are determined through grid search. 
In addition, notice that applying \deltagrad\ or \retrain\ to update the model parameters may lead to the termination of the training process at different epochs. Therefore, for fair comparison of the running time for \deltagrad\ and \retrain, we run SGD for fixed number of epochs and record the running time of the two methods. After the training process is done. we apply the early stopping on the model parameters cached at each SGD epoch to determine the model parameters. 


Recall that for \deltagrad, to balance between the approximation error and efficiency in \deltagrad, $\nabla_\w F (\iw_t,\miniB_{t})$ is explicitly evaluated in the first $j_0$ SGD iterations and every $T_0$ SGD iterations afterwards, where $T_0$ and $j_0$ are pre-specified hyper-parameters. Also, As Algorithm \ref{alg: deltagrad} indicates, the use of L-BFGS algorithm also requires last $m_0$ explicitly evaluated gradients and model parameters as the input. In the experiments, we set up the above three hyper-parameters as, 
$m=2$, $j_0=10$ and $T_0= 10$ for all six datasets.

\begin{table}[h]
    \centering
    \small
    \caption{The hyper-parameters for each dataset}\label{Table: hyper_params0}
    \vspace{-0.2cm}
    \begin{tabular}[!h]{|>{\arraybackslash}p{3cm}|>{\centering\arraybackslash}p{0.7cm}|>{\centering\arraybackslash}p{0.7cm}|>{\centering\arraybackslash}p{0.9cm}|>{\centering\arraybackslash}p{0.7cm}|>{\centering\arraybackslash}p{0.7cm}|>{\centering\arraybackslash}p{0.7cm}|} \hline
        \makecell{Dataset} &\mimic & \retina & \chexpert & \fashion & \fact & \twitter \\\hline
        Learning rate & 0.0005 & 0.001 & 0.02 &0.05 &0.005&0.01\\\hline
        L2 regularization & 0.05 & 0.05 & 0.05 &0.001 &0.01 &0.01\\ \hline
        \# of epochs & 150& 200 & 200 & 200& 150&400\\ \hline
    \end{tabular}
\end{table}

\subsection{Adapting \duti\ to handle probabilistic labels}\label{appendix-sec: duti}

According to \cite{zhang2018training}, the original version of \duti\ is as follows:
\begin{align}\label{eq: duti}
    \begin{split}
        &\min_{\textbf{Y}' = [\y_1',\y_2',\dots, \y_n'], \hat{\w}}[\frac{1}{|\mathcal{Z}_{\text{val}}|}\sum_{\z \in \mathcal{Z}_{\text{val}}} F(\hat{\w}, \z) + \frac{1}{n}\sum_{i=1}^n F(\hat{\w}, (\x, \y_i')) + \frac{\gamma}{n}\sum_{i=1}^n (1-\y_{i,y_i}')],\\
        &\text{s.t.}~\hat{\w} = \text{argmin}_{\w} \frac{1}{n}\sum_{i=1}^n F(\w, (\x_i, \y_i'))
    \end{split}
\end{align}

which is defined on the training dataset $\mathcal{Z} = \{(\x_i,\y_i)\}_{i=1}^n $ and the validation dataset $\mathcal{Z}_{\text{val}} = \{(\x_i,\y_i)\}_{i=1}^{|\mathcal{Z}_{\text{val}}|}$. In the formula above, each $\y_i'$ is a vector of length $C$ (recall that $C$ represents the number of classes) and the term $y_{i,y_i}'$ indicates the $(y_i)_{th}$ entry in the vector $\y_i'$, which implicitly suggests that each $\y_i$ should be a deterministic label.

Note that if $y_i$ is a probabilistic label (represented by a probabilistic vector of length $C$), we cannot calculate the term $y_{i,y_i}'$. Therefore, we replace $y_i$ in $y_{i,y_i}'$ by using the index with the largest entry in $y_i$.

\section{Supplementary experiments}\label{sec: exp_supple}

\subsection{Detailed experimental results of \textbf{Exp1}}\label{sec: detailed_exp1}
The detailed experimental results of \textbf{Exp1} are included in Table \ref{Table: prediction_strategy_b_1000} and Table \ref{Table: prediction_strategy_b_100} respectively for $b=100$ and $b=10$.

\begin{table*}
    \centering
    \small
    \caption{Comparison of the model prediction performance (F1 score) after 100 training samples are cleaned ($b=100, \gamma=0.8$)}\label{Table: prediction_strategy_b_1000}
    \vspace{-0.2cm}
    \begin{tabular}[!h]{>{\arraybackslash}p{1cm}||>{\arraybackslash}p{1.5cm}|>{\centering\arraybackslash}p{1.5cm}>{\centering\arraybackslash}p{1.5cm}>{\centering\arraybackslash}p{1.5cm}>{\centering\arraybackslash}p{1.5cm}>{\centering\arraybackslash}p{1.5cm}>{\centering\arraybackslash}p{1.5cm}>{\centering\arraybackslash}p{1.5cm}} \hline
         & uncleaned & \inflo & \ac\ & \actwo\ & \ou & \cleanone\ & \cleantwo\ & \cleanthree\ \\\hline
        \mimic&0.6284$\pm$0.0012 &0.6283$\pm$0.0011&0.6286$\pm$0.0008&0.6286$\pm$0.0008&0.1850$\pm$0.0006&0.6292$\pm$0.0005&\textBF{0.6293$\pm$0.0012}&0.6293$\pm$0.0008 \\
        \retina&0.5565$\pm$0.0019 &0.5556$\pm$0.0012&0.5566$\pm$0.0029&0.5566$\pm$0.0029&0.1331$\pm$0.0012&0.5580$\pm$0.0013&\textBF{0.5582$\pm$0.0011}&0.5581$\pm$0.0009\\ 
        \chexpert&0.5244$\pm$0.0016& 0.5244$\pm$0.0033&0.5248$\pm$0.0024&0.5248$\pm$0.0024
        &0.5276$\pm$0.0012&0.5286$\pm$0.0023&\textBF{0.5297$\pm$0.0022}&0.5289$\pm$0.0022 \\ 
        \fashion&0.5140$\pm$0.0142 &0.5143$\pm$0.0146&0.5145$\pm$0.0144&0.5145$\pm$0.0144&0.5148$\pm$0.0142&\textBF{0.5178$\pm$0.0132}&0.5177$\pm$0.0132&0.5177$\pm$0.0126 \\
        \fact& 0.6595$\pm$0.0017 &0.6596$\pm$0.0018&0.6598$\pm$0.0017&0.6598$\pm$0.0017&0.6599$\pm$0.0014&0.6601$\pm$0.0021&\textBF{0.6609$\pm$0.0021}&0.6603$\pm$0.0021 \\
        \twitter& 0.6485$\pm$0.0050&0.6530$\pm$0.0088&0.6540$\pm$0.0045&0.6540$\pm$0.0045&0.6481$\pm$0.0023&0.6594$\pm$0.0034&\textBF{0.6680$\pm$0.0044}&0.6594$\pm$0.0032 \\ \hline
    \end{tabular}
\end{table*}

\begin{table*}
    \centering
    \small
    \caption{Comparison of the model prediction performance (F1 score) after 100 training samples are cleaned ($b=10, \gamma=0.8$)}\label{Table: prediction_strategy_b_100}
    \vspace{-0.2cm}
    \begin{tabular}[!h]
    {c||c|ccccccc}
    \hline
         & uncleaned & \inflo & \ac\ & \actwo\ & \ou & \cleanone\ & \cleantwo\ & \cleanthree\ \\\hline
        \mimic&0.6284$\pm$0.0012 &0.6283$\pm$0.0011&0.6287$\pm$0.0005&0.6287$\pm$0.0005&0.1850$\pm$0.0008&0.6292$\pm$0.0007&\textBF{0.6293$\pm$0.0011}&0.6292$\pm$0.0008 \\
        \retina&0.5565$\pm$0.0019 &0.5556$\pm$0.0016&0.5568$\pm$0.0001&0.5568$\pm$0.0016&0.1314$\pm$0.0006&0.5579$\pm$0.0013&\textBF{0.5582$\pm$0.0003}&0.5581$\pm$0.0018\\ 
        \chexpert&0.5244$\pm$0.0016& 0.5246$\pm$0.0036&0.5246$\pm$0.0020&0.5246$\pm$0.0020&0.5281$\pm$0.0016&0.5287$\pm$0.0024&\textBF{0.5300$\pm$0.0024}&0.5291$\pm$0.0023 \\ 
        \fashion&0.5140$\pm$0.0142 &0.5143$\pm$0.0144&0.5145$\pm$0.0135&0.5145$\pm$0.0135&0.5152$\pm$0.0143
        &0.5178$\pm$0.0125&\textBF{0.5181$\pm$0.0131}&0.5180$\pm$0.0128 \\
        \fact& 0.6595$\pm$0.0017 &0.6596$\pm$0.0018&0.6600$\pm$0.0017&0.6600$\pm$0.0017&0.6598$\pm$0.0010&0.6601$\pm$0.0019&\textBF{0.6609$\pm$0.0020}&0.6602$\pm$0.0022 \\
        \twitter& 0.6485$\pm$0.0050&0.6518$\pm$0.0081&0.6515$\pm$0.0082&0.6515$\pm$0.0082&0.6490$\pm$0.0067&0.6578$\pm$0.0039&\textBF{0.6697$\pm$0.0058}&0.6586$\pm$0.0032 \\ \hline
    \end{tabular}
\end{table*}

\subsection{Comparing \infl\ against baseline methods with neural network models}\label{sec: exp_cnn_model}

In this section, we conduct some initial experiments when neural network models are used in the {\em Model constructor}. To goal is to compare \infl\ (with different strategies to clean labels) against all the baseline methods mentioned in Section \ref{sec: exp} (including \inflo, \ac, \actwo, and \ou) in this more general setting.
Specifically, for the image dataset, 
we applied the LeNet \cite{le1989handwritten} (a classical type of convolutional neural network structure) on the original image features (instead of features transformed by using transfer learning). For the text dataset, such as \fact\ and \twitter\ dataset, similar to Section \ref{sec: exp}, we still transform each plain-text sample into the corresponding embedding representations by using the pre-trained bert-based transformer and then applied one 1D convolutional neural network on the resulting embedding representations. We found that the performance of applying LeNet model on \fashion\ and \chexpert\ dataset is significantly worse than that when the pre-trained models are used, even when all the probabilistic labels are replaced with the ground-truth labels or the aggregated human annotated labels. Therefore, we only present the experimental results on \mimic, \retina, \fact\ and \twitter\ dataset, which are included in Table \ref{Table: strategy_compare_cnn}.

As Table \ref{Table: strategy_compare_cnn} indicates, \cleantwo\ can still achieve the best model performance for those four datasets, thus indicating the potential of applying \infl\ even when neural network model is used. Note that LeNet model is obviously less complicated than other large neural network models, such as ResNet50. Therefore, in the future, we would do more extensive experiments to evaluate the performance of \infl\ when those large neural network models are used. In addition, recall that unlike \infl, \deltagradg\ and \deltagradl\ are only applicable for strongly convex models such as logistic regression models. How to extend those two methods to handle neural network models will also be part of the future work.

\begin{table*}
    \centering
    \small
    \caption{Comparison of the model prediction performance (F1 score) after 100 training samples are cleaned (CNN)}\label{Table: strategy_compare_cnn}
    \vspace{-0.2cm}
    \begin{tabular}[!h]{>{\arraybackslash}p{1cm}||>{\arraybackslash}p{1cm}||>{\centering\arraybackslash}p{0.7cm}>{\centering\arraybackslash}p{0.7cm}>{\centering\arraybackslash}p{0.7cm}|>{\centering\arraybackslash}p{0.7cm}>{\centering\arraybackslash}p{0.7cm}>{\centering\arraybackslash}p{0.7cm}>{\centering\arraybackslash}p{0.7cm}>{\centering\arraybackslash}p{0.7cm}>{\centering\arraybackslash}p{0.7cm}>{\centering\arraybackslash}p{0.7cm}} \hline
    && \multicolumn{3}{c|}{b=100} & \multicolumn{7}{c}{b=10}\\ \hhline{~~----------}
         & uncleaned&\cleanone & \cleantwo & \cleanthree & \cleanone & \cleantwo & \cleanthree & \inflo & \ac & \actwo & \ou \\\hline
        \mimic&0.8897& 0.8895 &0.8883&0.8893&0.8893&\textBF{0.8908}&0.8904&0.8895&0.8895&0.8895&0.8873 \\
        \retina&0.5529&0.5577&0.5636&0.5602&0.5596&\textBF{0.5645}&0.5607&0.5556&0.5574&0.5574&0.5608\\ 
        \fact&0.7510&0.7536&0.8027&0.7636&0.7553&\textBF{0.8063}&0.7654&0.7560&0.7566&0.7516&0.7739 \\
        \twitter&0.6609&0.6727&0.6828&0.6783&0.6585&\textBF{0.6907}&0.6745&0.6642&0.6544&0.6610&0.6740 \\ \hline
    \end{tabular}
\end{table*}

\subsection{Comparing \infl\ against \tars}\label{sec: exp_tars}
As claimed in Section \ref{sec: exp}, similar to \infl, \tars\ \cite{dolatshah12cleaning} also targets prioritizing the most influential uncleaned training samples for cleaning. However, this method explicitly assumes that all the labels (no matter they are clean or not) are either 0 or 1 rather than probabilistic labels, thus indicating its inapplicability in the presence of the probabilistic labels. To facilitate a fair comparison between \infl\ and \tars, we round the probabilistic labels on the uncleaned training samples to the nearest deterministic labels and still regularize those samples. 

In addition, we notice that to determine the influence of each uncleaned training sample, \tars\ needs to estimate how each uncleaned label will be changed if it is to be cleaned. This depends on ``all'' the possible combinations of labels provided by ``all'' human annotators, which are thus exponential in the number of human annotators. Therefore, since the number of human annotators for \fact\ and \twitter\ dataset is not small (over 50), we only compare \infl\ against \tars\ on \cleanset\ and \fashion\ dataset. In this experiment, we still train logistic regression models on the features transformed by using the pre-trained models and use the same hyper-parameters as Section \ref{sec: exp}. In the end, we summarize the experimental results in Table \ref{Table: strategy_compare_tars_b_100}-\ref{Table: strategy_compare_tars_b_10}.


\begin{table*}
    \centering
    \small
    \caption{Comparison of the model prediction performance (F1 score) after 100 training samples are cleaned (against \tars, b=100)}\label{Table: strategy_compare_tars_b_100}
    \vspace{-0.2cm}
    \begin{tabular}[!h]{>{\arraybackslash}p{1cm}||>{\centering\arraybackslash}p{1.5cm}|>{\centering\arraybackslash}p{1.5cm}>{\centering\arraybackslash}p{1.5cm}>{\centering\arraybackslash}p{1.5cm}>{\centering\arraybackslash}p{1.5cm}>{\centering\arraybackslash}p{1.5cm}>{\centering\arraybackslash}p{1.5cm}>{\centering\arraybackslash}p{1.5cm}>{\centering\arraybackslash}p{1.5cm}} \hline
       &uncleaned  & \inflo  & \ac&\actwo & \ou& \tars & \cleanone & \cleantwo & \cleanthree  \\\hline
        \mimic&0.6413$\pm$0.0008&0.6647$\pm$0.0129&0.6569$\pm$0.0240&0.6569$\pm$0.0240&\textBF{0.6686$\pm$0.0016}&0.6022$\pm$0.0016&0.6606$\pm$0.0117&0.6375$\pm$0.0445&0.6600$\pm$0.0079 \\\hline
        \chexpert&0.5359$\pm$0.0040&0.5506$\pm$0.0015&0.5404$\pm$0.0025&0.5404$\pm$0.0025&0.5419$\pm$0.0008&0.5257$\pm$0.0031&0.5505$\pm$0.0022&\textBF{0.5671$\pm$0.0065}&0.5537$\pm$0.0037\\ \hline
        \retina&0.5702$\pm$0.0015&0.6077$\pm$0.0090&0.5910$\pm$0.0028&0.5910$\pm$0.0028&0.6021$\pm$0.0110&0.5573$\pm$0.0295&0.6015$\pm$0.0082&\textBF{0.6168$\pm$0.0063}& 0.6057$\pm$0.0076\\ \hline
        \fashion&0.6280$\pm$0.0035&0.6318$\pm$0.0047&0.6329$\pm$0.0070&0.6329$\pm$0.0070&0.6380$\pm$0.0027&0.4964$\pm$0.0022&0.6365$\pm$0.0022&\textBF{0.6460$\pm$0.0060}&0.6372$\pm$0.0026 \\ \hline
    \end{tabular}
\end{table*}

\begin{table*}
    \centering
    \small
    \caption{Comparison of the model prediction performance (F1 score) after 100 training samples are cleaned (against \tars, b=10)}\label{Table: strategy_compare_tars_b_10}
    \vspace{-0.2cm}
    \begin{tabular}[!h]{>{\arraybackslash}p{1cm}||>{\centering\arraybackslash}p{1.5cm}|>{\centering\arraybackslash}p{1.5cm}>{\centering\arraybackslash}p{1.5cm}>{\centering\arraybackslash}p{1.5cm}>{\centering\arraybackslash}p{1.5cm}>{\centering\arraybackslash}p{1.5cm}>{\centering\arraybackslash}p{1.5cm}>{\centering\arraybackslash}p{1.5cm}>{\centering\arraybackslash}p{1.5cm}} \hline
       &uncleaned  & \inflo  & \ac&\actwo & \ou& \tars & \cleanone & \cleantwo & \cleanthree  \\\hline
        \mimic&0.6413$\pm$0.0008&0.6874$\pm$0.0110&0.6815$\pm$0.0063&0.6815$\pm$0.0063&0.6698$\pm$0.0019&0.6025$\pm$0.0008&0.6867$\pm$0.0005&\textBF{0.6968$\pm$0.0220}&0.6925$\pm$0.0149 \\\hline
        \chexpert&0.5359$\pm$0.0040&0.5581$\pm$0.0119&0.5404$\pm$0.0136&0.5404$\pm$0.0136&0.5396$\pm$0.0045&0.5333$\pm$0.0029&0.5626$\pm$0.0061&\textBF{0.5863$\pm$0.0046}&0.5579$\pm$0.0128\\ \hline
        \retina&0.5702$\pm$0.0015&0.6098$\pm$0.0111&0.5989$\pm$0.0062&0.5989$\pm$0.0062&0.6026$\pm$0.0012&0.5572$\pm$0.0296&0.6136$\pm$0.0097&\textBF{0.6347$\pm$0.0034}& 0.6173$\pm$0.0070\\ \hline
        \fashion&0.6280$\pm$0.0035&0.6318$\pm$0.0047&0.6329$\pm$0.0070&0.6329$\pm$0.0070&0.6380$\pm$0.0027&0.4964$\pm$0.0022&0.6365$\pm$0.0022&\textBF{0.6460$\pm$0.0060}&0.6372$\pm$0.0026 \\ \hline
    \end{tabular}
\end{table*}

According to Table \ref{Table: strategy_compare_tars_b_100}-\ref{Table: strategy_compare_tars_b_10}, \infl\ still results in much better models than other baseline methods, including \tars. This thus demonstrates the performance advantage of \infl\ even when the uncleaned labels are all deterministic. So in comparison to \tars, \infl\ is not only suitable for more general scenarios, but also capable of producing higher-quality models in those scenarios.

\subsection{Vary the weight for the uncleaned training samples}\label{sec: exp_vary_weight}
We also repeat the \textbf{Exp1} in Section \ref{sec: exp} with varied weights on the uncleaned training samples, i.e. varied $\gamma$ in Equation \eqref{eq: obj_function}. 
Specifically, we use two different $\gamma'$s, 1 and 0. The results with $\gamma=0$ and $\gamma=1$ are included in Table \ref{Table: exp1_weight_1_b_1000}-\ref{Table: exp1_weight_1_b_100} and Table \ref{Table: exp1_weight_0_b_1000}-\ref{Table: exp1_weight_0_b_100} respectively.

First of all, when $\gamma=1$,  we can observe that either \cleanone\ or \cleantwo\ or \cleanthree\ achieves the best model performance. Since both \cleantwo\ and \cleanthree\ involve the labels suggested by \infl, it therefore again indicates those labels are reasonable.

It is also worth noting that when $\gamma$ is one where all the training samples are equally weighted. \duti\ performs worse than \infl. Based on our observations in the experiments, this phenomenon might be due to the difficulty in {\em exactly} solving the bi-optimization problem in \duti, thus producing sub-optimal selections of the influential training samples.

Plus, when $\gamma=1$, we also observe that \inflg\ performs worse than \infl. Recall that by comparing against \infl, \inflg\ quantifies the influence of each training sample without taking the magnitude of the label changes into the considerations. Since \inflg\ fails to outperform \infl, it thus justifies the necessity of explicitly considering the label changes in the influence function.

On the other hand, when $\gamma=0$, except \mimic\ and \retina, \infl\ can still beat other baseline methods, thus indicating that the potential of \infl\ when the uncleaned labels are not included in the training process. Note that for \mimic\ and \retina\ dataset, the performance of \infl\ is not ideal. One possible reason is that with $\gamma=0$, the samples with probabilistic labels are not included in the training process, meaning that only a small portion of samples (up to 100) are used for model training. 
Note that there are 100 samples cleaned in total, thus violating the small cleaning budge assumption. Plus, note that for the influence function method, due to the Taylor expansion in Equation \eqref{eq: infl_taylor}, one implicit assumption is thus the slight modification on model parameter after small amount of training samples are modified. However, we also observe that significant updates on the model parameters occur after the 100 samples are cleaned for \mimic\ and \retina\ dataset (due to the violation of the small cleaning budge assumption), thus leading to inaccurate estimate on the training sample influence. How to handle this pathological scenario will be also part of our future work.

Lastly, by comparing Table \ref{Table: exp1_weight_1_b_1000}-\ref{Table: exp1_weight_1_b_100} and Table \ref{Table: exp1_weight_0_b_1000}-\ref{Table: exp1_weight_0_b_100}, it is worth noting that with $\gamma=1$, the model performance is worse with respect to that with $\gamma=0$, thus implying the negative effect of the probabilistic labels. But as we can see, when $\gamma=1$, the strong negative effect of the probabilistic labels do not hurt the performance of \infl, thus suggesting the robustness of \infl\ when the probabilistic labels are not ideal.



\begin{table*}
    \centering
    \small
    \caption{Comparison of the model prediction performance (F1 score) after 100 training samples are cleaned ($b=100, \gamma=1$)}\label{Table: exp1_weight_1_b_1000}
    \vspace{-0.2cm}
    \begin{tabular}[!h]{>{\arraybackslash}p{1cm}||>{\arraybackslash}p{1.5cm}|>{\centering\arraybackslash}p{1.35cm}>{\centering\arraybackslash}p{1.35cm}>{\centering\arraybackslash}p{1.35cm}>{\centering\arraybackslash}p{1.35cm}>{\centering\arraybackslash}p{1.35cm}>{\centering\arraybackslash}p{1.35cm}>{\centering\arraybackslash}p{1.35cm}>{\centering\arraybackslash}p{1.35cm}>{\centering\arraybackslash}p{1.35cm}} \hline
         & uncleaned & \inflo&\inflg&\duti & \ac\ & \actwo\ & \ou & \cleanone\ & \cleantwo\ & \cleanthree\ \\\hline
        \mimic&0.6310$\pm$0.0010 &0.6310$\pm$0.0010&0.6310$\pm$0.0011&0.6310$\pm$0.0011&0.6314$\pm$0.0008&0.6314$\pm$0.0008&0.1278$\pm$0.0080&0.6320$\pm$0.0011&\textBF{0.6321$\pm$0.0011}&\textBF{0.6321$\pm$0.0010} \\
        \retina&0.5547$\pm$0.0020 &0.5543$\pm$0.0015&0.5546$\pm$0.0019&0.5558$\pm$0.0019&0.5554$\pm$0.0018&0.5554$\pm$0.0018&0.0940$\pm$0.0023&0.5564$\pm$0.0020&\textBF{0.5567$\pm$0.0021}&0.5565$\pm$0.0021\\ 
        \chexpert&0.5360$\pm$0.0123& 0.5354$\pm$0.0108&0.5360$\pm$0.0118&0.5361$\pm$0.0127&0.5366$\pm$0.0127
        &0.5366$\pm$0.0127&0.5282$\pm$0.0043&0.5403$\pm$0.0122&\textBF{0.5444$\pm$0.0080}&0.5403$\pm$0.0131 \\ 
        \fashion&0.5264$\pm$0.0078 &0.5262$\pm$0.0076&0.5265$\pm$0.0081&0.5238$\pm$0.0070&0.5267$\pm$0.0041&0.5267$\pm$0.0041
        &0.5277$\pm$0.0065&0.5297$\pm$0.0068&\textBF{0.5301$\pm$0.0053}&0.5299$\pm$0.0059 \\
        \fact& 0.6584$\pm$0.0015 &0.6587$\pm$0.0016&0.6584$\pm$0.0015&0.6582$\pm$0.0056&0.6580$\pm$0.0020&0.6580$\pm$0.0020&0.6587$\pm$0.0006&0.6586$\pm$0.0016&\textBF{0.6588$\pm$0.0016}&0.6585$\pm$0.0017 \\
        \twitter& 0.7034$\pm$0.0062&0.6853$\pm$0.0140&0.7102$\pm$0.0125&0.6230$\pm$0.0149&0.7051$\pm$0.0089&0.7051$\pm$0.0089&0.6401$\pm$0.0175&0.7164$\pm$0.0017&\textBF{0.7349$\pm$0.0290}&0.7200$\pm$0.0109 \\ \hline
    \end{tabular}
\end{table*}

\begin{table*}
    \centering
    \small
    \caption{Comparison of the model prediction performance (F1 score) after 100 training samples are cleaned ($b=10, \gamma=1$)}\label{Table: exp1_weight_1_b_100}
    \vspace{-0.2cm}
    \begin{tabular}[!h]{>{\arraybackslash}p{1cm}||>{\arraybackslash}p{1.5cm}|>{\centering\arraybackslash}p{1.5cm}>{\centering\arraybackslash}p{1.5cm}>{\centering\arraybackslash}p{1.5cm}>{\centering\arraybackslash}p{1.5cm}>{\centering\arraybackslash}p{1.5cm}>{\centering\arraybackslash}p{1.5cm}>{\centering\arraybackslash}p{1.5cm}>{\centering\arraybackslash}p{1.5cm}} \hline
         & uncleaned & \inflo& \inflg & \ac\ & \actwo\ & \ou & \cleanone\ & \cleantwo\ & \cleanthree\ \\\hline
        \mimic&0.6310$\pm$0.0010 &0.6310$\pm$0.0010&0.6310$\pm$0.0011&0.6316$\pm$0.0007&0.6316$\pm$0.0007&0.1284$\pm$0.0077&\textBF{0.6321$\pm$0.0010}&\textBF{0.6321$\pm$0.0010}&\textBF{0.6321$\pm$0.0010} \\
        \retina&0.5547$\pm$0.0020 &0.5543$\pm$0.0015&0.5547$\pm$0.0018&0.5552$\pm$0.0018&0.5552$\pm$0.0018&0.0934$\pm$0.0021&0.5564$\pm$0.0018&\textBF{0.5567$\pm$0.0019}&0.5564$\pm$0.0019\\ 
        \chexpert&0.5360$\pm$0.0123& 0.5355$\pm$0.0112&0.5359$\pm$0.0121&0.5368$\pm$0.0127&0.5368$\pm$0.0127&0.5281$\pm$0.0036&0.5402$\pm$0.0117&\textBF{0.5412$\pm$0.0120}&0.5406$\pm$0.0124 \\ 
        \fashion&0.5264$\pm$0.0078 &0.5263$\pm$0.0076&0.5265$\pm$0.0081&0.5268$\pm$0.0041&0.5268$\pm$0.0041
        &0.5275$\pm$0.0065&0.5298$\pm$0.0068&\textBF{0.5303$\pm$0.0053}&0.5300$\pm$0.0059 \\
        \fact& 0.6584$\pm$0.0015 &0.6585$\pm$0.0015&0.6585$\pm$0.0017&0.6584$\pm$0.0021&0.6584$\pm$0.0021&0.6586$\pm$0.0020&0.6585$\pm$0.0016&\textBF{0.6588$\pm$0.0016}&0.6585$\pm$0.0017 \\
        \twitter& 0.7034$\pm$0.0062&0.5854$\pm$0.0099&0.6999$\pm$0.0020&0.7030$\pm$0.0109&0.7030$\pm$0.0109&0.6268$\pm$0.0069&0.7153$\pm$0.0019&\textBF{0.7420$\pm$0.0019}&0.7171$\pm$0.0029 \\ \hline
    \end{tabular}
\end{table*}

\begin{table*}
    \centering
    \small
    \caption{Comparison of the model prediction performance (F1 score) after 100 training samples are cleaned ($b=100, \gamma=0$)}\label{Table: exp1_weight_0_b_1000}
    \vspace{-0.2cm}
    \begin{tabular}[!h]{>{\arraybackslash}p{1cm}||>{\centering\arraybackslash}p{1.5cm}||>{\centering\arraybackslash}p{1.5cm}>{\centering\arraybackslash}p{1.5cm}>{\centering\arraybackslash}p{1.5cm}>{\centering\arraybackslash}p{1.5cm}>{\centering\arraybackslash}p{1.5cm}>{\centering\arraybackslash}p{1.5cm}>{\centering\arraybackslash}p{1.5cm}} \hline
         &uncleaned & \inflo & \ac&\actwo & \ou &\cleanone & \cleantwo & \cleanthree \\\hline
        \mimic&0.6323$\pm$0.0041 & 0.8880$\pm$0.0174 & 0.8641$\pm$0.0348 &0.8641$\pm$0.0348&0.8933$\pm$0.0082&0.8740$\pm$0.0306&\textBF{0.8989$\pm$0.0008}&0.8853$\pm$0.0196 \\
        \retina&0.5614$\pm$0.0018 & 0.6728$\pm$0.0138 & \textBF{0.6898$\pm$0.0038}&\textBF{0.6898$\pm$0.0038}&0.6707$\pm$0.0340&0.6756$\pm$0.0133&0.6894$\pm$0.0048&0.6884$\pm$0.0077\\ 
        \chexpert&0.5231$\pm$0.0009& 0.5709$\pm$0.0059 & 0.5752$\pm$0.0512&0.5752$\pm$0.0512&0.5825$\pm$0.0195&0.5853$\pm$0.0238&\textBF{0.5924$\pm$0.0171}&0.5878$\pm$0.0188 \\ 
        \fashion&0.5057$\pm$0.0163 & 0.5789$\pm$0.0437 & 0.5593$\pm$0.0089&0.5593$\pm$0.0089&0.5830$\pm$0.0021&0.5905$\pm$0.0212&\textBF{0.6020$\pm$0.0431}&0.5975$\pm$0.0139\\
        \fact&0.6604$\pm$0.0020 & 0.6584$\pm$0.0221 & 0.6723$\pm$0.0185&0.6723$\pm$0.0185&0.6553$\pm$0.0232&0.6618$\pm$0.0192&\textBF{0.6691$\pm$0.0159}&0.6616$\pm$0.0221\\ 
        \twitter&0.6335$\pm$0.0301 & 0.6534$\pm$0.0248 &0.6324$\pm$0.0024&0.6324$\pm$0.0112&\textBF{0.7703$\pm$0.0210}&0.6029$\pm$0.0689&0.6739$\pm$0.0403&0.6232$\pm$0.0560\\ \hline
    \end{tabular}
\end{table*}

\begin{table*}
    \centering
    \small
    \caption{Comparison of the model prediction performance (F1 score) after 100 training samples are cleaned ($b=10, \gamma=0$)}\label{Table: exp1_weight_0_b_100}
    \vspace{-0.2cm}
    \begin{tabular}[!h]{>{\arraybackslash}p{1cm}||>{\centering\arraybackslash}p{1.5cm}||>{\centering\arraybackslash}p{1.5cm}>{\centering\arraybackslash}p{1.5cm}>{\centering\arraybackslash}p{1.5cm}>{\centering\arraybackslash}p{1.5cm}>{\centering\arraybackslash}p{1.5cm}>{\centering\arraybackslash}p{1.5cm}>{\centering\arraybackslash}p{1.5cm}} \hline
         &uncleaned & \inflo & \ac&\actwo & \ou &\cleanone & \cleantwo & \cleanthree \\\hline
        \mimic&0.6323$\pm$0.0041 & \textBF{0.9013$\pm$0.0006} & 0.9008$\pm$0.0007 &0.9007$\pm$0.0007&0.9010$\pm$0.0008&0.9007$\pm$0.0004&0.8991$\pm$0.0012&0.9008$\pm$0.0010 \\
        \retina& 0.5614$\pm$0.0018& 0.6550$\pm$0.0613 &0.6916$\pm$0.0038&0.6916$\pm$0.0038&\textBF{0.6962$\pm$0.0193}&0.6902$\pm$0.0007&0.6904$\pm$0.0254&0.6802$\pm$0.0108\\ 
        \chexpert&0.5231$\pm$0.0009& 0.5736$\pm$0.0100 & 0.5890$\pm$0.0190&0.5890$\pm$0.0190&0.5775$\pm$0.0271&0.6275$\pm$0.0329&\textBF{0.6382$\pm$0.0225}&0.6254$\pm$0.0206 \\ 
        \fashion&0.5057$\pm$0.0163 & 0.5956$\pm$0.0296 & 0.5651$\pm$0.0407&0.5651$\pm$0.0407&0.5498$\pm$0.0486&0.6449$\pm$0.0126 &\textBF{0.6926$\pm$0.0148}&0.6706$\pm$0.0174   \\
        \fact&0.6604$\pm$0.0020 & 0.6511$\pm$0.0306 & 0.6532$\pm$0.0305&0.6532$\pm$0.0305&0.6829$\pm$0.0157&0.6576$\pm$0.0283&\textBF{0.6871$\pm$0.0801}&0.6582$\pm$0.0229\\ 
        \twitter&0.6335$\pm$0.0301 & 0.6140$\pm$0.0226 & 0.6891$\pm$0.0024&0.6891$\pm$0.0086&0.7547$\pm$0.0087&0.7406$\pm$0.1798&\textBF{0.8184$\pm$0.0446}&0.7711$\pm$0.0258\\ \hline
    \end{tabular}
\end{table*}

\subsection{Vary the size of $b$}\label{sec: vary_b}

As the first step toward determining an appropriate $b$ to balance the model performance and the running time given a fixed cleaning budget, we set up the clean budget as 1000 and vary $b$ from 10 to 1000. All the other hyper-parameters are the same as that in Section \ref{sec: exp}.
The experimental results are provided in Table \ref{Table: exp1_varied_b}. As this table shows, roughly speaking, when the cleaning budget is 1000 and $b$ is 100, i.e. roughly 10\% of the cleaning budget, the model performance is close to the peak performance. After $b$ becomes even smaller, the model performance will not be significantly improved but will increase the overall running time. Therefore, to balance between the model performance and the running time, setting $b$ as the 10\% of the cleaning budget would be recommended.

\begin{table*}
    \centering
    \small
    \caption{Comparison of the model prediction performance (F1 score) with varied $b$ on \twitter\ dataset (\cleantwo)}\label{Table: exp1_varied_b}
    \vspace{-0.2cm}
    \begin{tabular}[!h]{>{\arraybackslash}p{1.2cm}||>{\centering\arraybackslash}p{1.2cm}||>{\centering\arraybackslash}p{1.2cm}>{\centering\arraybackslash}p{1.2cm}>{\centering\arraybackslash}p{1.2cm}>{\centering\arraybackslash}p{1.2cm}>{\centering\arraybackslash}p{1.2cm}>{\centering\arraybackslash}p{1.2cm}>{\centering\arraybackslash}p{1.2cm}>{\centering\arraybackslash}p{1.2cm}>{\centering\arraybackslash}p{1.2cm}} \hline
         &uncleaned &b=1000 & b=500 & b=200& b=100 & b=50 & b=20 & b=10  \\\hline
        \twitter &0.6509 & 0.8672 & 0.8932&0.8939&0.9149&0.9105&0.9046&0.9064 \\\hline
        \fashion &0.6605 & 0.6942 & 0.6993&0.6990&0.7042&0.7065&0.7089&0.7082 \\\hline
    \end{tabular}
\end{table*}

\end{document}